\documentclass[aps,prb,amsmath,superscriptaddress,amssymb,longbibliography,twocolumn]{revtex4-2}
\usepackage{amsmath}
\usepackage{graphicx}
\graphicspath{ {images/}}
\usepackage{dcolumn}
\usepackage{bm}
\usepackage[	
citecolor=magenta, colorlinks, linkcolor=magenta, urlcolor=magenta]{hyperref}
\usepackage{braket}
\usepackage{appendix}
\usepackage{wasysym}
\usepackage{slashed}
\usepackage[table, svgnames, dvipsnames]{xcolor}
\usepackage{textcomp}
\usepackage{amsmath}
\usepackage{array, multirow, bigdelim, makecell, booktabs} 
\usepackage{soul}
\usepackage{ulem}
\usepackage{graphicx}
\usepackage{subfigure}
\usepackage{caption}
\usepackage{ragged2e}
\DeclareCaptionJustification{justified}{\justifying}
\usepackage{setspace}
\DeclareUnicodeCharacter{2212}{-}
\begin{document}
	\title{Inter-chain Interactions, Multi-magnon condensation and Strain effect in chain compound NaVOPO$_4$}
	\author{Manoj Gupta$^{a}$}
	\thanks{$a$ The authors contributed equally to this work.}
	\affiliation{Department of Condensed Matter and Materials Physics,
		S. N. Bose National Centre for Basic Sciences, Kolkata 700106, India}
	\author{Manodip Routh$^{a}$}
	\affiliation{Department of Condensed Matter and Materials Physics,
		S. N. Bose National Centre for Basic Sciences, Kolkata 700106, India}
	\author{Manoranjan Kumar}
	\email{manoranjan.kumar@bose.res.in}
	\affiliation{Department of Condensed Matter and Materials Physics,
		S. N. Bose National Centre for Basic Sciences, Kolkata 700106, India}
	\author{Tanusri Saha Dasgupta}
	\email{t.sahadasgupta@gmail.com}
	\affiliation{Department of Condensed Matter and Materials Physics,
		S. N. Bose National Centre for Basic Sciences, Kolkata 700106, India}
	\date{\today}
	
	\begin{abstract}
		Employing first-principles modelling and many-body methods, the magnetic properties of spin-1/2 chain compound NaVOPO$_4$ are explored. The extensive first-principles calculations establish an intricate three-dimensionally coupled model that consists of weakly alternating
		$J$-$J^{\prime}$ antiferromagnetic chains running along cris-cross directions between two consecutive $ab$ planes, connected via two subleading couplings, a ferromagnetic exchange along the $c$ direction ($J_c$) and a weaker antiferromagnetic exchange ($J_a$) along the body diagonal direction. The exact diagonalization and density matrix renormalized group study has been carried out on a two-dimensional spin model with $J$-$J^{\prime}$-$J_c$ and effective $J_d$ couplings, constructed based on the full model, for numerical
		ease. The $J_c$-$J_d$ phase diagram is found to host a {\it disorder} phase with a finite spin gap for comparable values of $J_c$ and $J_d$, arising out of the competing nature of these two interactions, other than two ordered phases. 
		The calculated thermodynamic properties of this model provide a fair description of experimentally measured data.
		The predominant manifestation of $J_c$ and $J_d$ in the disorder phase happens in the stabilisation of a multi-magnon condensed phase,  upon gap closing by application of an external magnetic field. We further explore the effect of tensile uniaxial strain, which is found to drive the system from gapful to gapless ground state.
		
	\end{abstract}
	\maketitle
	\section{Introduction}
	
	Over the years, one-dimensional antiferromagnetic S=1/2 chain compounds have attracted the interest of researchers\textcolor{black}{. They} offer a fascinating playground to study properties like spin-gap states, spin-charge separation, and quantum criticality, dictated by the quantum fluctuation effect arising from the low dimensionality and smallness of spin \cite{Review_on_spin_chain, science.288.5465.475, sachdev2008quantum, PhysRevLett.109.177206}.  Many of these
	one-dimensional spin compounds, host a spin gap in the spin excitation spectra
	between the singlet ground state and the triplet excited state, which may arise from
	different competing exchanges. As expected, these spin chains with spin gaps are sensitive to perturbations like subleading inter-chain interactions, strain and/or external magnetic field. In particular,  the spin gap can be reduced or even closed by applying an external field above a threshold value, which leads to the emergence of a multitude of field-induced phases~\cite{PhysRevB.55.5816, PhysRevLett.101.137207, PhysRevB.79.020408, PhysRevLett.103.207203, PhysRevLett.105.137207, PhysRevB.91.060407}.
	
	\textcolor{black}{In this context, the family of vanadate compounds with V$^{4+}$ ions separated by PO$_4$ or AsO$_4$ tetrahedral units have drawn attention in recent time~\cite{PhysRevB.83.144412,PhysRevB.100.104422,PhysRevB.99.014421,PhysRevB.100.144433}. Among them, in the present study, we focus on NaVOPO$_4$ (NVOPO). Although this compound was synthesized in the early 90's \cite{+1991+67+73,benhamada1992synthese}, there is a renewed interest due to the possible realization of a spin-1/2 chain system~\cite{PhysRevB.100.144433} as well as for its potential use as cathode material~\cite{cathoda_appl}.}
	Low-temperature X-ray diffraction measurements rule out possible Peierls transition~\cite{PhysRevB.100.144433}. 
	The magnetic susceptibility data~\cite{PhysRevB.100.144433} confirms the nominal 4$^{+}$ valence of V, with S=1/2 spin, and dominance of antiferromagnetic coupling. The magnetic susceptibility data could be fitted with an alternating-chain spin model with an extremely weak alternation parameter close to 1, suggesting the compound to be at the boundary of the uniform chain and the alternating-chain model. Magnetization isotherm measurement
	~\cite{PhysRevB.100.144433} established the existence of a spin gap with the critical field of 
	gap closing $\le$ 2T leading to a possible Bose-Einstein condensed (BEC) phase~\cite{RevModPhys.86.563, PhysRevLett.84.5868, giamarchi2008bose, rice2002condense} at low temperature. The absence of long-range order in the absence of an external magnetic field was further confirmed in NMR measurements~\cite{PhysRevB.100.144433}. While
	the above observations
	support the spin-gapped ground state of NVOPO and an external field-driven possible condensed phase, 
	the detailed \textcolor{black}{analysis} of the \textcolor{black}{underlying} spin model and its implication have not been explored. The theoretical modeling reported in the same work~\cite{PhysRevB.100.144433} finds evidence of subleading inter-chain interactions\textcolor{black}{. However,} the experimental results are interpreted as a bond-alternating spin chain model with an extremely weak alternation parameter of 0.98 and negligible inter-chain couplings. \textcolor{black}{The issue of inter-chain interactions thus remains unresolved.}
	
	\begin{figure*}[t]
		\centering
		\includegraphics[width=0.8\textwidth]{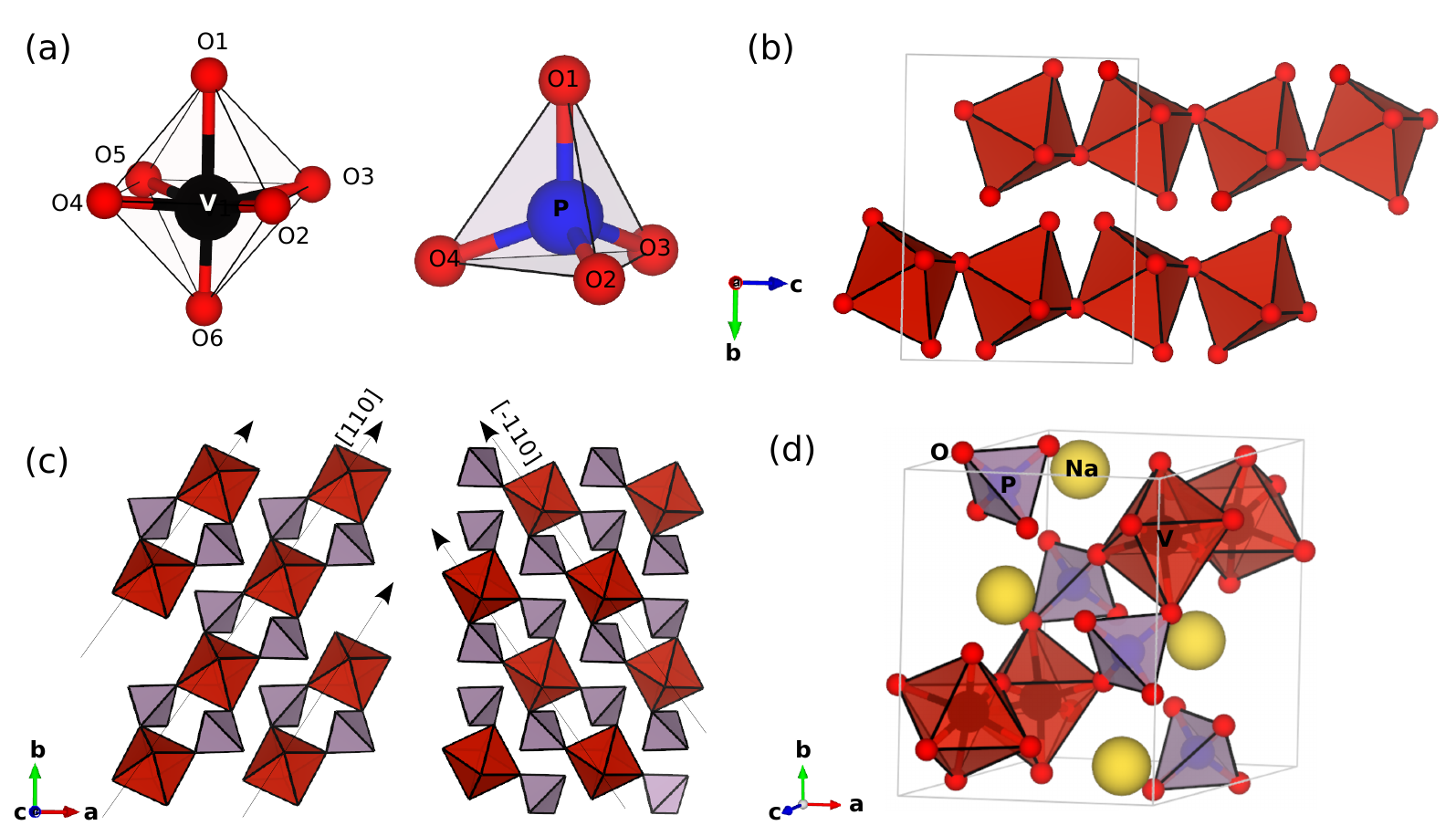}
		\caption{Crystal structure of NaVOPO$_4$.(a) Structural units of distorted VO$_6$ octahedra surrounded by six inequivalent oxygen atoms (O1-O6) and PO$_4$ square pyramid formed by O1-O4 inequivalent oxygen atoms.
			(b) Corned shared VO$_6$ octahedra forming structural chains along the crystallographic $c$ axis. The rectangular box represents the unit cell. (c) VO$_6$ octahedra connected by pair of PO$_4$ square forming VO$_6$-PO$_4$-PO$_4$-VO$_6$ chains running along [110] and [$\Bar{1}$10] in two consecutive $ab$ layers. (d) Three dimensional
			network formed by cris-cross running chains and Na atoms.}
		\label{fig: crystal structure}
	\end{figure*}
	
	\textcolor{black}{It is to be noted here,} the simple ﬁtting treatment to magnetic susceptibility data, as often adapted, makes it hard to predict details
	of the microscopic model. \textcolor{black}{Such a model would typically} involve the prediction of exchange paths and the relative magnitudes of various magnetic interactions. This in turn
	calls for the need for microscopic investigation in terms of ab initio calculations to derive the underlying spin model coupled with many-body methods for calculation of magnetic properties. Very often, the true nature of the exchange networks in these compounds differs from the mere expectation from the crystal structure.
	A famous example in this context is that of (VO)$_2$P$_2$O$_7$~\cite{PhysRevLett.79.745}, which turns out to be an alternating spin chain compound while
	originally it was thought of as a two-leg spin ladder system
	based on structural considerations.

	In this study, we thus \textcolor{black}{revisit the modelling of the compound through rigorous and extensive \textcolor{black}{Density Functional Theory (DFT)} calculations\textcolor{black}{. Our calculations are} based on the construction of three-orbital, low-energy Wannier Hamiltonian and solution of 
		total energy calculations of a large number of equations, to rule out the dependency of the derived model
		on the chosen spin configurations. This rigorous \textcolor{black}{study} established a robust nature of intra-chain couplings with an alternation ratio of 0.92 as opposed to 0.98 as predicted from susceptibility fit~\cite{PhysRevB.100.144433}. Most importantly, this \textcolor{black}{investigation} resulted in two additional subleading inter-chain interactions, the values of which are found to depend
		strongly on the chosen spin configurations.} Employing the
	exact diagonalization and density matrix renormalization group study of a
	DFT-inspired model spins Hamiltonian, we establish the important contribution
	of these sub-leading interactions in introducing \textcolor{black}{\it frustration} in the system,
	in terms of the competing nature of these interactions\textcolor{black}{. We} map out a \textcolor{black}{intriguing} phase diagram in the phase space of subleading inter-chain interactions, rationalized by large variations in their values in first-principles estimates.  
	Due to the competing nature, interestingly at the point when the two subleading terms become equal, thereby compensating each other, the spin gap value can be described by a single alternating chain model. The gap value reduces in moving away from the compensation point, finally leading to ordered states when one subleading term becomes substantial and completely dominates the other. These competing subleading interactions\textcolor{black}{, however,} manifest themselves in the excited state properties upon gap closing by the application of an external magnetic field, and a multi-magnon condensed phase\cite{Nishimoto_2012, 2015Satoshi, 2009Sudan, 2008Hikihara, 2018Parvej, PhysRevB.76.060407, PARVEJ201696, Shanon_2006_square_lattice, Shanon_2006_triangular_lattice} is observed. \textcolor{black}{\it Importantly, at the compensation point, although the gap value can be mimicked by a single alternating chain model, the same model will fail to describe the multi-magnon condensed phase.} \textcolor{black}{We compute different thermodynamic properties and compare them with experimental measurements. We demonstrate the goodness of the model with sub-leading interactions in capturing the experimental data.}
	We further explore the effect of strain. The introduction of a moderate amount of uniaxial strain is found to tune the subleading interactions, driving a strain-assisted quantum transition
	from gapful to a gapless phase. Our prediction may be verified in future experiments.
	
	\begin{figure}
		\centering
		\includegraphics[width=0.8\linewidth]{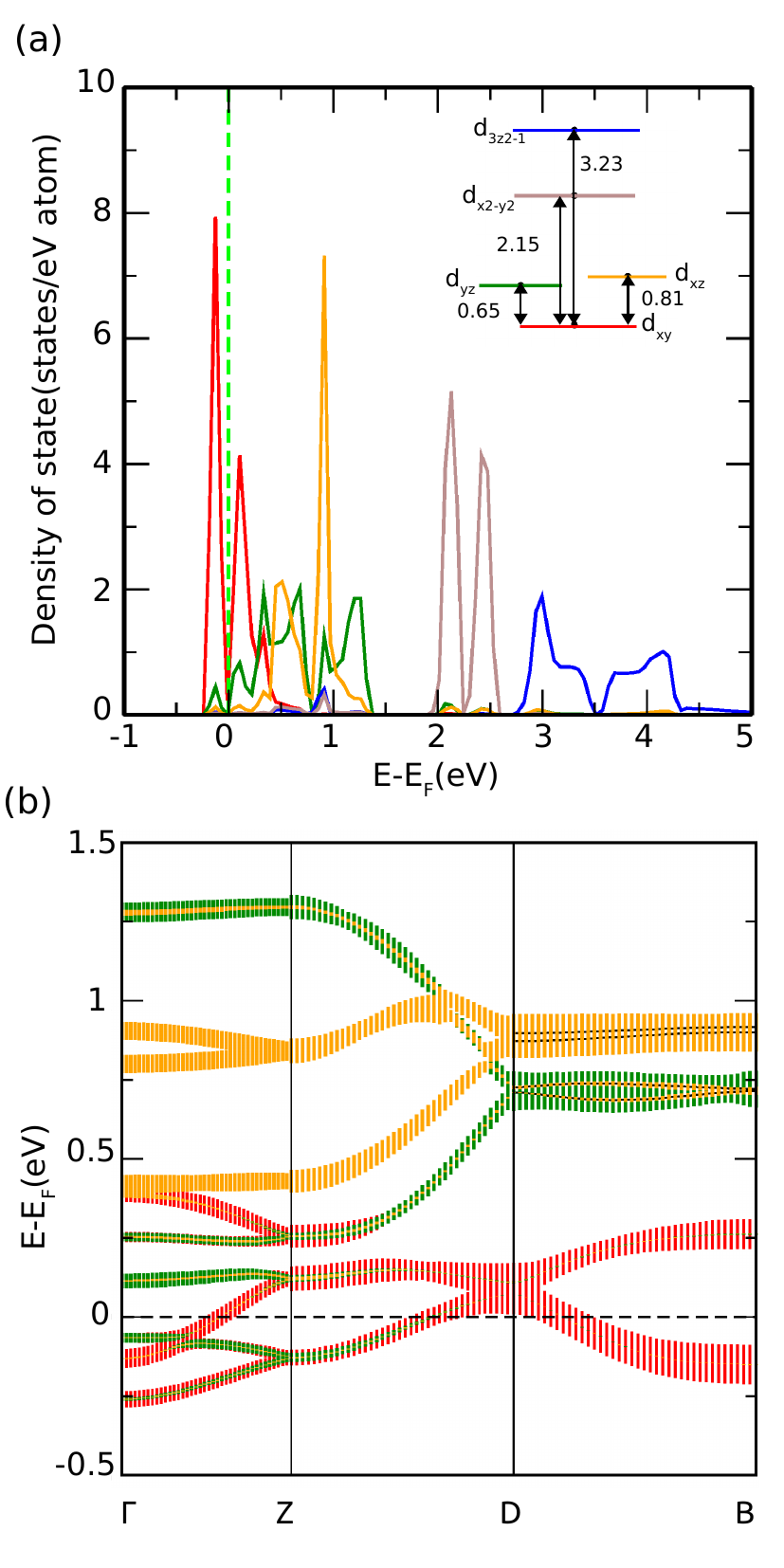}
		\caption{The non-spin-polarized GGA density of states and band-structure of NVOPO. (a) Density of states, projected to different V-$d$ characters, $d_{xy}$ (red), $d_{yz}$ (green), $d_{xz}$ (orange), $d_{x2-y2}$ (brown), and $d_{3z^2-r^2}$ (blue). Zero of the energy is set at E$_F$. Inset shows the crystal field splitting (in eV) of the V-$d$ levels. (b) Orbital projected band structure, plotted in an energy range close to E$_F$. Fatness of the bands denote the associated orbital character ($d{_{xy}}$ - red, $d_{yz}$ - green, and $d_{xz}$ -orange).}
		\label{fig:dos}
	\end{figure}

	\section{\label{method} METHOD}
	
	The first-principles DFT calculations were carried out in two different basis sets, (a) pseudo-potential, plane-wave basis and (b) muffin-tin
	orbital basis. The consistency of the two different basis set calculations has
	been cross-checked in terms of band structure and density of state plots. The 
	construction of low-energy Hamiltonian was achieved in muffin-tin orbital basis while for accurate total energy calculations, and structural relaxation, plane wave basis was used.
	
	The plane wave calculations were carried out employing 
	projector augmented-wave potential~\cite{perdew1996generalized}, as implemented in Vienna Ab-initio Simulation package~\cite{PhysRevB.50.17953,TACKETT2001348,10.1063/1.1926272}. The convergence of energies and forces \textcolor{black}{were} ensured by using a plane-wave energy cutoff of 600 eV and BZ sampling with 6 $\times$ 6 $\times$ 6 Monkhorst-Pack grids. During the structural relaxation, the ions were allowed to move until the atomic forces became lower than 0.0001 eV/Å.
	
	The Perdew-Burke-Ernzerhof generalized gradient approximation (GGA)~\cite{PhysRevLett.77.3865} was used to approximate the exchange-correlation functional. To check the influence of correlation effect at transition metal site, beyond GGA, GGA+$U$ with supplemented Hubbard $U$ correction was carried out~\cite{dudarev1998electron}.
	
	The muffin-tin orbital (MTO) basis was used in deriving low energy, few band Hamiltonian in the effective $t_{2g}$ Wannier basis of the transition metal ions. For this purpose NMTO-downfolding technique~\cite{andersen2000muffin} of integrating out degrees of freedom that are not of interest, starting from the all-orbital DFT band-structure, was employed. The self-consistent potentials required for these calculations were generated through Stuttgart implementation of Linear-MTO (LMTO) package~\cite{andersen1984explicit}. For muffin-tin orbital calculations,
	the MT radii were chosen as 1.89 $\AA$, 1.25 $\AA$, 1.161 $\AA$, and 0.86 $\AA$ for Na, V, P, and O respectively. 
	
	State of art numerical techniques of exact diagonalization~\cite{DAVIDSON197587, davidson1993monster, MURRAY1992382} and density matrix renormalization group  (DMRG)\cite{white1992density,white1993density,schollwock2005density}
	were used to solve the DFT-derived spin model. \textcolor{black}{While for small system sizes, exact diagonalization (ED) was used, for larger system sizes 
		the DMRG method of systematic truncation of irrelevant degrees of freedom and renormalization of coupling parameters for growing the system sizes was used.} 
	The accuracy in DMRG is known to decrease for systems with long bonds or multiple times renormalized operators used in the construction of Hamiltonian. To reduce the number of multiple times renormalized operators used in Hamiltonian, four new site algorithm is used. It demonstrates better convergence in the $J$- $J^{\prime}$ model with the same computational cost \cite{Mkumar2010}. In particular, in the present work, we used this modified DMRG method \cite{Mkumar2010} on a \textcolor{black}{rectangular} lattice with cylindrical geometry. Cylinders are labelled XC or YC, corresponding to open boundary conditions (OBC) along the larger lattice direction (X-direction) and periodic boundary conditions (PBC) along the smaller direction (Y-direction). We retain up to $m=700$ density matrix eigenstates during the renormalization process. We perform $\sim$ 10-12 sweeps until the ground-state energy converges within an error of $\sim$ $10^{-5}J$.
	
	\section{Crystal Structure}
	
	NaVOPO$_4$ compound crystallizes in the monoclinic structure with $P21/c$ (No. 14) space group and $Z$ = 4. \textcolor{black}{Starting from the experimentally determined crystal structure~\cite{PhysRevB.100.144433}, for accurate determination of the 
		positions of light atoms like O, we optimize the structure keeping
		the lattice parameters as well as high symmetry atomic positions fixed, under the selective dynamics scheme.} In the optimized geometry, the magnetic
	ion in the structure, V$^{4+}$ is surrounded by six inequivalent oxygen atoms (O1-O6) with four equatorial V-O bonds in the range 
	1.98–2.01 \AA, one long and one short apical bonds of lengths 2.13 \AA  ~and  1.62 \AA, respectively. \textcolor{black}{This results in} a distorted VO$_6$ octahedron
	of $C_{2h}$ symmetry, instead of $O_h$ symmetry of an ideal octahedra.  
	The P$^{5+}$ ions, on the other hand, form a nearly regular PO$_4$ tetrahedron with P–O bond lengths of $\sim1.54$ \AA (cf Fig.~\ref{fig: crystal structure}(a)). The corner-shared VO$_6$ octahedra form structural chains, running along the crystallographic $c$-axis, as shown in Fig.~\ref{fig: crystal structure}(b). 
	A pair of VO$_6$ octahedra bridged by a pair of PO$_4$ tetrahedra also form chains running along [110] and [-110] axes in two consecutive 
	$ab$ plane, as shown in Fig.~\ref{fig: crystal structure}(c). A three-dimensional cris-cross connected network is formed by corner sharing of VO$_6$ octahedra 
	and PO$_4$ tetrahedra, in which Na$^{+}$ ions sit in the hollows to
	bring cohesion in the structure, as shown in Fig.~\ref{fig: crystal structure}(d).
	
	\begin{figure*}
		\centering
		\includegraphics[width=0.85\linewidth]{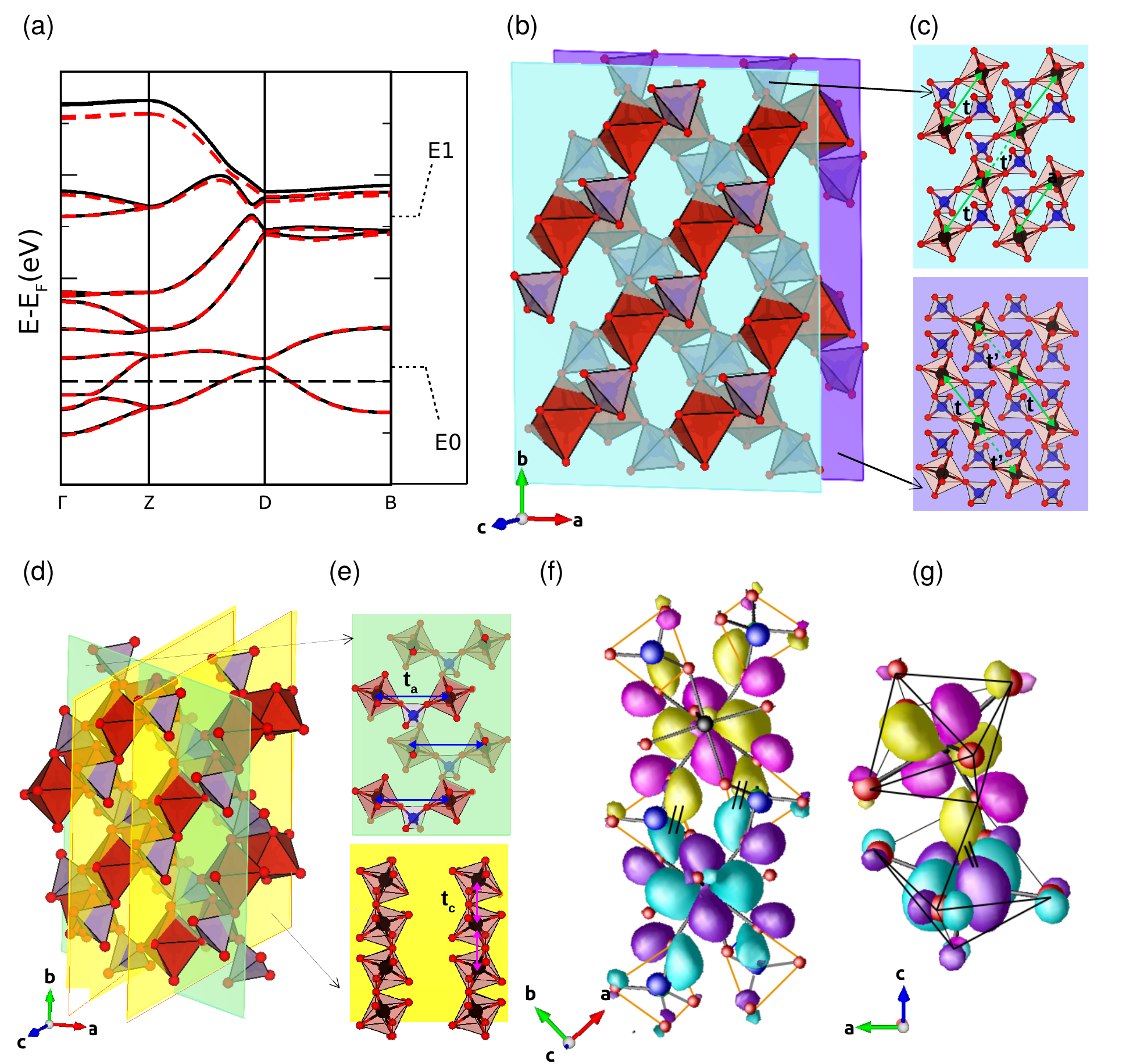}
		\caption{Low energy three-orbital model. (a) Downfolded band structure (dotted red lines) in comparison to full DFT band structure (solid black lines) plotted along the high symmetry directions of monoclinic BZ. Zero of the energy is set at E$_F$. The energy points [E0 and E1] used for energy selective NMTO-downfolding calculation have been marked.
			(b) and (c) The alternating chains of V ions, connected by hopping interactions $t$ and $t^{'}$.
			The blue and magenta shaded planes denote two successive $ab$ planes. (d) and (e) V ions
			connected by $t_c$ and $t_a$ hoppings. Different $ab$ and $ac$ planes are shaded to highlight the complex connectivity. (f) Overlap of effective Wannier functions $d_{xy}$-$d_{xy}$ placed at neighboring V sites
			connected by hopping $t$. Shown are the constant value isosurfaces with lobes of opposite signs colored as magenta and yellow for one site and blue and purple for neighbouring site. (g)
			Same as (f) but between $d_{xy}$ and $d_{xz}$ placed at V sites connected through hopping $t_c$.}
		\label{wannier_plots}
	\end{figure*}

	\section{DFT essentials}
	
	\subsection{Basic Electronic Structure}
	
	The non-spin-polarized GGA-PBE density of states (DOS) of NVOPO, projected onto V-$d_{xy}$, $d_{yz}$, $d_{xz}$, $d_{x2-y2}$ and $d_{3z^2-r^2}$ orbital characters,
	\textcolor{black}{is shown in Fig.~\ref{fig:dos}(a). The local coordinate system is chosen with local $z$-axis and $x$-axis pointing along V-O6 and approximately along V-O3 bond, respectively.}
	Following the nominal $d^{1}$ valence of V$^{4+}$ ion the \textcolor{black}{V-$d$} states are found to be mostly unfilled with 1/6 of the states being filled. The crystal
	field splitting of the V-$d$ states computed from the real-space representation of V-$d$ only Hamiltonian obtained from NMTO-downfolding calculations by integrating all other degrees of freedom is shown in the inset of Fig.~\ref{fig:dos}(a). As is seen,  the octahedral  environment pushes the $d_{x2-y2}$ and $d_{3z^2-r^2}$ levels, belonging to $e_g$ manifold 2-3 eV away from $d_{xy}$, $d_{yz}$, $d_{xz}$ levels,  forming $t_{2g}$ manifold. The octahedral distortion brings in further splitting of about 1 eV between $d_{x2-y2}$ and $d_{3z^2-r^2}$ and 0.7-0.8 eV between $d_{xy}$ and $d_{yz}/d_{xz}$ along with a tiny splitting between $d_{yz}$ and $d_{xz}$. The crystal field splitting between $d_{xy}$ and $d_{yz}/d_{xz}$ being smaller than the dispersional width associated with $d_{xy}$, $d_{yz}$ and $d_{xz}$, the $t_{2g}$ manifold consisting of 
	12 bands arising out of 4 V ions in the unit cell (cf Fig.~\ref{fig:dos}(b)) cross the Fermi level (E$_F$), separated from the $e_g$ manifold by about 0.5 eV. The metallic nature of the GGA-PBE non-spin-polarized electronic structure, contrary to the insulating
	character of the compounds suggests the inadequacies of GGA to capture the strong correlation effect at the V site. Inclusion of Hubbard correction ($U$), 
	within the GGA calculation supplemented with $U$ (GGA+$U$), pushes the $d_{xy}$ bands below E$_F$, well separated from other $t_{2g}$ bands, opening a gap
	at E$_F$. See Supplementary Materials (SM)\cite{SM}. The non-spin-polarized GGA electronic structure though
	serves as a good starting point for identifying the predominant V-V hopping
	pathways in a low-energy representation of the problem that is responsible for the magnetic exchanges.
	
	\textcolor{black}{Although the basic GGA electronic structure shows an overall similarity with the LDA electronic structure of
		one of the related compounds, AgVOAsO$_4$, studied in past literature \cite{PhysRevB.83.144412} in terms of the presence of partially filled
		low-lying $d_{xy}$ bands strongly hybridizing with higher-lying $d_{yz}$/$d_{xz}$ bands, there are subtle differences in terms of crystal field
		splittings and bandwidths, which is expected to bring in crucial differences in the spin model of the two compounds.}

	\subsection{Low-Energy Hamiltonian and effective V-V exchange pathways}

	In an attempt to derive V-$d$ only low-energy Hamiltonian 
	out of DFT calculations, we resort to the energy-selective Nth order muffin-tin orbital-based (NMTO-based) downfolding technique~\cite{andersen2000muffin}. Starting from a self-consistent DFT calculation in the linear muffin-tin orbital basis~\cite{andersen1984explicit}, an NMTO-downfolding calculation arrives at a low-energy V-$d$ Hamiltonian by integrating out degrees that are not of interest. This defines effective V 
	Wannier functions with the head of the function shaped as V-$d$ orbital characters, and tails shaped as integrated out orbitals having appreciable
	hybridization with V-$d$.
	
	To construct the low-energy model of NVOPO, we consider a three orbital model in which V-$d_{xy}$, $d_{yz}$, and $d_{xz}$ are kept active and rest all, Na-$s$, O-$p$, P-$p$ as well as V-$e_g$ degrees of freedom are integrated out. \textcolor{black}{Construction of three-orbital $t_{2g}$
		only model is justifiable by the fact that DFT estimated $t_{2g}$-$e_g$ crystal field
		of 2-3 eV, gets renormalized to 4-5 eV upon application of missing correlation effect (see SM\cite{SM}), which would weaken the $t_{2g}$-$e_g$ hybridization. A very similar approach was also adapted in Ref.~\cite{PhysRevB.83.144412} for modelling AgVOAsO$_4$.} 
	
	\textcolor{black}{We would like to stress, in contrast to DFT results of Ref.\cite{PhysRevB.100.144433} in which a $d_{xy}$-only model was proposed, we find such one orbital description does not work satisfactorily due to strong hybridization between three $t_{2g}$’s and the entangled nature of these bands.}

	The three-orbital downfolded band structure provides a faithful representation of the DFT bands, as shown in Fig.~\ref{wannier_plots}(a). The real-space
	representation of the three-orbital model leads to a 3 $\times$ 3 onsite and hopping matrices joining the two V sites with a connecting vector. 
	\textcolor{black}{The detailed results are presented in Appendix A.} The two leading hopping pathways appear to
	be V-V interactions through the PO$_4$ bridges, $t$ and $t^{'}$ forming
	alternating chains running along [110] and [-110] directions in $ab$ planes, as shown in
	Figs.~\ref{wannier_plots}(b) and (c). The next subleading interactions appear to be between
	corner-shared VO$_6$ running along [001], $t_c$ and the V-V pairs in the body diagonal positions connecting two $ab$ planes, $t_a$, as shown in Figs.~\ref{wannier_plots}(d) and (e). 
	
	To understand the dominating effect of $t$ and $t^{'}$ connecting 
	V pairs with no shared O, over $t_c$ between V's with corner shared connectivity, we plot the overlap of V effective Wannier functions, placed
	at 1NN and 3NN V sites (cf Figs.~\ref{wannier_plots}(f) and (g)). As is seen, for $t_c$, 
	due to the titled geometries of the octahedra and their corner-shared nature, the overlap of the two functions is minimal at the connecting O site. On the other hand, due to finite hybridization between O-$p$ and P-$p$, the O-$p$ like tails of Wannier functions bend and make a connection at intervening P site for V pairs, separated by PO$_4$ bridges, resulting in a well-defined hopping path for $t$. \textcolor{black}{This general aspect is found to be similar to that found for AgVOAsO$_4$~\cite{PhysRevB.83.144412}, although the details are different.}
	
	Starting from the information of the hopping integrals, a second-order perturbation theory can be employed to provide a rough estimate of the exchange-integrals ($J$), in the limit Coulomb interaction $U$ much larger than the hopping integrals. Considering the $d_{xy}$ orbitals to be majorly half-filled and $d_{xz}/d_{yz}$ orbitals to be majorly empty, leads to a) an antiferromagnetic interaction due to  the hopping of an up electron from occupied $d_{xy}$ state to neighbouring $d_{xy}$ state, occupied by a down electron, costing $U$ amount of energy and b)
	a ferromagnetic interaction arising from electronic hopping from occupied $d_{xy}$ to unoccupied $d_{yz}$,$d_{xz}$ states, separated by charge transfer energy of $\Delta$, as shown in schematic figure in SM~\cite{SM}. The latter process involves the energy gain between electrons in parallel spin configuration by Hund's coupling
	($J_{H}$). Thus the effective exchange interaction can be written \textcolor{black} {following the formulation of Kugel-Khomskii~\cite{Kliment}, which
		has been applied successfully to several cases~\cite{PhysRevB.73.014418,PhysRevLett.100.186402,PhysRevB.83.144412} as,}
	\begin{align}
		J=&J_{AFM}+J_{FM} = \frac{4t_{xy,xy}^{2}}{U} \nonumber\\
		- &\sum_{\alpha=xy,\beta={yz,xz}} \frac{4t_{\alpha,\beta}^{2} J_{H}}{(U+\Delta_{\alpha,\beta})(U+\Delta_{\alpha,\beta}-J_{H})}
	\end{align}
	
	\textcolor{black}{In the above we have assumed, intra-atomic Coulomb interaction and Hund’s coupling do not depend on the particular orbital, {\it i.e.}, the repulsion between different orbitals ($U_{mm’}$) is same as between the same orbital ($U_{mm}$) = $U$, and $J_H^{mm’}$ = $J_H$, where $m$ and $m’$ are orbital indices that run over the three t2g orbitals.}
	
	The exchange interactions estimated employing the above formula, and using DFT inputs for $\Delta$
	and $t_{\alpha,\beta}$'s with choice of $U$ = 4 eV and $J_H$ = 1 eV are listed in \textcolor{black}{Table~\ref{tab:t2g_table} of Appendix A}. This
	gives rise to the antiferromagnetic nature of the two dominating exchanges, $J$ and $J^{\prime}$, related
	to hopping integrals $t$ and $t^{'}$, two sub-leading exchanges, one of ferromagnetic nature, $J_c$ related
	to $t_c$, and another of antiferromagetic nature, $J_a$ related to $t_a$.

	
	\subsection{Exchange Interactions from total energy calculations}

	The second-order perturbative formula employing DFT inputs of effective hopping integrals and onsite matrix elements, as discussed above, provides a rough estimate of the relative importance of different exchanges\textcolor{black}{. This step therefore serves} as an important step for the identification of the underlying spin model in the complex geometry of NVOPO. However, once this identification is done, in the next step accurate determination of exchanges is required. For this purpose, we employ the total
	energy method, in which various $J's$ are evaluated by mapping the GGA+$U$ total 
	energies of different possible spin configurations of V ions, to an effective $J$-$J^{\prime}$-$J_c$-$J_a$ spin Hamiltonian. \textcolor{black}{As opposed to earlier calculations in this context\cite{PhysRevB.100.144433}
		our total energy calculations are carried out by solving a large number of equations ($\sim$ $^{11}C_{4}$)  
		obtained from GGA+$U$ total calculations of 12 different spin configurations, to avoid any bias on the chosen
		spin configurations. The details of the results are presented in Appendix B. Our rigorous analysis established a 
		robust alternation ratio of 0.92 in comparison to 0.98 extracted from susceptibility fit~\cite{PhysRevB.100.144433}.
		The obtained alternation ratio however is in good agreement with the DFT value of 0.93, as reported in Ref.\cite{PhysRevB.100.144433}. Our analysis, importantly uncovered a strong dependency of the \textcolor{black}{values of 
			$J_a$ and $J_c$ on the chosen spin configurations. The AFM and FM
			nature of $J_a$ and $J_c$, though,} is maintained in each choice. Thus the subleading terms $J_c$ and especially 
		$J_a$ show a large variation (cf. Table~\ref{tab:table-III} in Appendix B).}

	\section{Solution of the Spin-Hamiltonian}
	\begin{figure}
		\centering
		\includegraphics[width=0.8\columnwidth]{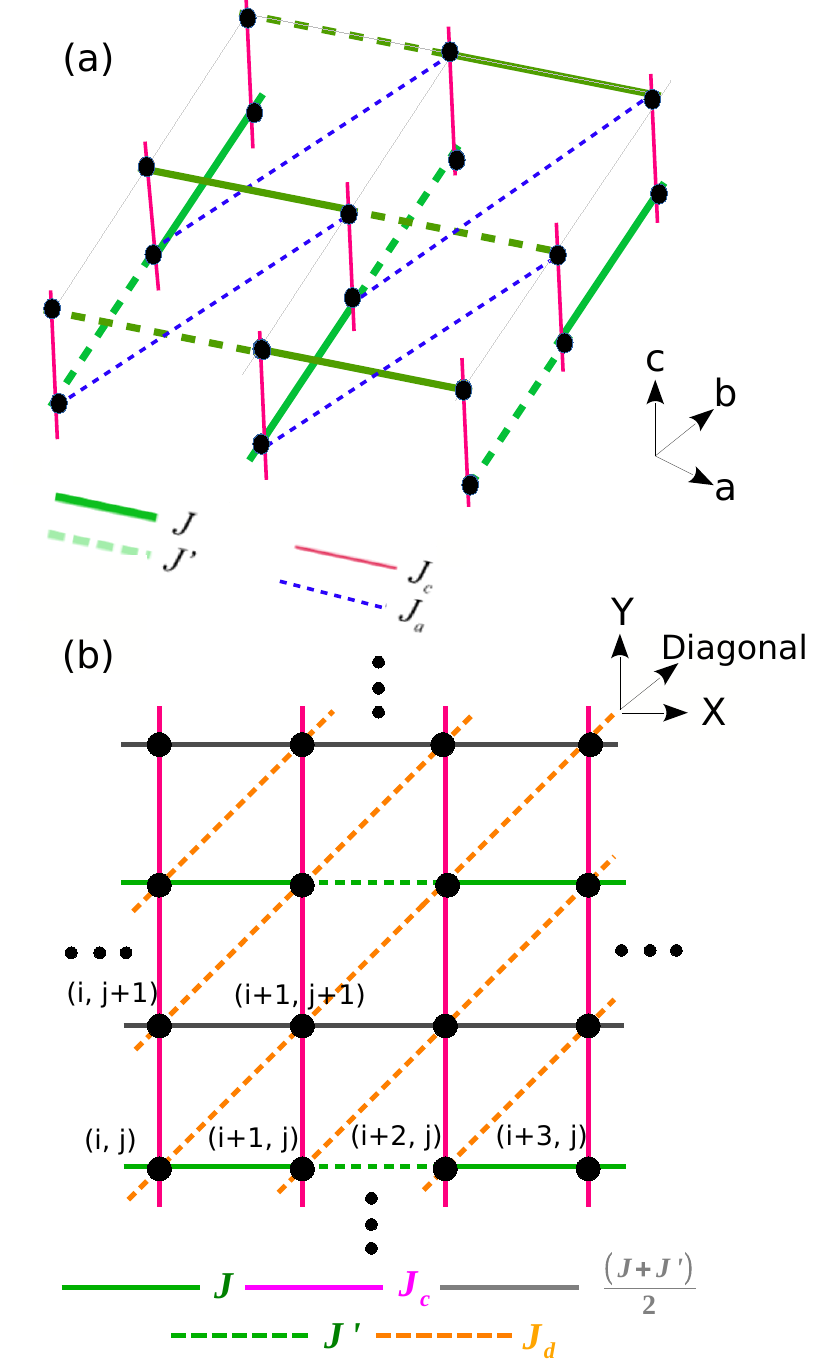}
		\caption{(a) The 3D spin model of NVOPO, where solid green, dashed green, red and dotted blue lines represent V-V magnetic exchanges, $J$, $J^{\prime}$, $J_c$ and $J_a$, respectively. (b) Effective 2D spin model, with solid green, dashed green, black, and dashed orange representing, $J$, $J^{\prime}$, $\tilde{J} = 
			\frac{(J+J^{\prime})}{2}$ and $J_d$ interactions, respectively. While the exchanges
			$J$ and $J^{\prime}$ are same as in 3D model, $\tilde{J}$ and $J_d$ are effective
			interactions arising from 3D to 2D mapping. See text and \textcolor{black}{Appendix C} for details.}
		\label{effective_spin_model}
	\end{figure}
	We employed the many-body techniques of exact diagonalization and density matrix renormalization group to solve the ab initio derived spin Hamiltonian. However, the three-dimensionally coupled network of alternating chains poses a numerical challenge \textcolor{black}{in solving the Hamiltonian.} To make the problem numerically tractable, we consider a two-dimensional (2D) model that captures the essential features of the full 3D problem. Figs.~\ref{effective_spin_model}(a) and (b) show the full 3D model and the 2D model derived from the 3D model, respectively. \textcolor{black}{See Appendix C for details.} In the full model,
	a parallel array of $J$-$J^{\prime}$ chains run along two perpendicular directions, in two
	consecutive $ab$ planes, stacked along $c$ direction, connected by $J_c$. In its two-dimensional projection, this is mapped onto  $J$-$J^{\prime}$ chain, alternating with uniform chain of average
	interaction $(J+J')/2$, connected via $J_c$. 
	Furthermore, in order to capture the effect of $J_a$ present on the body diagonal of the 3D Hamiltonian, an effective,
	renormalized interaction $J_d$, which is a combination of $J_a$ and $J_c$, is considered on the phase diagonal of the 2D plane. \textcolor{black}{While the projection establishes a 2D model of $J$-$J^{\prime}$ chains that alternate with a uniform chain of average interaction $(J+J^{\prime})/2$, connected via $J_c$ (see Fig.~\ref{fig:2D-3D} in Appendix C), the exact estimate of $J_d$ is non-trivial, given the complex intricacy of the 3D model, as is evident from the 3D and 2D structures, presented in
		Fig.~\ref{fig:frustration} of Appendix C. We thus approximate it as 
		$J_a +\gamma J_c$, where $\gamma$ is a variational parameter, chosen to 1.5 to reproduce the correct ground state of NVOPO. \textcolor{black}{It is to be noted that
			both 3D and the constructed 2D models are frustrated, as demonstrated in Appendix C.
			$J_a$, $J_c$ in the 3D model and $J_d$ in the 2D model are primary factors for introducing the frustration in the system. In 3D model this involves one FM and
			three AFM couplings of four bond units and in 2D  model this involves two FM and one AFM couplings of three bond unit.}}
	Thus, two main aspects are ensured in this construction, (i) the coordination number of each interaction is the same as in the full model. (ii) The frustrated nature arising due to competing antiferromagnetic and ferromagnetic interactions is kept intact.
	
	The Hamiltonian \textcolor{black}{for the 2D model of spin-1/2 isotropic Heisenberg system} in an axial magnetic field $B$, is given by,
	
	\begin{align}
		\mathcal{H} & = \sum_{j=1,3, \cdots, }^{L_Y} \sum_{i=1}^{L_X}  (J \Vec{S}_{i,j} \cdot \Vec{S}_{i+1,j} + J^{\prime} \Vec{S}_{i+1,j} \cdot \Vec{S}_{i+2,j} \nonumber \\ &
		+ \frac{J+J^{\prime}}{2} \Vec{S}_{i,j+1} \cdot \Vec{S}_{i+1,j+1}) \nonumber \\ &
		+ \sum_{j=1}^{L_Y} \sum_{i=1}^{L_X} ({J_c} \Vec{S}_{i,j} \cdot \Vec{S}_{i,j+1} + J_{d}\Vec{S}_{i,j} \cdot \Vec{S}_{i+1,j+1} - B S_{i,j}^{z})
		\label{Eq:Heff}
	\end{align}
	
	$\Vec{S}_{i,j}$ is the spin-$1/2$ operator at site having coordinates ($i$, $j$) 
	in a 2D geometry ($L_{X} \times L_{Y}$), as shown in Fig.~\ref{effective_spin_model}(b). The first and second terms of the Hamiltonian represent the odd-numbered and even-numbered spin chains in the 2D plane, with alternating \textcolor{black}{$J$-$J^{\prime}$} antiferromagnetic exchanges and uniform antiferromagnetic exchanges, respectively. The third term represents the subleading competing ferromagnetic exchanges $J_c$ along the Y-direction and $J_d$ along the diagonal between spins on nearest neighbour chains as shown in Fig.~\ref{effective_spin_model}(b). The model exhibits a high degree of frustration due to ferromagnetic sub-leading exchanges $J_{c}$ and $J_{d}$. Based on the inputs of \textit{ab initio} calculations, in all subsequent calculations, we fix the dominant exchange ratio $J^{\prime}/J$ to 0.92. Keeping in mind the large variation in $J_c$ and $J_a$ values with $\vert J_c \vert $ 2-4 times larger than $\vert J_a \vert$, the sub-leading exchange ratios $J_{c}/J$ and $J_{d}/J$ are varied within the ranges $-0.5 \leq J_{c}/J \leq 0$ and $-1 \leq J_{d}/J \leq 0$, respectively.
	
	\section{\label{ZeroField} Quantum phases at $B=0$}
	
	In this section, we study the ground state (GS)  properties of the Hamiltonian given in Eq.~\eqref{Eq:Heff} \textcolor{black}{in the $J_c/J$ and $J_d/J$ parameter space}. In order to characterize the quantum phases and possible boundaries between them, we employ  two \textcolor{black}{numerical} techniques: exact diagonalization and density matrix renormalization group. The second derivative of energy $-E''/N$ was calculated from ED for 24-site system, and spin correlations $C(r)=<S_{i} \cdot S_{i+r}>$ in the GS were calculated from DMRG for relatively larger system sizes of 80-120 sites.
	
	\subsection{\label{EDPD} Second derivative of energy}
	\begin{figure}
		\centering
		\includegraphics[width=1.0\columnwidth]{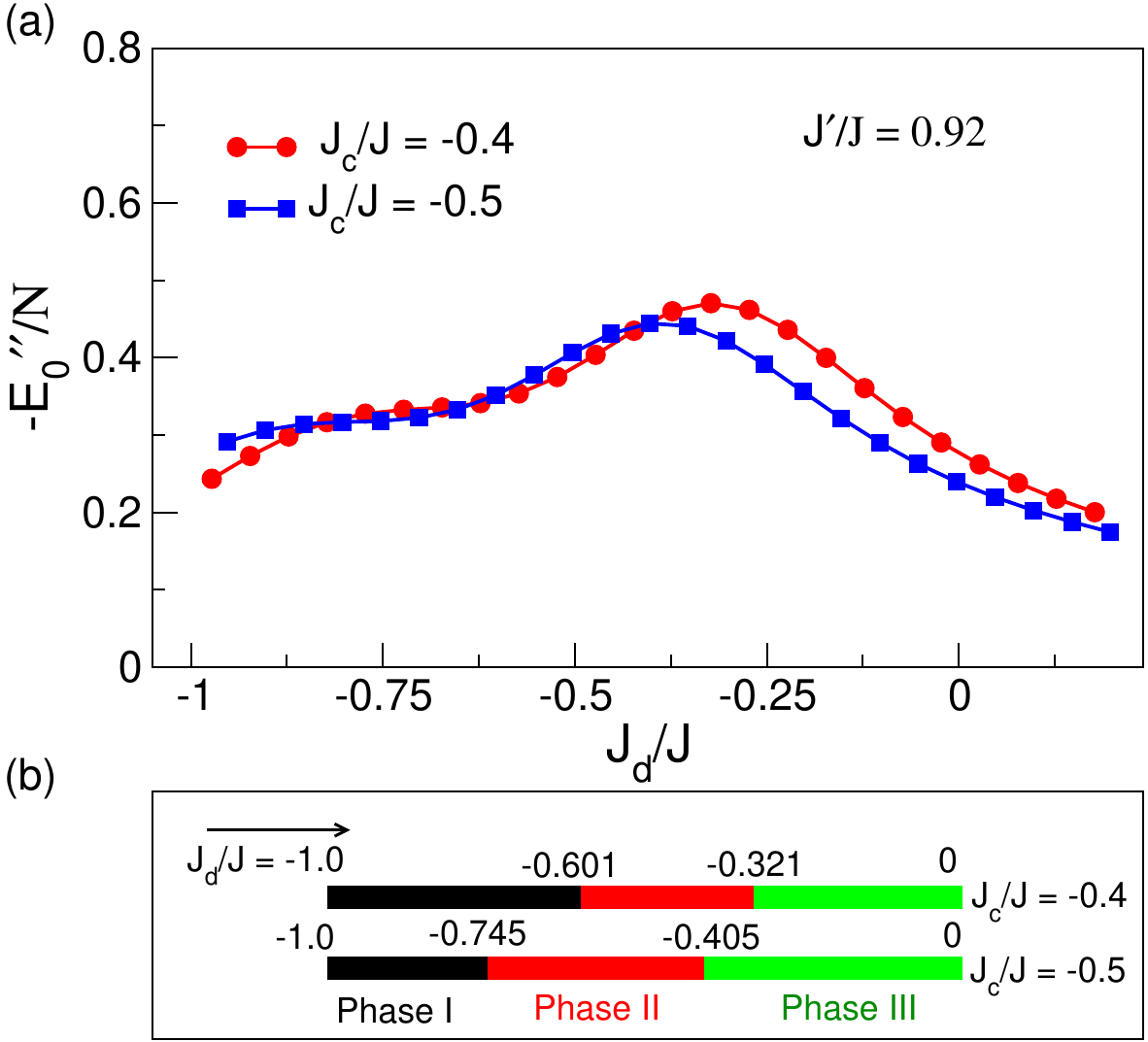}
		\caption{(a) Second-order derivative of the ground state energy per site, \( -E_{0}^{''}/N \), calculated from ED on a \( 24 \)-site torus as a function of $J_d/J$ for different $J_c/J$ values. The two prominent peaks in \( -E_{0}^{''}/N \) serve as indicators of the phase boundaries separating Phase I and Phase II, and Phase II and Phase III, as shown in (b).}
		\label{fig:nrgder}
	\end{figure}
	
	The second derivative of the GS energy is expected to show maxima or discontinuity at the transition points between different phases and serves as an indicator of the existence of different phases. Fig.~\ref{fig:nrgder}(a) shows $-E''/N$ as a function of $J_{d}/J$ for two representative values of \textcolor{black}{$J_{c}/J = -0.4, -0.5$} with fixed value of $J^{\prime}/J=0.92$. \textcolor{black}{A similar nature was found for $J_{c}/J = -0.1, -0.2, -0.3$ (not shown in the figure for clarity).}
	The two maxima in $-E''/N$, as seen in the figure, suggest there are three distinct phases. Fig.~\ref{fig:nrgder}(b) shows the transition points between three phases for $J_{c}/J = -0.4, -0.5$.  We characterize these phases,  phase I, phase II, and phase III in the following section through spin correlations obtained from DMRG for larger system sizes.
	
	\subsection{\label{spincorr} Spin correlations $C(r)$}
	To further characterize the three phases found in ED, we have used DMRG to analyze the nature of the ground state of $\mathcal{H}$ \textcolor{black}{\textcolor{black}{on} cylindrical geometries, \textcolor{black}{which is discussed in details in Appendix D}.}
	We computed spin correlations in the three different phases identified in ED,
	(i) weak $J_{c}$, strong $J_{d}$ (Phase I: $J_{c}/J=-0.3$, $J_{d}/J=-0.8$), (ii) comparable $J_{d}$, $J_{c}$ (Phase II: $J_{c}/J=J_{d}/J=-0.3$), and (iii) strong $J_{c}$, weak $J_{d}$ (Phase III: $J_{c}/J=-0.5$ and $J_{d}/J=-0.1$). Since the Hamiltonian is isotropic, the correlations $S^{x}$, $S^{y}$ and $S^{z}$ are identical, we computed the total correlation, defined as,
	
	\begin{align}
		C(r) & = <\Vec{S}_{0} \cdot \Vec{S}_{r}> \nonumber \\
		& = <S^{x}_{0} \cdot S^{x}_{r}> +<S^{y}_{0} \cdot S^{y}_{r}> +<S^{z}_{0} \cdot S^{z}_{r}>
	\end{align}
	
\begin{figure}[t]
		\centering
		\includegraphics[width=1.0\columnwidth]{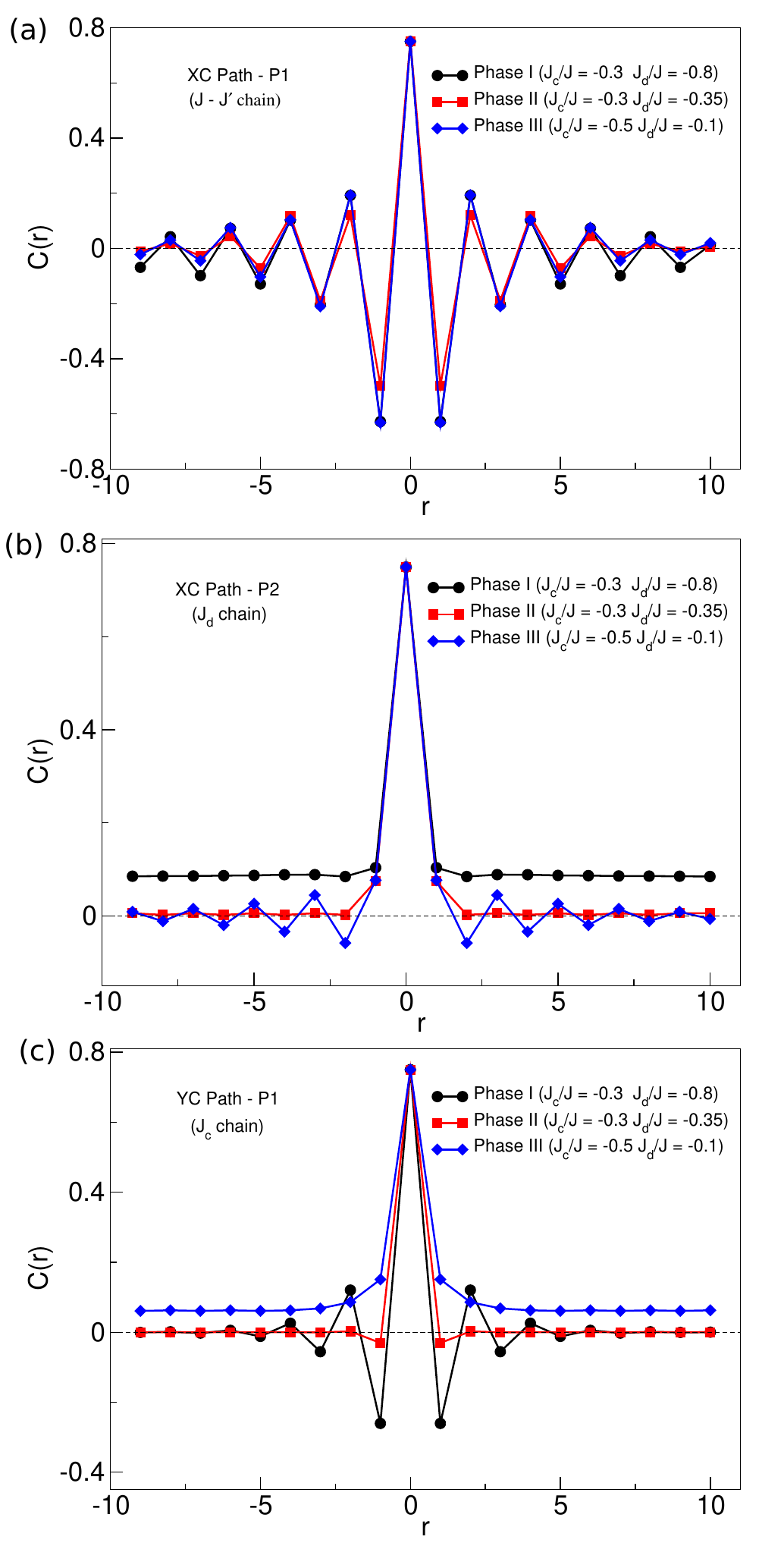} 
		\caption{\textcolor{black}{Correlations \( C(r) \), computed for different paths and geometries in the DMRG calculation (see Appendix D) in three different phases. (a) along path P-1 in XC geometry, (b) along path P-2 in XC geometry, and (c) along path P-1 in YC geometry.}}
		\label{fig:XCDMRG}
	\end{figure}

	\textcolor{black}{Figs.~\ref{fig:XCDMRG}(a) and (b)  present the spin correlations $C(r)$ on the 80-site XC-cluster along the path P-1 and P-2 (cf Appendix D for reference).} Since the path P-1 is considered along the 
	alternating AFM $J$-$J^{\prime}$ chain, the spin correlations in all three phases mentioned above, exhibit short-range AFM correlation with exponential decay. On the other hand, the spin correlations along path P-2 show different
	behaviour in different phases. We note this path includes the sub-leading FM exchange $J_d$. In Phase I, correlations indicate FM long-range order ($m_{FM} \approx 0.08$) while in Phase III short-range AFM correlation with exponential decay is observed. The AFM correlations in phase III though decay faster than those of path P-1. However, in phase II, correlations vanish even at the next-nearest neighbour, which suggests spins are highly disordered in this phase.
	
	\textcolor{black}{The spin correlations on the 80-site YC-cluster along the path P-1 (cf Appendix D), which includes the sub-leading FM exchange $J_c$ is shown in Fig.~\ref{fig:XCDMRG}(c).} In Phase I and Phase III, correlations exhibit short-range AFM order with exponential decay and FM long-range order ($m_{FM} \approx 0.05$), respectively. This is opposite to that found for the path P-2 in XC geometry. However in Phase II, similar to path P-2 in XC-cluster, the spin correlations vanish at the next-nearest neighbour.
	
	\begin{table*}[]
		\centering	
		\begin{tabular}{p{4.5cm} p{3cm} p{5cm} p{4cm}}
			\hline
			\hline
			Path     & Phase I &  Phase II & Phase III   \\
			& (Staggered) &  (Disorder) & (Stripy)   \\
			\hline
			XC P-1  & AFM order  & AFM order & AFM order  \\
			(along X-direction) & (short-range) & (short-range) & (short-range) \\
			($J-J^{\prime}$ chain) & & \\
			\hline
			XC P-2   & FM order & Nearest-neighbour correlation &  AFM order \\
			(along diagonal direction) & (long-range) & ($C(r) = 0$ for $r >1$) & (short-range) \\
			($J_d$ chain)  & & \\
			\hline
			YC P-1  & AFM order & Nearest-neighbour correlation & FM order \\
			(along Y-direction) & (short-range) & ($C(r) = 0$ for $r >1$) & (long-range) \\
			($J_c$ chain) & &\\
			\hline		
			\hline		
		\end{tabular}
		\caption{Characterization of three different phases, based on the analysis of spin correlation, presented in Section VI.B.}
		\label{tab:phases}
	\end{table*}

	\begin{figure}
		\centering
		\includegraphics[width=1.0\columnwidth]{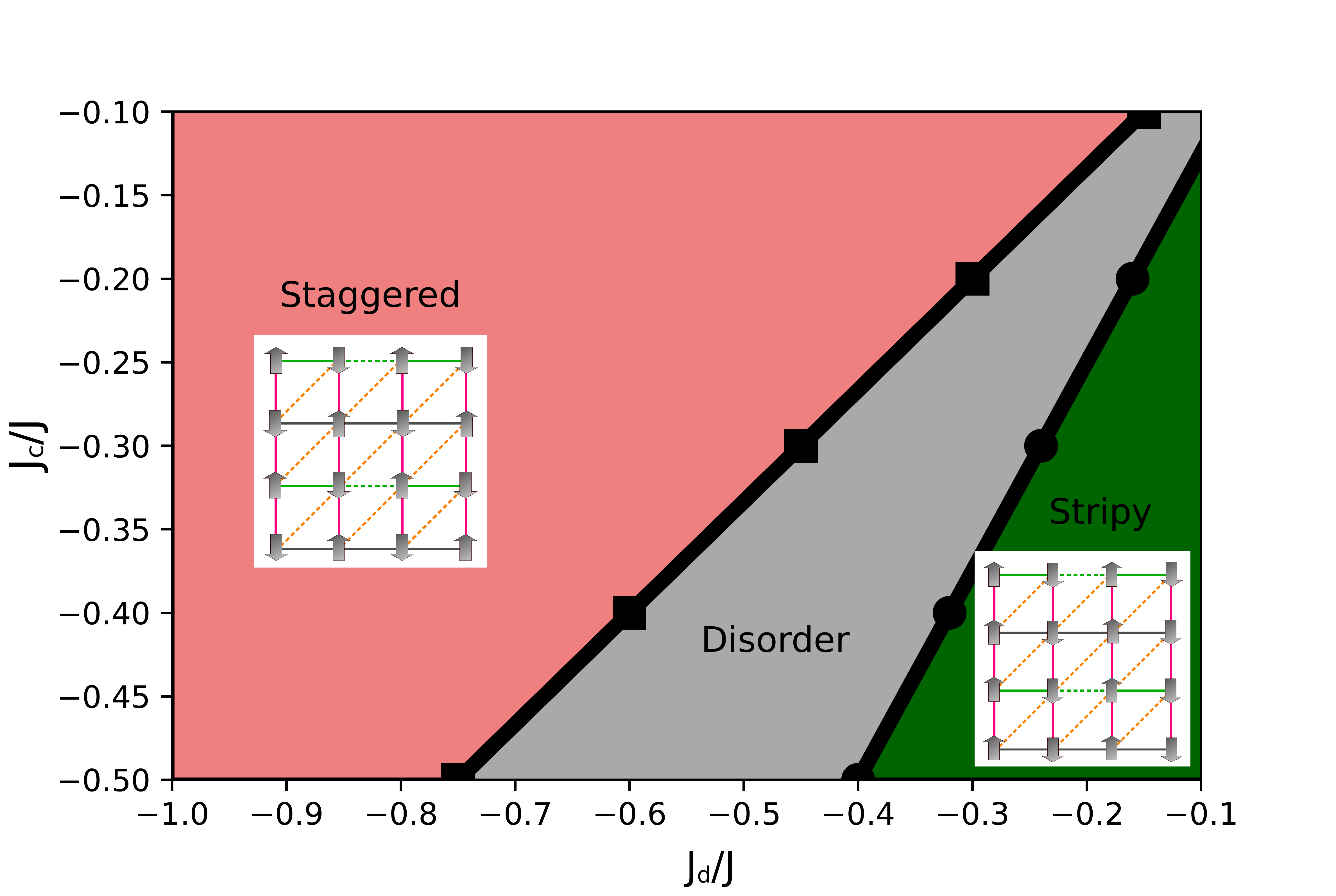}
		\caption{Phase diagram of the 2D effective model Hamiltonian 
			$\mathcal{H}$ in0 Eq.~\eqref{Eq:Heff} in parameter space of $J_d/J$ and $J_c/J$,
			constructed out of ED total energy and DMRG computed spin correlations.
			The inset shows then local spin orientations in the {\it staggered} and {\it stripy} phases.}
		\label{fig:ZeroFieldPD}
	\end{figure}
	
	The findings of spin correlation calculations are summarized in Table~\ref{tab:phases}. As is evident, Phase I
	characterized by a strong $J_d$, exhibits short-range AFM correlation both in $J-J^{\prime}$ (X-direction) as well as $J_c$ (Y-direction) direction, which makes correlation along $J_d$ FM with long range order. In Phase III, characterized by a strong $J_{c}$, on the other hand, correlation is of AFM nature in $J-J^{\prime}$ (X-direction) and $J_d$ (diagonal) direction, making the correlation along $J_c$ to be FM. On the other hand, in Phase II, with comparable values of $J_c$ and $J_d$ the spins are disordered
	along both $J_c$ and $J_d$ with short-range AFM correlation along the leading $J-J^{\prime}$ chain direction. This
	suggests a staggered and stripy spin orientation in Phase I and Phase III, respectively. In Phase II, where both sub-leading terms are comparable (\( J_{c} \approx J_{d} \)), the system experiences significant frustration in terms of 
	fierce competition between $J_c$ and $J_d$. This results in vanishing correlations along the Y and diagonal directions, leading to what we term as `\textit{disorder}' phase. It is worth noting that in this `\textit{disorder}' phase though there exists weak short-range order along the X-direction due to the presence of weakly alternating AFM chains.
	
	Armed with information on spin correlations, which characterizes three different phases, we construct the zero-field phase diagram of this model as shown in Fig.~\ref{fig:ZeroFieldPD} in the parameter space of $J_{c}/J$ and $J_{d}/J$ for fixed $J^{\prime}/J = 0.92$ \textcolor{black}{based on the energy derivative}. The local spin orientations in the {\it staggered} (phase I) and {\it stripy} (phase III) are also shown in the inset of the phase diagram (Fig.~\ref{fig:ZeroFieldPD}).

	\section{\label{FieldResults} Effect of axial magnetic field}
	\begin{figure}
		\centering
		\includegraphics[width=1.0\columnwidth]{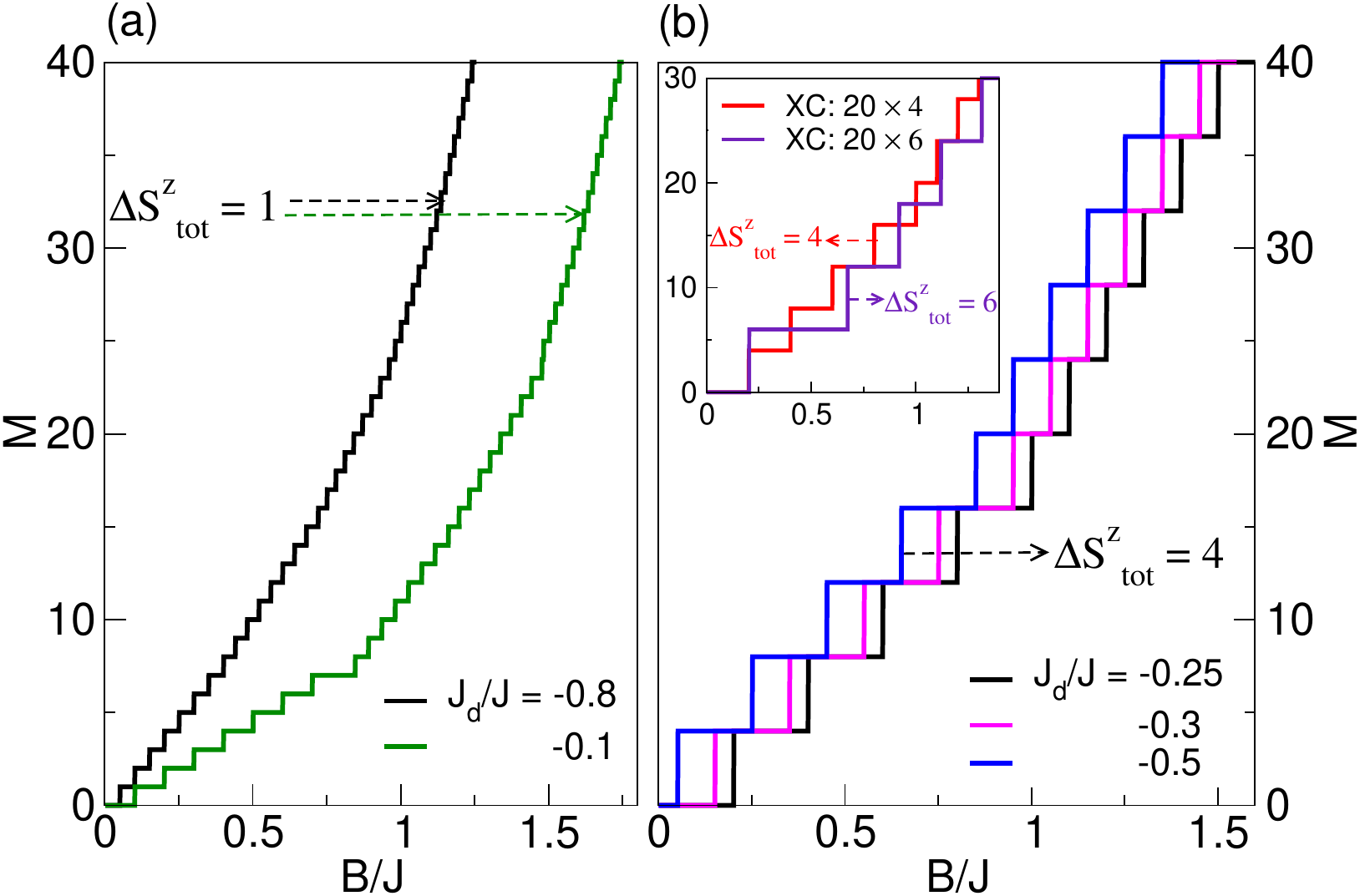}
		\caption{DMRG computed magnetization in XC geometry with system size $20 \times 4$ in (a) staggered and stripy phase, (b) disorder phase plotted as a function axial magnetic field. In all the calculations, $J^{\prime}/J$ and $J_{c}/J$ are fixed at $0.92$,
			and $-0.3$, respectively. $J_{d}/J$ values are chosen to be $-0.8$, and
			$-0.1$ in (a) and $-0.25$, $-0.3$ and $-0.5$ in (b). The inset in (b) shows the system size dependence of magnetization in disorder phase, shown for the representative case of $J_d/J=-0.3$.}
		\label{fig:MHDMRG}
	\end{figure} 
	
	Upon clarification on the nature of the GS in the absence of any external perturbation, the next interesting question is to ask the effect of an external magnetic field on the GS of the Hamiltonian in Eq.~\eqref{Eq:Heff}. In
	the `\textit{disorder}' phase, the lowest \textcolor{black}{singlet-triplet (ST)} gap ($E_m = E(S^z=1) - E(S^z=0)$) is found to be finite, though the gap value is found to vary within the phase space of Phase II. The gap value
	is found to be largest along the $J_c=J_d$ line and decreases upon moving away from this line. For fixed choice of $J^{\prime}/J=0.92$, the gap value for $J_c/J$ = -0.3
	and $J_d/J$ = -0.3 is found to be $\approx$ 0.2$J$ while that reduces to $\approx$ 0.05$J$ for  $J_c/J$ = -0.3 and $J_d/J$ = -0.5. 
	
	We note that at the $J_c$=$J_d$ point due to the competing nature
	of $J_c$ and $J_d$, they compensate. \textcolor{black}{This results in a system that can be mimicked by decoupled alternating chains with finite ST gap.} This can be appreciated by considering the fact that the spin gap for weakly dimerized chain is given by $\Delta(\delta) = A \delta^{\beta}$, where $\delta$ is dimerization,
	connected to alternation ratio by the relationship, $J^{\prime}$/$J$ = $\frac{1-\delta}{1+\delta}$
	and $A = 2.0375 $,  $\beta = 0.7475$~\cite{PhysRevB.75.052404}. Putting $J^{\prime}/J$ = 0.92, this gives a values 0.189$J$, very close
	to our estimate of $\sim$ 0.2$J$ for the case of $J_c$=$J_d$. Moving away from
	the $J_c=J_d$ line, the compensation point, the resultant uncompensated
	subleading interactions suppress the gap value from the single chain limit.
	The estimated value of 0.05 $J$ for $J_c/J$ = -0.3 and $J_d/J$ = -0.5 is very
	close to the experimental estimate of 2K for the compound and is also supported by the DFT estimates of exchanges.  Thus the
	spin-gap can be reproduced either by considering
	a single alternating chain with an alternation parameter of 0.98, as employed in susceptibility fitting,~\cite{PhysRevB.100.144433} or by considering a stronger alternation parameter of
	$\sim$ 0.92 as found in the DFT study, but including uncompensated subleading interactions. 
	\textcolor{black}{The latter scenario turns out to be the realistic representation. 
		The non-uniqueness of the fitting process of the susceptibility fit is demonstrated in section IX. In comparison to the DFT estimate of $J^{\prime}/J$ is found to be robust, consistently obtained in a large number of solutions.} It is to be stressed that the effect of subleading interactions, either compensated or uncompensated, will be manifested in the excited state properties. \textcolor{black}{This would enable the difference between a single alternating chain model, as invoked in Ref.~\cite{PhysRevB.100.144433} and the inter-chain coupled chain model of the discussed system. This will be taken up in the following section.} 
	
	\textcolor{black}{The system size dependence of the spin gap is shown in SM\cite{SM}.
		This shows a gaped  solution in the disorder phase survives even in asymptotic limit
		(see Fig 3(a) in SM).} The gap can be closed at a critical field $B=B_{c1}$. For the parameters $J_c = -0.3$ and $J_d = -0.5$ the critical field is found to be $\approx 1.8$ T, close to experimentally found gap of $\le$ 2T~\cite{PhysRevB.100.144433}. However, in the staggered and stripy phases, the dominant ferromagnetic $J_d$ and $J_c$ lead to ordering along the diagonal and Y-direction, respectively, which result in the closing of the spin gap, upon extrapolation 
	to infinite size (see SM~\cite{SM}).
	
	To understand the magnetization processes, we further computed the magnetization as a function of $B$. The results for the staggered and stripy phases are shown in Fig.~\ref{fig:MHDMRG}(a), and that for the disordered phase is shown
	in Fig.~\ref{fig:MHDMRG}(b). The results for 20 $\times$ 4 lattice in 
	XC geometry reveals while in the staggered and stripy phases, the magnetization $M = S^{z}_{total}=\sum_{i}S^{z}_{i}$ increases with $B$ in steps of $\Delta S^{z}_{total}=1$, in the disordered phase, the magnetization ($M$) changes in steps of $\Delta S^{z}_{total} = 4$. Interestingly
	throughout Phase II, $M$ changes in the step of $\Delta S^{z}_{total} = 4$,
	as in see in Fig.~\ref{fig:MHDMRG}(b) for a range of  $J_{d}/J$ values
	of -0.25, -0.3, -0.5, although the critical field for closing the gap varies. To probe this further, we carried out calculations on 20 $\times$ 6 sites system in XC geometry, thereby increasing the system along the $J_c$ bond direction,
	and focusing on the disordered phase with $J_d/J=-0.3$. We notice while the critical field $B_{c1}$ for the $20\times 4$ and $20\times 6$ XC clusters remains nearly the same ($\approx 0.2J$), the magnetization changes in the step of $\Delta S^{z}_{total} = 6$ in comparison to $4$, as found in $20\times 4$ sites system. Thus, the magnetization step scales with the system size along the $J_c$ bond direction, signaling multi-magnon condensation in Phase II. \textcolor{black}{We call this phenomenon magnon condensation, as the quasi-particles generated in the weakly dimerized chains are identified as magnons \cite{footnote}. The magnons on each leg develop attractive interactions with the magnons on the neighboring chains. Consequently, a multi-magnon condensate is formed in a finite field that regulates the magnon density within the system.}

	\section{Multi-magnon condensation}
	
\begin{figure}[h]
		\centering
		\includegraphics[width=1.0\columnwidth]{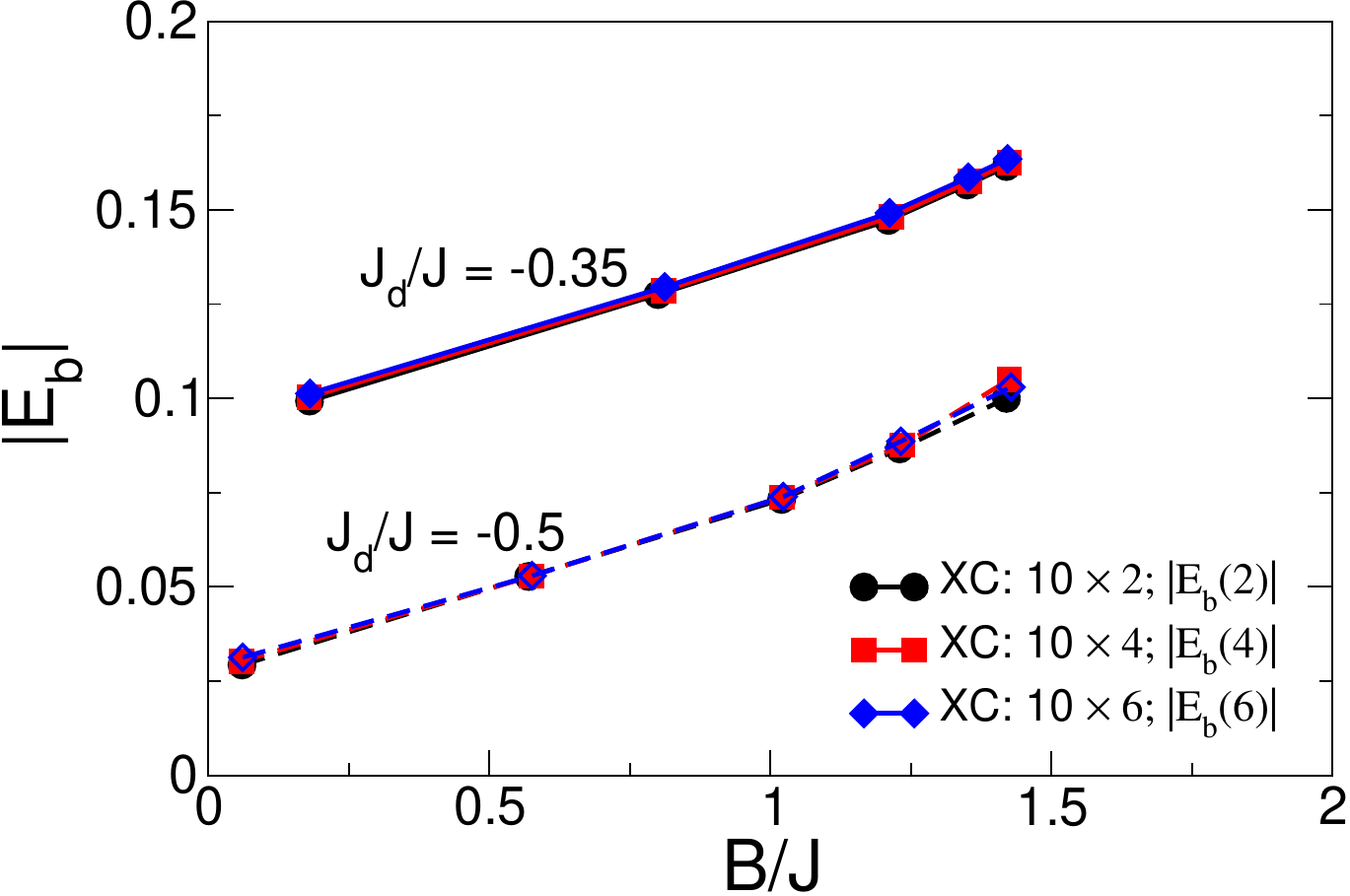}
		\caption{DMRG computed absolute value of the binding energy per magnon ($|E_{b}(n)|$) as a function of external magnetic field ($B/J$) for $10 \times 2$, $10 \times 4$ and $10 \times 6$ system sizes in XC geometry. The parameters of the spin model are chosen to belong to disorder phase with choice of $J_{d}/J=-0.35$, $-0.5$ for fixed $J^{\prime}/J=0.92$, $J_{c}/J=-0.3$.}
		\label{fig:BE}
	\end{figure}

	\textcolor{black}{In the multi-magnon condensation phase, a large number of magnons form a bound state, and there are two quantities to characterize this phase. The first is the order parameter, which is discussed in Appendix E, and the second is the binding energy of the bound state. We computed the binding energy of $n$-magnon condensed state \cite{2015Satoshi, Nishimoto_2012}. The per magnon binding energy of an nn-magnon bound state can be defined as,}
	\begin{align}
		E_{b}(n) = \frac{1}{n} [(E(S^{z}-n) - E(S^{z})) - n(E(S^{z} -1) - E(S^{z} ))]
	\end{align}
	\textcolor{black}{where $E(S^{z}=S)$ is the GS energy in $S^{z}=S$ sector. For stabilization of $n$-magnon condensed state, the corresponding binding energy $E_{b}(n)$ should be negative. We computed the binding energy of 2-magnon, 4-magnon, and 6-magnon states in the disordered phase in 10 $\times$ 2, 10 $\times$ 4, and 10 $\times$ 6 XC geometry for the representative cases of $J_d/J = -0.35$ and $-0.5$, with fixed $J^{\prime}/J = 0.92$ and $J_c/J = -0.3$ values. Fig.~\ref{fig:BE} shows the plot of $|E_{b}(n)|$ as a function of the external magnetic field $B/J$. We find that the binding energy increases with the magnetic field, suggesting increased stabilization of a multi-magnon bound state with increasing field strength. The binding energy is found to be maximum at $J_c = J_d$ when the system is highly frustrated and decreases away from the $J_c = J_d$ line.}
	
	\textcolor{black}{The finite binding energy indicates a multimagnon condensed state, which can be rationalised as follows. The ground state of the disordered phase exhibits a singlet state characterised by short-range AFM correlations along the $J-J^{\prime}$ chain, while only a weak nearest neighbour ferromagnetic correlation persists along the Y and diagonal directions. To understand the attractive nature of the magnon, we use the mean-field approximation for weak sub-leading exchange interactions $J_c$ and $J_d$. In this limit, these sub-leading terms act as an effective magnetic field generated by the magnetic moments on neighbouring chains, and the field strength is proportional to $J_c$ and $J_d$. The mean field model Hamiltonian is the sum over all $J-J'$ spin-1/2 chains under the effective field of the magnetic moments of other neighbouring chains and the external magnetic $B$ and the Hamiltonian can be expressed as,}
	
	%
	\begin{align}
		\mathcal{H}_{\rm MF} & \equiv  \sum_{i} \sum_{j} J \Vec{S}_{i,j} \cdot \Vec{S}_{i+1,j} + J^{\prime} \Vec{S}_{i+1,j} \cdot \Vec{S}_{i+2,j} \nonumber \\ 
		& - \Vec{B}_{\rm eff}(J_{c}, J_{d}, B) \cdot \Vec{S}_{i,j}
	\end{align}

	\textcolor{black}{where $\vec{S}_{i,j}$ represents the spin at the $i$-th site of the $j$-th effective single $J-J'$ chain model, and $J$ and $J^{\prime}$ are effective exchange interactions of dimerized chains. The total effective field, $B_{\rm eff}$, is given by $(B - J_{c}(\langle\Vec{S}{i,j+1}\rangle + \langle\Vec{S}{i,j-1}\rangle) - J_d(\langle\Vec{S}{i+1,j+1}\rangle + \langle\Vec{S}{i-1,j-1}\rangle))$, where $B$ is the external field and $\langle\Vec{S}_{i,j}\rangle$ is the average magnetic moment generated in the 2D model at the site with coordinates ($i$, $j$) at field $B$. We notice that both the ferromagnetic interactions $J_c$ and $J_d$ induce an effective field which aligns the $ij$-th spin along the moment of the neighboring spin, i.e., induces attractive interactions between the magnons on the neighboring chains.}
	
	
	Our analysis thus highlights the importance of the sub-leading ferromagnetic exchanges $J_c/J$ and $J_d/J$ in the stabilization of the multi-magnon condensed phase. The presence of these two subleading interactions helps the process in two ways. Firstly presence of competing $J_c$ and $J_d$ causes frustration which significantly reduces the correlation length along the Y and diagonal directions. Secondly, ferromagnetic subleading exchanges between the chains are responsible for giving rise to finite binding energy between the magnons on different chains, acting as a glue for the multi-magnon condensation. 

	\textcolor{black}{\section{Calculated Thermodynamic Properties: Comparison to Experimental Measurements}}

	\textcolor{black}{We next compute the measurable thermodynamic properties, taking into account the sub-leading inter-chain interactions, dealt in a mean-field manner, and compare with the published literature~\cite{PhysRevB.100.144433}}.
	
\begin{figure}
		\centering
		\includegraphics[width=1.0\columnwidth]{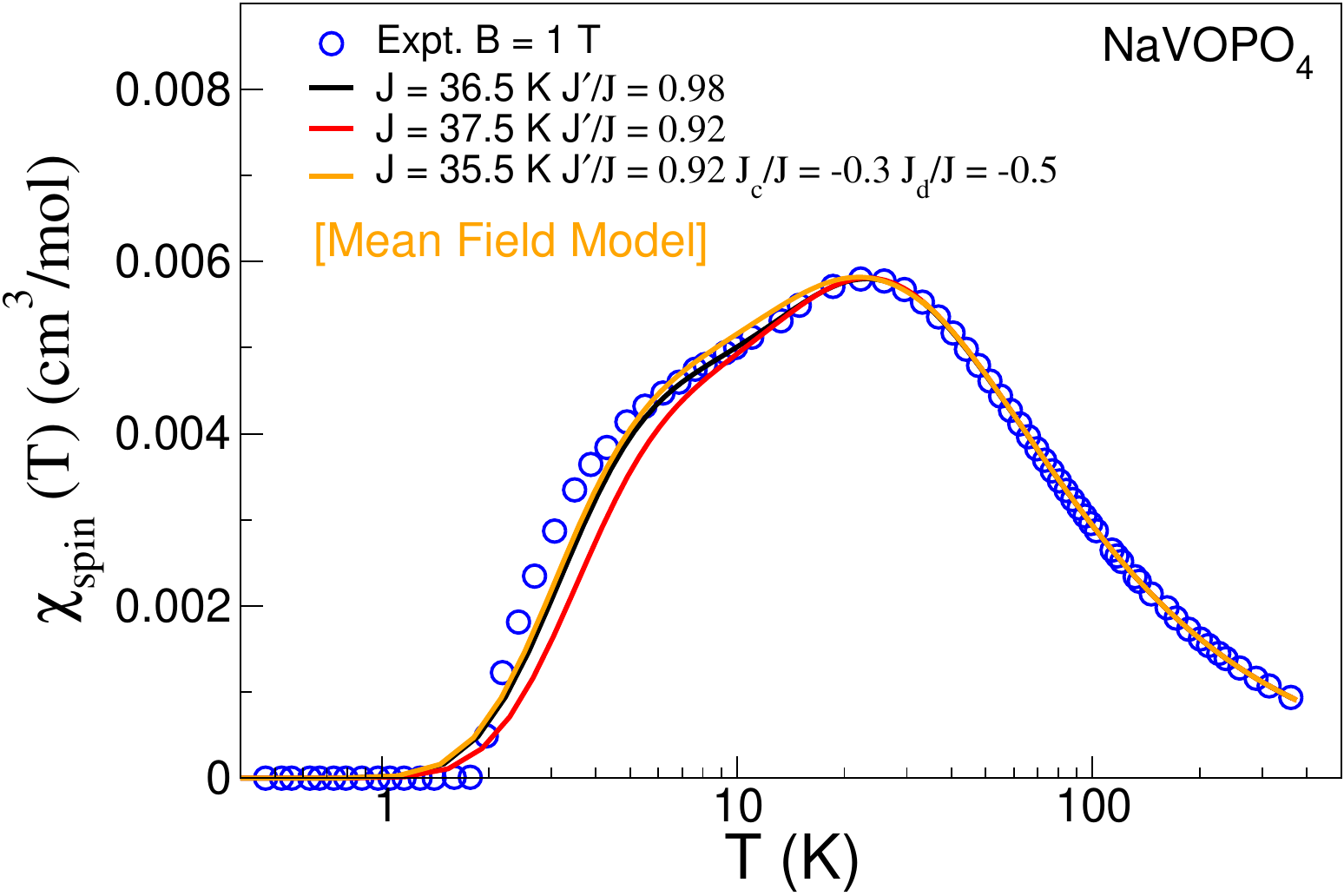}
		\caption{\textcolor{black}{Magnetic susceptibility of NaVOPO$_4$ in applied field of 1T, subtracting the
				impurity contribution. The experimental data is reproduced from Ref.\cite{PhysRevB.100.144433}. The theoretically computed
				suscepbility for various models are compared.}}
		\label{fig:chi}
	\end{figure}

	\textcolor{black}{For this purpose, we first compute the magnetic susceptibility.
		The comparison of different models with experimentally measured susceptibility~\cite{PhysRevB.100.144433} in the presence of the magnetic field
		of 1T after subtracting the impurity contribution is shown in Fig.~\ref{fig:chi}. As is seen from the plot, an equally good fit
		of the measured susceptibility can be obtained either by a 1D alternating chain model with an alternation parameter of 0.98 or by
		a model including subleading inter-chain interactions with a slightly stronger alternation parameter of 0.92, upon slight variation
		in value of $J$ (36.5 K vs 35.5 K).}
	
	\begin{figure}
		\centering
		\includegraphics[width=1.0\columnwidth]{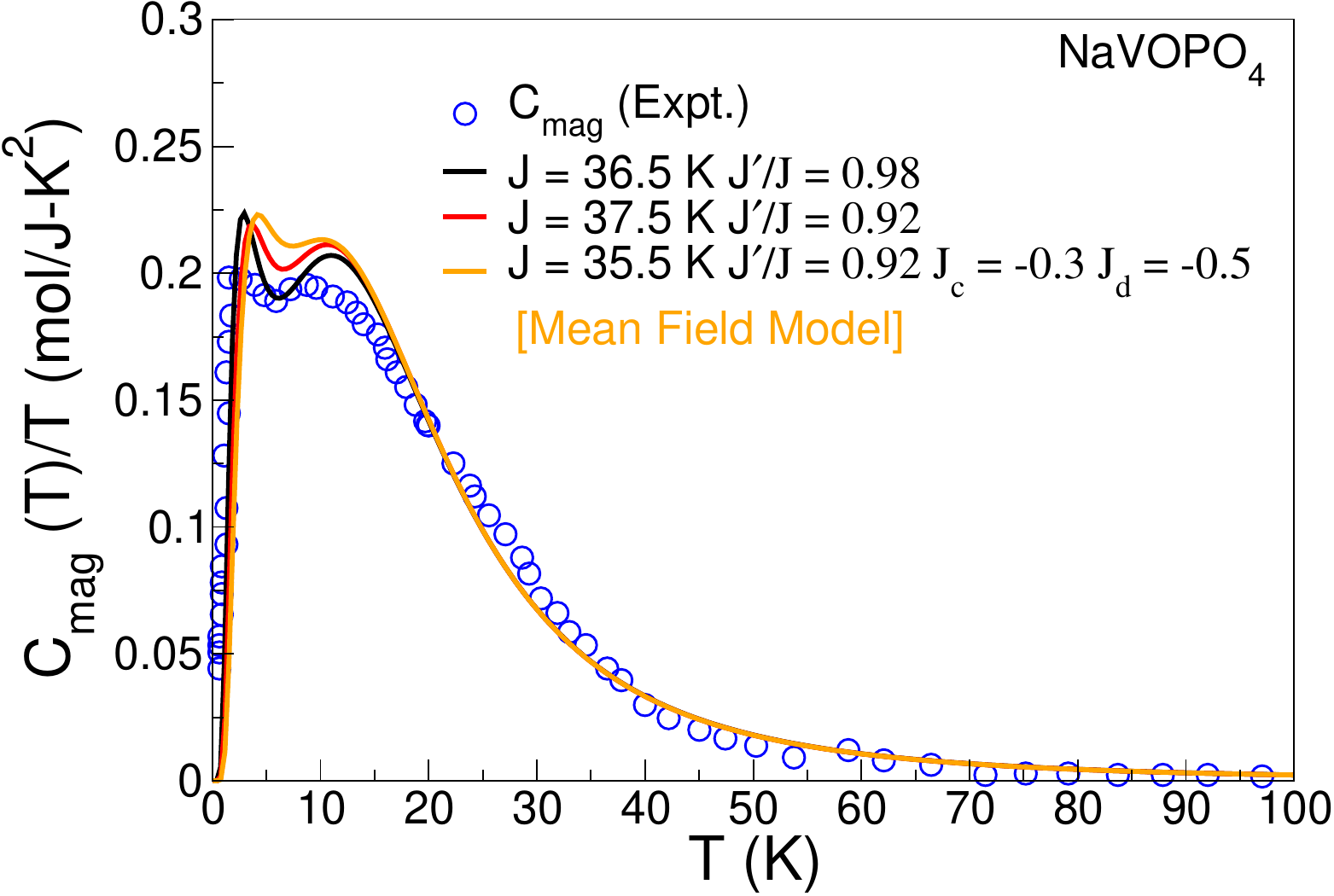}
		\caption{\textcolor{black}{Magnetic contribution of the specific heat. The theoretically computed
				results for various models are compared with the experimetal data from Ref.\cite{PhysRevB.100.144433}.}}
		\label{fig:sp}
	\end{figure}
	
	\textcolor{black}{We also compute the specific heat and compare it with the experimental data from Ref~\cite{PhysRevB.100.144433} at zero magnetic
		field. It is to be noted that the measured specific heat always has a dominating phonon contribution, and subtraction of phonon
		contribution is not straightforward. We consider the magnetic contribution of specific heat as published in Ref~\cite{PhysRevB.100.144433}.
		The 1D alternating chain model as well as that in our proposed model produce comparable results.} 
	
	\section{The Materials Perspective - Effect of Strain}

	Following the construction of a $J_c/J$-$J_d/J$ phase diagram, through a solution of the model spin Hamiltonian, it is a worthwhile exercise to put the material's perspective in it. Using
	the DFT derived average value of parameters, as given in Table~\ref{tab:table-III}, the positioning of NVOPO in the
	the phase diagram is marked with the red symbol while the box depicts
	the probable position taking into the standard deviation in estimates
	of exchanges (cf Table~\ref{tab:table-III}).
	As is seen, the box falls in Region II, characterized by multi-magnon
	condensation, separated from the one-magnon state.
	
	Considering the important effect of competing ferromagnetic couplings $J_c$ and $J_d$ on driving the multi-magnon condensed phase, we next investigate the effect of uniaxial tensile strain applied along the $c$-axis in tuning these exchanges. The system was subject to 1 - 3 $\%$ tensile
	strain. The system was fully relaxed for
	the in-plane $a$ and $b$ lattice parameters, keeping the volume intact, and for
	the free atomic positions. \textcolor{black}{As a first approximation, for a 3D solid, without significant anisotropy, the assumption of fixed volume is expected to hold well. We have explicitly checked this by carrying out c-axis relaxation without the constraint of constant volume. The fixed volume constraint is found to have minimal effect with c-axis parameter and the free atomic positions differing by less than 1$\%$.} 
	
	The application of strain is found to affect mostly the structural parameters related to V-V chains running along the c-axis. In particular, for
	1$\%$ and 3$\%$ tensile strain, 1NN V-V bond length is found to change from 
	3.56$\AA$ to 3.60$\AA$ to 3.70 $\AA$, while the V-O-V bond-angle is found decrease slightly from 144.9$^o$ to 144.8$^o$ to 144.2 $^o$. 
	
\begin{figure}[h]
		\centering
		\includegraphics[width=1.0\columnwidth]{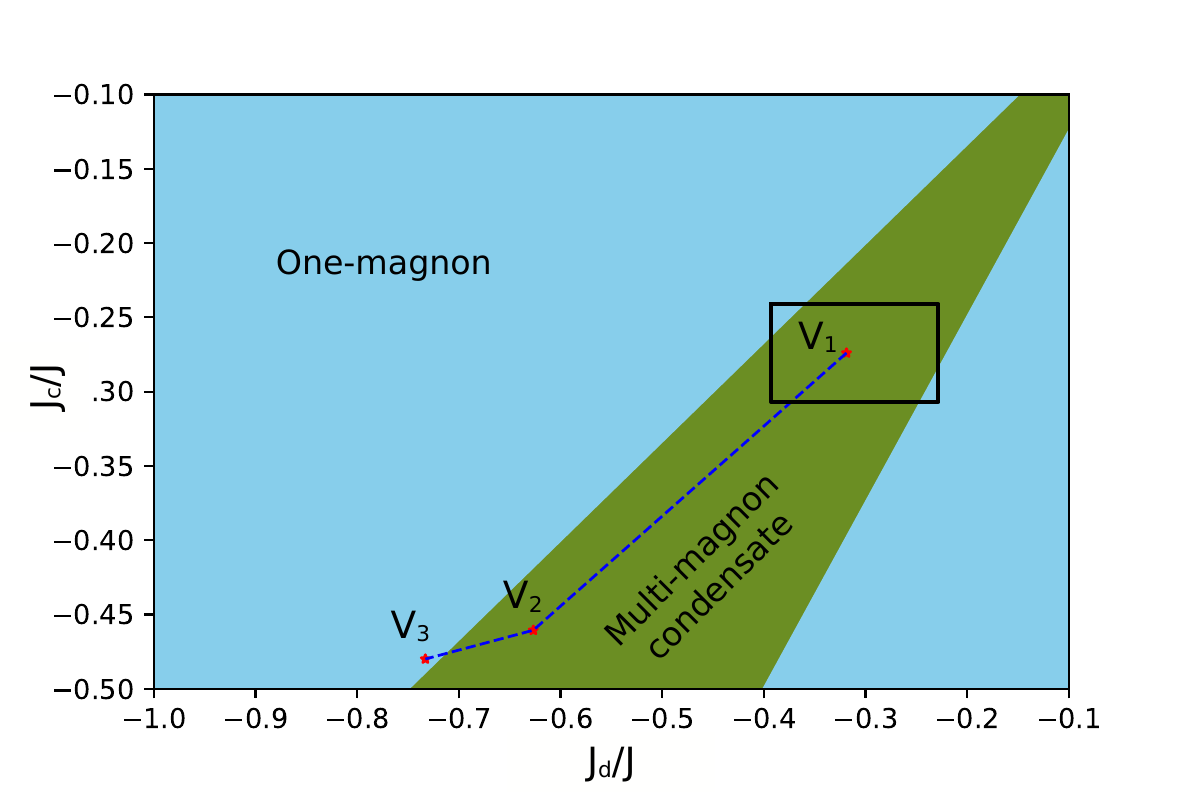}
		\caption{Phase-Diagram: $J_c/J$-$J_d/J$ phase diagram divided into multi-magnon condensed phase (marked in green) and the one-magnon, dipolar phase (marked in light blue). The data-point marked as V1 represents the
			positioning of NVOPO in the phase diagram, based on DFT estimates
			of the average value of exchanges. The box represents the positioning taking into account the standard devitations in DFT estimated exchanges. The data points V2 and V3 show the positioning upon application of 1$\%$ and 3$\%$ tensile uniaxial strain along c-axis, respectively.}
		\label{fig:FiniteFieldPD}
	\end{figure}

	The structural changes upon application of 1$\%$ strain resulted in an increase in \textcolor{black}{average value of $J_c$} by 67.5$\%$, a decrease in $J_a$ by 30.24$\%$, and thus increment in $J_d$ of 116.11$\%$, \textcolor{black}{compared to corresponding $J$ values in unstrained condition}. Interestingly, the alternation parameter $\alpha$ was found to remain more or less unchanged. Thus,
	application of 1$\%$ strain, as seen in Fig.\ref{fig:FiniteFieldPD},
	pushes the system further deeper into the disordered phase.
	Further increase of strain value to 3$\%$ increases the $J_c$ with a substantial increase in $J_d$, thus driving a quantum phase transition from gapful to gapless state.
	
	\section{Summary and Discussion}
	
	The physics of quantum spin chain compounds has been discussed at length in the literature. NaVOPO$_4$ compound, in this context, has been recently studied experimentally by x-ray diffraction, magnetic susceptibility, high-field magnetization, specific heat, electron spin resonance, as well as nuclear magnetic resonance measurements~\cite{PhysRevB.100.144433}. 
	The fit of susceptibility resulted in a $J$-$J^{\prime}$ model with an extremely weak alternation parameter $\alpha$ of 0.98~\cite{PhysRevB.100.144433}, in agreement with an estimated spin gap value of 2 K. The DFT study~\cite{PhysRevB.100.144433} indicated the possible presence of other subleading interactions. 
	\textcolor{black}{The effect of the subleading interactions in the description of the field-induced excited state properties remains to be explored.}
	
	In our study, we first revisit the first-principles modeling of NVOPO\textcolor{black}{. In this context, we extend beyond Ref.~\cite{PhysRevB.100.144433}} through two complementary techniques, a) \textcolor{black}{application of downfolding technique} to define low-energy \textcolor{black}{three orbital} model Hamiltonian in effective Wannier function basis, and b) mapping of GGA+$U$ total energy to spin model.
	The tight-binding representation of low-energy Hamiltonian suggests that 1st, 3rd, 4th and 8th neighbouring hopping pathways are enough to capture the low-energy physics. Following this finding, we construct a minimal spin-Hamiltonian with four exchange integrals\textcolor{black}{. We estimate the average values and the standard deviations of these four exchanges}, by
	solving a large number of equations ($\sim$ $^{11}C_{4}$) obtained
	from GGA+$U$ total calculations of 12 different spin configurations.
	This led to a spin model with the hierarchy of exchange coupling, two
	major, similar AFM interactions, $J$ and $J^{\prime}$, as found previously,~\cite{PhysRevB.100.144433} and two other sub-leading interactions, 
	FM $J_c$ and AFM $J_a$ with $J_c$ $>$ $J_a$. This led to an intricate 3D spin model of a parallel array of $J$-$J^{\prime}$ chains running along two perpendicular directions, in two consecutive $ab$ planes, stacked along c direction, connected by $J_c$ and a body-diagonal interaction of $J_a$.
	\textcolor{black}{Our rigorous first-principles modelling establishes a robust $J$-$J^{\prime}$ alternation parameter of $\sim$ 0.92, and a large variation in the values of the 
		two sub-leading interactions, dependent rather strongly on the chosen spin configuration.}

	The constructed spin model is solved using ED and DMRG. To make the problem numerically tractable, a 2D approximation of the full 3D model is used, with $J$-$J^{\prime}$ AFM chains alternating with uniform AFM chains of $(J+J^{\prime})/2$ exchanges,  connected via an FM $J_c$ and a renormalized FM exchange, $J_d$. \textcolor{black}{The large variation in the DFT estimated values of sub-leading
		inter-chain interactions prompted us to construct a phase diagram in the parameter space of $J_c$ and $J_d$.} The constructed \textcolor{black}{$J_c/J$-$J_d/J$} phase diagram with $J^{\prime}/J$ fixed at DFT derived value, results in three distinct phases, two long-range ordered phases with staggered and stripy local spin orientations, and a spin-gapped {\it disorder} phase, characterized by short-range correlation along \textcolor{black}{$J$-$J^{\prime}$} chain, and vanishing correlation along $J_c$ and $J_d$. Our analysis brings
	out the competing nature of the two subleading interactions, $J_c$ and $J_d$. 
	\textcolor{black}{This competing nature leads to a cancellation of interchain interactions at $J_c$=$J_d$ line in the GS. The system thus can be represented as a collection of decoupled weakly dimerized chains having finite spin gap.} 
	Away from the $J_c$=$J_d$ line, the gap is suppressed due to partial cancellation between $J_c$ and $J_d$, and finally ordered state prevails driven by the dominance of one competing interaction over the other. The computed spin gap value for choices of $J_c$ and $J_d$ in the disordered phase agrees well with the experimentally measured value. \textcolor{black}{The calculated thermodynamic properties like magnetic susceptibility and specific heat, also provide a faithful representation of the experimental data.\cite{PhysRevB.100.144433}}
	
	The disorder phase, stabilized by comparable values of competing interactions $J_c$ and $J_d$, as opposed to that of a single alternating dimerized chain, is found to host a multi-magnon condensed phase upon gap closing by a critical magnetic field. The competing sub-leading exchanges $J_c$ and $J_d$ are found to play a key role in driving the observed field-induced multi-magnon condensation state, by providing the glue to
	the magnetic field generated magnons on different chains \cite{2008Hikihara, 2009Sudan, 2018Parvej}. \textcolor{black}{In contrast to the dipolar XY order in the BEC~\cite{giamarchi2008bose}, the multi-magnon condensed phase exhibits finite multi-magnon binding energy and hosts multipolar order \cite{2008Hikihara, PARVEJ201696, Shanon_2006_square_lattice, Shanon_2006_triangular_lattice}. We computed the multipolar order parameter and found it to be finite in the multi-magnon condensed phase.} \textcolor{black}{Note,} very recently, the Tomonaga-Luttinger-liquid physics of NVOPO has been studied experimentally~\cite{PhysRevB.109.L060406}. This is a finite temperature phase, with a temperature scale above the scale of inter-chain interactions. Our focus is the low T behaviour, below the scale of sub-leading interactions $J_a$ ($J_d$), and $J_c$, that leads to the condensed phase rather than the liquid phase.
	
	A transition from gapful to gapless state is envisaged upon the application
	of moderate strength of the uniaxial strain, driving a quantum phase transition.
	\textcolor{black}{We further note that among the class vanadate compounds with
		chain-like structures, NVOPO is unique due to the weak nature of the alternation.
		The underlying physics upon application of magnetic field is expected to be different for AgVOAsO$_4$\cite{PhysRevB.83.144412} or NaVOAsO$_4$\cite{PhysRevB.99.014421} with significantly stronger alternation ratio of $\sim$ 0.6, and thus stronger dimerization. \textcolor{black}{In particular, the scenario of multi-magnon condensation in the disorder phase proposed in the present study relies on the phenomena of weak dimerization where magnons from each $J$ - $J^{\ast}$ chain generate n-magnon bound state. Note, the observed disorder phase in the present system arises due to frustration in the system not because of the dimerization.  This scenario of course is thus expected change in the presence of strong dimerization. The detailed study on this will be
			taken up in future.}}
	
	\textcolor{black}{Finally, while the 2D model was devised to overcome
		the numerical restrictions, the 2D model studied in this work motivated 
		by the electronic aspects of NVOPO, will in general be important in understanding the physics of the 2D frustrated model with possible implications to a variety of gaped quantum magnets. In particular, the
		stablization of multi-magnon condensed phase, driven by the frustration
		in weakly dimerized systems will be of potential interest to the community.}

	\acknowledgements
	
	MG acknowledges CSIR, India, for the senior research fellowship [grant no. 09/575 (0131) 2020-EMR-I. MR acknowledges DST-India for financial help. MK acknowledges support from SERB through Grant (Grant sanction No. CRG/2020/000754). T.S.D. acknowledges JC Bose National Fellowship (Grant No. JCB/2020/000004) for funding.

	\appendix
	
	\textcolor{black}{\section{Three-orbital Model}}
	
	Table~\ref{tab:t2g_table} lists the non-negligible hopping, connecting first neighbour (1NN) ($t_c$), 3NN ($t$), 4NN ($t^{'}$) and 8NN ($t_a$) V pairs. The onsite matrix elements show the presence of small off-diagonal elements due to angular distortion of O-V-O bond angles in the VO$_6$ octahedra.
	
	\begin{table*}[]
		\centering	
		\begin{tabular}{p{1.5cm} p{2.5cm}p{4cm} p{1cm} p{0.1cm} p{2cm} p{2cm} p{1.5cm}p{0.1cm}p{2cm} }
			\hline
			\hline
			$t_i$& Bond-length($\AA$) & Connecting vector &  & & $d_{xy}$&$d_{yz}$&$d_{xz}$ & & $J_{i}(K)$  \\
			\hline
			\\
			Onsite & &[0.000, 0.000, 0.000]  &$d_{xy}$ & \hspace{-1em}\rdelim\{{3}{*}[{}]& -1.057&      0.048  &   -0.001 &\hspace{-1em}\rdelim\}{3}{*}[{}] &   \\

			&   &  &$d_{yz}$ & &0.048 &    -0.432  &   -0.018 &  & \\
			
			&  & &$d_{xz}$ & & -0.001 &    -0.018 &     -0.373  & & \\
			\\

			$t_1$=$t_c$ & 3.565 &[0.000, -0.024, -0.421]   & $d_{xy}$&\hspace{-1em}\rdelim\{{3}{*}[{}] & 0.000&      0.043&     -0.007 & \hspace{-1em}\rdelim\}{3}{*}[{}]& -5.336  \\
			
			&     & & $d_{yz}$ & &-0.073 &    -0.261  &    0.128&  & \\
			
			&   &  & $d_{xz}$ & & -0.031  &   -0.152  &   -0.075 & &\\
			\\

			$t_3$=$t$ &5.337   &[-0.354, 0.476, 0.216]  & $d_{xy}$&\hspace{-1em}\rdelim\{{3}{*}[{}] &0.089    &  0.014    &  0.009&\hspace{-1em}\rdelim\}{3}{*}[{}] & 91.212 \\
			
			&  &   &$d_{yz}$ & &-0.014  &  0.005  &   -0.002 & & \\
			
			&  &  &$d_{xz}$& &0.009     & 0.002&      0.003 & & \\
			\\
			
			$t_4$=$t'$ &5.385   & [0.344, -0.524, -0.114]   & $d_{xy}$ & \hspace{-1em}\rdelim\{{3}{*}[{}] & 0.120 &    -0.035    &  0.028 & \hspace{-1em}\rdelim\}{3}{*}[{}] &161.665\\
			
			& &   &$d_{yz}$ & &0.035 &    -0.003   &   0.003 & & \\
			
			&  & &$d_{xz}$& & 0.028 &    -0.003  &    0.002 & & \\
			\\

			$t_8$=$t_a$ & 5.952 & [-0.698, -0.024, -0.091]     & $d_{xy}$ & \hspace{-1em}\rdelim\{{3}{*}[{}]&0.048 &    -0.002  &    0.003& \hspace{-1em}\rdelim\}{3}{*}[{}] & 25.120 \\
			
			&  &  &$d_{yz}$ & &0.024      &0.010 &    -0.004 &  &\\
			
			&   &  &$d_{xz}$ & & -0.002    &0.004  &   -0.000   & &\\

			\\
			\hline		
			\hline		
		\end{tabular}
		\caption{3 $\times$ 3 onsite matrices and hopping integrals
			of the V $t_{2g}$ only three-orbital model. The matrix elements are given in eV. 
			Listed are also the $J$ values in K, obtained employing the perturbative approach.}
		\label{tab:t2g_table}
	\end{table*}
	
	\textcolor{black}{\section{Details of Total energy calculations and exchanges}}
	
	\begin{figure}[h]
		\centering
		\includegraphics[width=0.6\columnwidth]{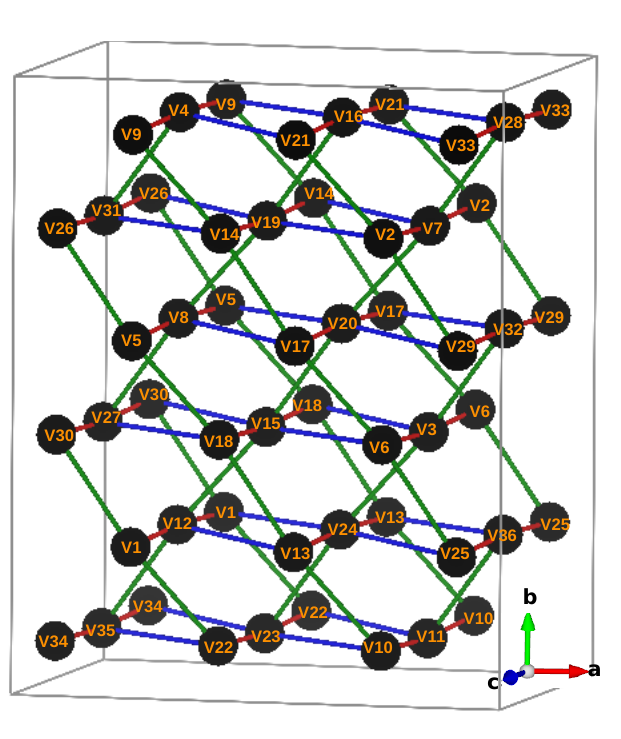}
		\caption{3 $\times$ 3 $\times$ 1 supercell containing 36 V ions, employed for
			total energy calculations. The numbering of V atoms are used to represent the magnetic arrangements given in Table~\ref{tab:table-III}. The V ions connected via $J_c$, $J_a$ and $J$-$J^{\prime}$ are shown as magenta, blue and green bonds, respectively.}
		\label{GGA+U_energy_mapping}
	\end{figure} 
	
	We consider a 3 $\times$ 3 $\times$ 1 supercell, containing 36 V ions (cf Fig.~\ref{GGA+U_energy_mapping}). \textcolor{black}{There should thus be 2$^{36}$ configurations of V-spins possible, out of which 12 were found to be independent (distinct), including the FM arrangement.} This leads to 11 different energy differences,
	measured from FM energy, and thus 11 different equations with four unknown $J$'s. Choosing four
	of the 11 available equations leads to 330 possible sets of values of $J$. Table~\ref{tab:table-III} shows the
	spin arrangements, energy differences and exchange values in four representative sets, obtained
	from GGA+$U$ calculations with $U$ = 4 eV. We note the $U$ values estimated by linear response for vanadium oxides \textcolor{black}{lie} in range of 4 to 5 eV \cite{Uvalue}.
	In all the cases, $J$ and $J^{\prime}$ turn out to be the dominant exchanges of antiferromagnetic nature of comparable magnitude, along with sub-leading exchanges, $J_c$ and $J_a$ of antiferro- and ferro-magnetic nature, in agreement with the perturbative estimates. The average along with the percentage standard deviation of $J$ values, considering all the sets, are also shown in Table~\ref{tab:table-III}. In addition to the estimates obtained from total energies calculated with $U$ = 4 eV, those obtained with the choice of $U$ = 5 eV are presented for comparison. The qualitative trend remains the same.  This leads to a spin model of three-dimensionally coupled, weakly alternating spin chain.

	\begin{table*}
		\centering
		\begin{tabular}{p{1.2cm}p{0.7cm}p{0.7cm}p{0.7cm}p{0.7cm}p{0.7cm}p{0.7cm}p{0.7cm}p{0.7cm}p{0.9cm}p{0.7cm}p{0.7cm}p{0.7cm}p{0.7cm}p{0.7cm}p{0.7cm}p{0.7cm}p{0.7cm}p{0.9cm}p{1.2cm}}
			\hline
			\hline
			$AFM{i}$&1-8 & 9 &10& 11& 12& 13& 14& 15& 16-19& 20& 21& 22& 23& 24& 25& 26& 27& 28-36&$\Delta$E(meV)\\
			\hline
			$AFM1$& +& +& +& +& -& +& +& -& +& +& +& +& +& +& +& +& +&+&−2.218\\
			
			$AFM2$& +& -& -& -& +& +& +& +& +& +& +& +& +& +& +& +& +&+&−10.989\\
			
			$AFM3$ &+& +& +& +& +& + &+& + &+& -& -& -& +& +& +& +& +&+&−5.598 \\
			
			$AFM4$ &+ &+ &+ &+ &+ &+ &+ &+ &+& +& +& +& +& +& -& -& -&+&−9.243\\
			
		\end{tabular} \\
		Set I: Exchange integral (K) $J$ =45.77, $J'$ =42.30, $J_c$ =-10.14, and $J_a$=1.86 \\

		\begin{tabular}{p{1.2cm}p{0.7cm}p{0.7cm}p{0.7cm}p{0.7cm}p{0.7cm}p{0.7cm}p{0.7cm}p{0.7cm}p{0.9cm}p{0.7cm}p{0.7cm}p{0.7cm}p{0.7cm}p{0.7cm}p{0.7cm}p{0.7cm}p{0.7cm}p{0.9cm}p{1.2cm}}

			$AFM1$& +& -& -& -& +& +& +& +& +& +& +& +& +& +& +& +& +&+&−10.989\\
			
			$AFM2$ &+& +& +& +& +& + &+& + &+& -& -& -& +& +& +& +& +&+&−5.598 \\
			
			$AFM3$ &+ &+ &+ &+ &+ &+ &+ &+ &+& +& +& +& +& +& -& -& -&+&−9.243\\
			
			$AFM4$& +& +& +& +& -& +& +& -& +& -& +& +& +& +& +& +& +&+&−1.730\\
			
		\end{tabular} \\
		Set II: Exchange integral (K) $J$ =44.89, $J'$ =42.30, $J_c$ =-10.14, and $J_a$=2.30 \\
		
		\begin{tabular}{p{1.2cm}p{0.7cm}p{0.7cm}p{0.7cm}p{0.7cm}p{0.7cm}p{0.7cm}p{0.7cm}p{0.7cm}p{0.9cm}p{0.7cm}p{0.7cm}p{0.7cm}p{0.7cm}p{0.7cm}p{0.7cm}p{0.7cm}p{0.7cm}p{0.9cm}p{1.2cm}}
			\\

			$AFM1$ &+ &+ &+ &+& -& -& -& +& +& +& +& +& +& +& +& +& +&+&−8.582 \\
			
			$AFM2$ & +& +& +& +& +& +& +& +& +& +& -& -& -& +& +& +& +&+&−7.482 \\
			
			$AFM3$ & +& +& +& +& +& +& +& +& +& +& +& +& -& -& -& +& +&+&−5.067 \\
			
			$AFM4$& +& +& +& +& -& +& +& -& +& -& +& +& +& +& +& +& +&+&−1.730\\
			
		\end{tabular} \\
		Set III: Exchange integral (K)   $J$ = 43.20, $J'$ =40.78, $J_c$ = -11.77, and $J_a$= 4.47\\

		\begin{tabular}{p{1.2cm}p{0.7cm}p{0.7cm}p{0.7cm}p{0.7cm}p{0.7cm}p{0.7cm}p{0.7cm}p{0.7cm}p{0.9cm}p{0.7cm}p{0.7cm}p{0.7cm}p{0.7cm}p{0.7cm}p{0.7cm}p{0.7cm}p{0.7cm}p{0.9cm}p{1.2cm}}
			\\
			$AFM1$ &+& +& +& +& +& + &+& + &+& -& -& -& +& +& +& +& +&+&−5.598 \\
			
			$AFM2$ & +& +& +& +& +& +& +& +& +& +& +& +& -& -& -& +& +&+&−5.067 \\
			
			$AFM3$ &+ &+ &+ &+ &+ &+ &+ &+ &+& +& +& +& +& +& -& -& -&+&−9.243\\
			
			$AFM4$& +& +& +& +& -& +& +& -& +& -& +& +& +& +& +& +& +&+&−1.730\\

		\end{tabular} \\
		Set IV: Exchange integral (K)   $J$ = 44.89, $J'$ =42.30, $J_c$ =-13.99, and $J_a$= 6.16\\

		\begin{tabular}{p{2.5cm}p{2.5cm}p{3.cm}p{3.cm}p{3.cm}p{3.cm}}
			\\
			\hline

			$U$ (eV) & J$_{H}$ (eV)  &$J_{c}$(K)  & $J(K)$ & $J'(K)$ & $J_{a}$(K)  \\
			\hline
			4 &1 & -12.17 $ \pm$ 12.56 $\%$ &44.51,$ \pm$ 2.07$\%$&40.59 $ \pm$ 2.41$\%$ &4.71 $ \pm$ 31.34$\%$ \\
			5 &1  & -11.13 $ \pm$ 11.02 $\%$ & 36.33 $ \pm$ 2.03$\%$ & 32.84 $\pm$ 2.38$\%$& 3.89 $ \pm$ 30.30$\%$ \\
			\hline
			\hline
		\end{tabular}\\
		\captionof{table}{The spin arrangements (cf Fig.~\ref{GGA+U_energy_mapping} for
			numbering of the V ions) with "+" and "-" denoting up and down spins, total energy difference $\Delta E$, exchange integral estimates for four representative sets. The bottom table lists
			the average values and standard deviation of $J's$ obtained from GGA+$U$ calculations with choice of $U$ = 4eV and 5eV, and Hund's coupling of J$_H$ = 1 eV, as appropriate for 3d transition metal.}
		\label{tab:table-III}
		

\end{table*}

\textcolor{black}{\section{Construction of an effective 2D model from the 3D model}}
\begin{figure}
\centering
\includegraphics[width=0.9\columnwidth]{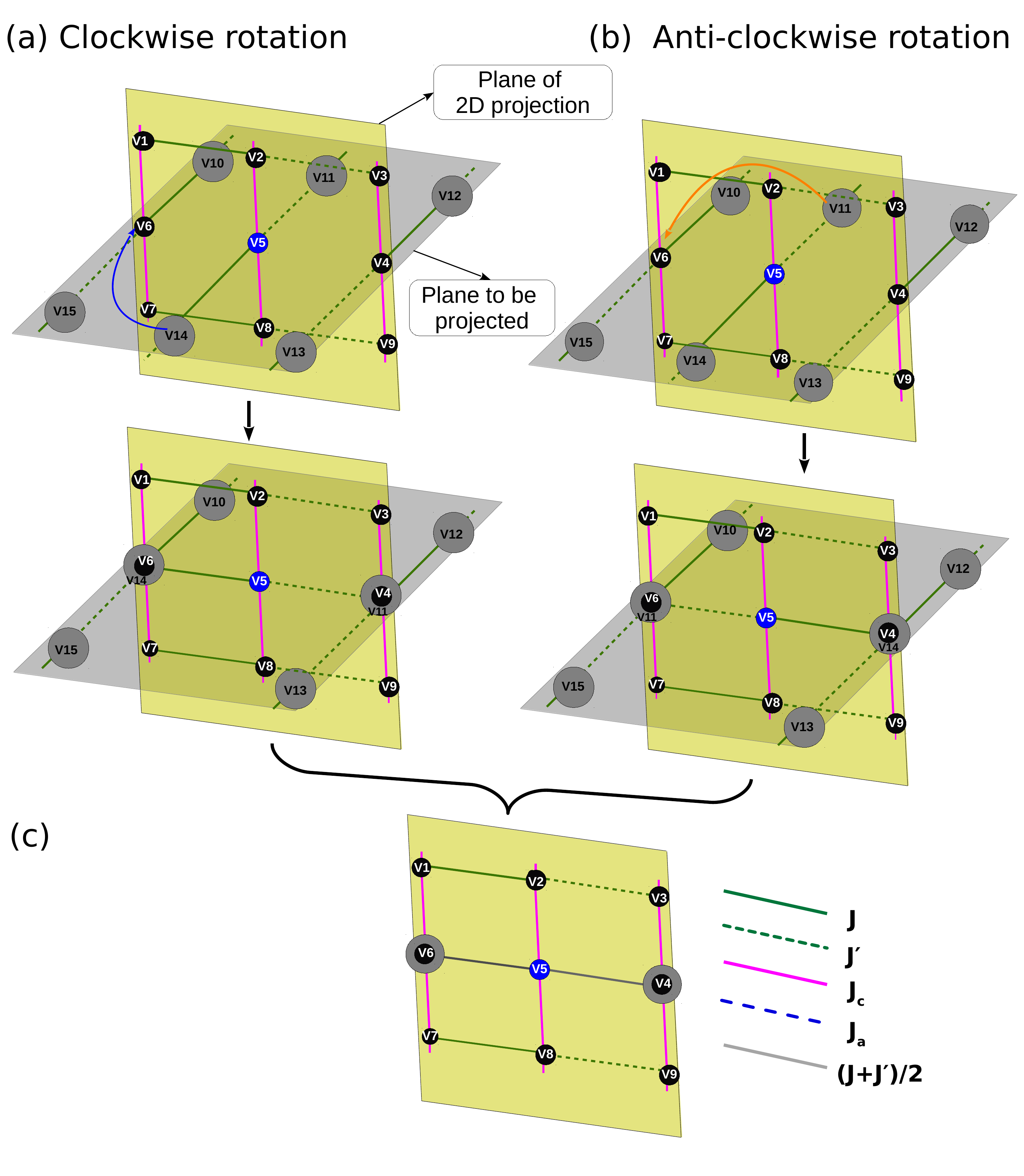}
\caption{\textcolor{black}{Mapping of 3D model of J-J$^{\prime}$ chains to an effective 2D model. (a) Following clockwise rotation and (b) following anti-clockwise rotation. The resultant 2D model is shown in (c).}}
\label{fig:2D-3D}
\end{figure}

\begin{figure}
\centering
\includegraphics[width=0.9\columnwidth]{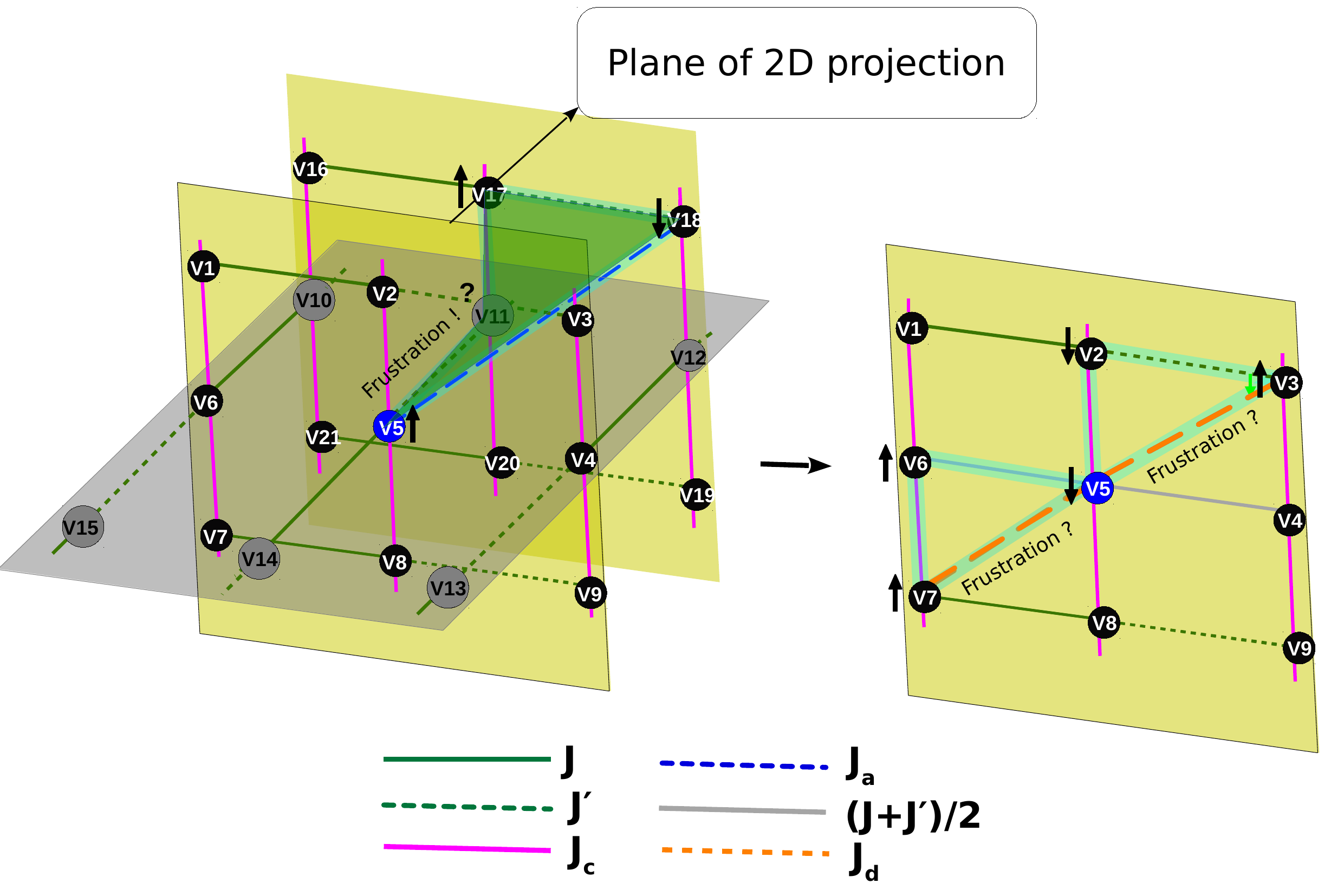}
\caption{\textcolor{black}{Frustration effect in 3D model (left) and in the effective 2D model (right).}}
\label{fig:frustration}
\end{figure}

\textcolor{black}{Construction of an effective 2D model from the 3D model involves rotation of planes. As shown in Fig 1(c),
the alternating chains belong to two perpendicular planes, rotated by 90$^{0}$ with respect to each other, giving rise to a 3D
character, even in the absence of the diagonal interaction, $J_a$ and out-of-plane interaction, $J_c$. The 2D mapping of the 3D model
containing two perpendicular planes of alternating chains is illustrated in Fig.~\ref{fig:2D-3D}. As is seen
from the figure, the site V5 (marked in blue) is common between the two planes, the plane of 2D projection and the plane
to be projected. The plane can be projected through clockwise (cf panel (a)) and anti-clockwise (cf panel (b)) rotation
which maps sites V14 and V11 from the plane to be projected to sites V6 and V4 in the plane of 2D projection. The clockwise
(anti-clockwise) rotations connect V6 to V14 (V11) and V4 to V11(V14). While in the 3D model, the site
V5 is not connected to V4 and V6, in the effective 2D model, a $J$ or $J^{\prime}$ bond is established following either
clockwise or anti-clockwise rotation. The resultant one, combining the two paths of rotation is shown in panel (c) of
the figure, where the site V5, gets connected to effective sites V6 and V4, via an average interaction $\frac{(J+J^{\prime})}{2}$.}

\textcolor{black}{Finally, the frustration in a 3D model involving three AFM exchanges and an FM exchange, and that in an effective 2D model involving
one AFM and two FM exchanges is illustrated in Fig.~\ref{fig:frustration}. As in seen, frustration originates from $J_a$ and $J_c$
in the 3D model, while it originates from $J_d$ in 2D model. $J_d$ is thus approximated as $J_a$ + $\gamma$ $J_c$, with a variational parameter $\gamma$. With $\vert J_c \vert $ (FM) 2-4 times larger than $\vert J_a \vert$ (AFM), J$_d$ is of FM nature, driving
the frustration in effective 2D model.}

\textcolor{black}{\section{DMRG details}}
\begin{figure}
\centering
\includegraphics[width=0.9\columnwidth]{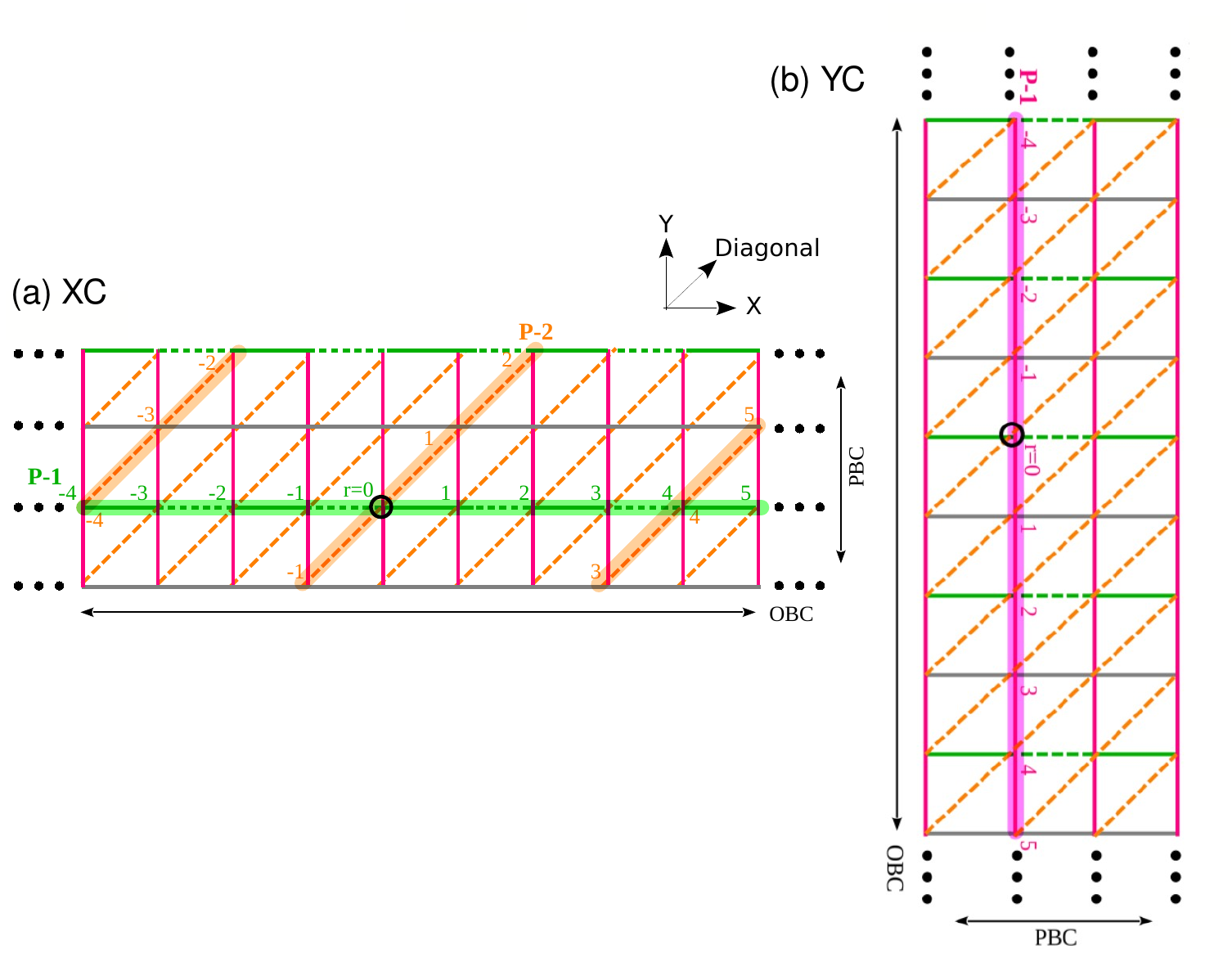}
\caption{\textcolor{black}{ XC/YC geometry used in DMRG calculations, as shown in (a)/(b) with PBC along Y/X and OBC along X/Y.}}
\label{fig:XCYCDMRG-supp}
\end{figure}
For DMRG calculations we considered cylinders of two geometries XC and YC. In XC geometry, shown in Fig.~\ref{fig:XCYCDMRG-supp}(a), periodic boundary condition (PBC) is applied along $J_c$ bonds, while open boundary condition (OBC) is applied along the $J-J^{\prime}$ chain.
This geometry enables the calculation of spin correlation along X ($J-J^{\prime}$ chain direction) and diagonal ($J_d$ chain direction) paths, P-1 and P-2, highlighted by green and orange lines in the figure. The YC geometry is a 90$^\circ$ rotation of the XC geometry, with periodic boundary condition along $J-J^{\prime}$ chain and open boundary condition along $J_c$, as shown in 
Fig.~\ref{fig:XCYCDMRG-supp}(b). In this geometry, we consider only one path P-1 along the X-direction ($J_c$ chain direction) highlighted by the magenta line.

\textcolor{black}{\section{Multipolar Order}}
\begin{figure}[h]
\centering
\includegraphics[width=1.0\columnwidth]{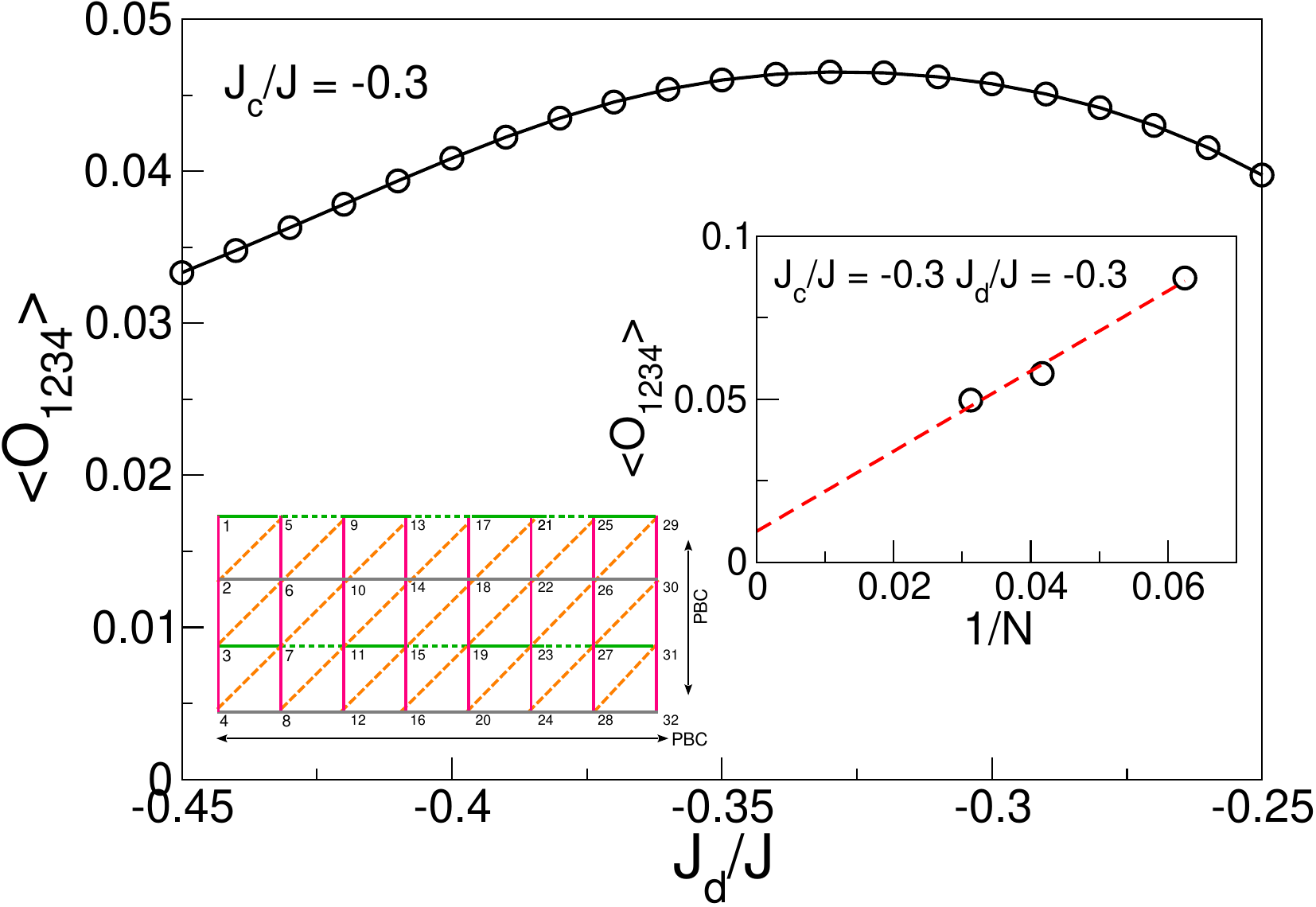}
\caption{\textcolor{black}{Multipolar order parameter ($\hat{O}_{1234}$) as a function of $J_{d}/J$ in the multi-magnon condensed phase with system size 8$\times$ 4. In all the calculations, $J^{\prime}/J$ and $J_{c}/J$ are fixed at 0.92 and -0.3, respectively. The inset shows the system size dependence of multipolar order for the representative case of $J_{d}/J = -0.3$.}}
\label{fig:multipolar}
\end{figure}
\textcolor{black}{In the presence of competing exchange interactions, magnons can bind together and form a bound state. In some cases, only two magnons pair up and form a quadrupolar state in the presence of a axial field \cite{2008Hikihara, Shanon_2006_square_lattice} and have the symmetric order parameter defined as, $Q_{ij}^{x^{2} - y^{2}} = \frac{1}{2} (S_{i}^{+}S_{j}^{+} + S_{i}^{-}S_{j}^{-})$, where $S_{i}^{+}$/$S_{j}^{-}$ is raising/lowering spin operator at site i/j. For three magnon bound pair the order parameter~\cite{Shanon_2006_triangular_lattice} is defined as, $\hat{O}_{123} = (S_{1}^{-}S_{2}^{-}S_{3}^{-} + h.c)$. This order parameter can be generalized to $n$-magnon bound state as,
\begin{eqnarray} 
	\hat{O}_{123 ... n}= (S_{i,1}^{-}S_{i,2}^{-}S_{i,3}^{-}S_{i,4}^{-} ... S_{i,n}^{-} + h.c)
\end{eqnarray}}

\textcolor{black}{We notice that in Fig.~\ref{fig:MHDMRG}(b) the step jump in magnetization curve is proportional to the width of the 2D geometry. At the point of jump the GS state is two-fold degenerate and GS energy of $S^{z}$ and $S^{z}=n$ sectors are same, i,e., $E(s^{z})=E(S^{z}+n)$. Therefore the expectation value of $\hat{O}_{1234 ... n}$ can be calculated as eigenvalue of $2\times 2$ matrix with elements $\langle  \psi_{\alpha}| \hat{O}_{123 ... n} | \psi_{\beta} \rangle$, where $|\psi_{\alpha/\beta}\rangle$ are GS wave-function with spin multiplicity $\alpha / \beta$, where $\alpha / \beta$ can take the value of $S^{z}$ or $S^{z}+n$. Thus one can write the $2 \times 2$ matrix as,}

\textcolor{black}{\begin{eqnarray}
	\hat{O}_{123 ... n}  = 
	\begin{bmatrix}
		\langle  \psi_{S^{z}}| \hat{O}_{123 ... n} | \psi_{S^{z}} \rangle & \langle  \psi_{S^{z}}| \hat{O}_{123 ... n} | \psi_{S^{z}+n} \rangle \\
		\langle  \psi_{S^{z}+n}| \hat{O}_{123 ... n} | \psi_{S^{z}} \rangle & \langle  \psi_{S^{z}+n}| \hat{O}_{123 ... n} | \psi_{S^{z}+n} \rangle
	\end{bmatrix} \nonumber \\
\end{eqnarray}
}

\textcolor{black}{where, $\langle  \psi_{S^z/S^z+n}| \hat{O}_{123 ... n} | \psi_{S^z/Sz+n} \rangle  = 0$ and $\langle  \psi_{S^z/S^z+n}| \hat{O}_{123 ... n} | \psi_{S^z+n/Sz`} \rangle  \ne 0$. }

\textcolor{black}{In this case, the order parameter for a system of width four sites along the \( J_c \) bond is \(\hat{O}_{1234} = (S_{i,1}^{-}S_{i,2}^{-}S_{i,3}^{-}S_{i,4}^{-} + h.c)\), where \( S_{i}^{-} \) acts at site \( i \) along the \( J_c \) bond. We computed the order parameter (\(\hat{O}_{1234}\)) in the multi-magnon condensed phase in \( 4 \times 4 \), \( 6 \times 4 \), and \( 8 \times 4 \) toroidal geometries for fixed values of \( J^{\prime}/J = 0.92 \) and \( J_c/J = -0.3 \). Fig.~\ref{fig:multipolar} shows the plot of the order parameter (\(\hat{O}_{1234}\)) as a function of \( J_d/J \) for the \( 8 \times 4 \) toroidal geometry. The \( J_d/J \) values were chosen over a range ensuring the system remains in the disordered phase. We find that the order parameter is maximum at \( J_c/J = J_d/J \), where the four-magnon binding energy is also maximum. The inset of Fig.~\ref{fig:multipolar} shows the system size dependence of the order parameter (\(\hat{O}_{1234}\)). The system size dependence confirms that the order remains finite ($\sim 0.01$) in the thermodynamic limit.}

\bibliography{ref}

\begin{thebibliography}{54}%
\makeatletter
\providecommand \@ifxundefined [1]{%
 \@ifx{#1\undefined}
}%
\providecommand \@ifnum [1]{%
 \ifnum #1\expandafter \@firstoftwo
 \else \expandafter \@secondoftwo
 \fi
}%
\providecommand \@ifx [1]{%
 \ifx #1\expandafter \@firstoftwo
 \else \expandafter \@secondoftwo
 \fi
}%
\providecommand \natexlab [1]{#1}%
\providecommand \enquote  [1]{``#1''}%
\providecommand \bibnamefont  [1]{#1}%
\providecommand \bibfnamefont [1]{#1}%
\providecommand \citenamefont [1]{#1}%
\providecommand \href@noop [0]{\@secondoftwo}%
\providecommand \href [0]{\begingroup \@sanitize@url \@href}%
\providecommand \@href[1]{\@@startlink{#1}\@@href}%
\providecommand \@@href[1]{\endgroup#1\@@endlink}%
\providecommand \@sanitize@url [0]{\catcode `\\12\catcode `\$12\catcode
  `\&12\catcode `\#12\catcode `\^12\catcode `\_12\catcode `\%12\relax}%
\providecommand \@@startlink[1]{}%
\providecommand \@@endlink[0]{}%
\providecommand \url  [0]{\begingroup\@sanitize@url \@url }%
\providecommand \@url [1]{\endgroup\@href {#1}{\urlprefix }}%
\providecommand \urlprefix  [0]{URL }%
\providecommand \Eprint [0]{\href }%
\providecommand \doibase [0]{https://doi.org/}%
\providecommand \selectlanguage [0]{\@gobble}%
\providecommand \bibinfo  [0]{\@secondoftwo}%
\providecommand \bibfield  [0]{\@secondoftwo}%
\providecommand \translation [1]{[#1]}%
\providecommand \BibitemOpen [0]{}%
\providecommand \bibitemStop [0]{}%
\providecommand \bibitemNoStop [0]{.\EOS\space}%
\providecommand \EOS [0]{\spacefactor3000\relax}%
\providecommand \BibitemShut  [1]{\csname bibitem#1\endcsname}%
\let\auto@bib@innerbib\@empty
\bibitem [{\citenamefont {A.~N.~Vasil’ev}\ and\ \citenamefont
  {Popova}(2005)}]{Review_on_spin_chain}%
  \BibitemOpen
  \bibfield  {author} {\bibinfo {author} {\bibfnamefont {M.~M.~M.}\
  \bibnamefont {A.~N.~Vasil’ev}}\ and\ \bibinfo {author} {\bibfnamefont
  {E.~A.}\ \bibnamefont {Popova}},\ }\bibfield  {title} {\bibinfo {title} {Spin
  gap in low-dimensional magnets},\ }\href {https://doi.org/10.1063/1.1884423}
  {\bibfield  {journal} {\bibinfo  {journal} {Low Temp. Phys.}\ }\textbf
  {\bibinfo {volume} {31}},\ \bibinfo {pages} {203} (\bibinfo {year}
  {(2005})}\BibitemShut {NoStop}%
\bibitem [{\citenamefont {Sachdev}(2000)}]{science.288.5465.475}%
  \BibitemOpen
  \bibfield  {author} {\bibinfo {author} {\bibfnamefont {S.}~\bibnamefont
  {Sachdev}},\ }\bibfield  {title} {\bibinfo {title} {Quantum criticality:
  Competing ground states in low dimensions},\ }\href
  {https://doi.org/10.1126/science.288.5465.475} {\bibfield  {journal}
  {\bibinfo  {journal} {Science}\ }\textbf {\bibinfo {volume} {288}},\ \bibinfo
  {pages} {475} (\bibinfo {year} {2000})}\BibitemShut {NoStop}%
\bibitem [{\citenamefont {Sachdev}(2008)}]{sachdev2008quantum}%
  \BibitemOpen
  \bibfield  {author} {\bibinfo {author} {\bibfnamefont {S.}~\bibnamefont
  {Sachdev}},\ }\bibfield  {title} {\bibinfo {title} {Quantum magnetism and
  criticality},\ }\href {https://doi.org/10.1038/nphys894} {\bibfield
  {journal} {\bibinfo  {journal} {Nature Physics}\ }\textbf {\bibinfo {volume}
  {4}},\ \bibinfo {pages} {173} (\bibinfo {year} {2008})}\BibitemShut {NoStop}%
\bibitem [{\citenamefont {Mukhopadhyay}\ \emph {et~al.}(2012)\citenamefont
  {Mukhopadhyay}, \citenamefont {Klanj\ifmmode~\check{s}\else \v{s}\fi{}ek},
  \citenamefont {Grbi\ifmmode~\acute{c}\else \'{c}\fi{}}, \citenamefont
  {Blinder}, \citenamefont {Mayaffre}, \citenamefont {Berthier}, \citenamefont
  {Horvati\ifmmode~\acute{c}\else \'{c}\fi{}}, \citenamefont {Continentino},
  \citenamefont {Paduan-Filho}, \citenamefont {Chiari},\ and\ \citenamefont
  {Piovesana}}]{PhysRevLett.109.177206}%
  \BibitemOpen
  \bibfield  {author} {\bibinfo {author} {\bibfnamefont {S.}~\bibnamefont
  {Mukhopadhyay}}, \bibinfo {author} {\bibfnamefont {M.}~\bibnamefont
  {Klanj\ifmmode~\check{s}\else \v{s}\fi{}ek}}, \bibinfo {author}
  {\bibfnamefont {M.~S.}\ \bibnamefont {Grbi\ifmmode~\acute{c}\else
  \'{c}\fi{}}}, \bibinfo {author} {\bibfnamefont {R.}~\bibnamefont {Blinder}},
  \bibinfo {author} {\bibfnamefont {H.}~\bibnamefont {Mayaffre}}, \bibinfo
  {author} {\bibfnamefont {C.}~\bibnamefont {Berthier}}, \bibinfo {author}
  {\bibfnamefont {M.}~\bibnamefont {Horvati\ifmmode~\acute{c}\else
  \'{c}\fi{}}}, \bibinfo {author} {\bibfnamefont {M.~A.}\ \bibnamefont
  {Continentino}}, \bibinfo {author} {\bibfnamefont {A.}~\bibnamefont
  {Paduan-Filho}}, \bibinfo {author} {\bibfnamefont {B.}~\bibnamefont
  {Chiari}},\ and\ \bibinfo {author} {\bibfnamefont {O.}~\bibnamefont
  {Piovesana}},\ }\bibfield  {title} {\bibinfo {title} {Quantum-critical spin
  dynamics in quasi-one-dimensional antiferromagnets},\ }\href
  {https://doi.org/10.1103/PhysRevLett.109.177206} {\bibfield  {journal}
  {\bibinfo  {journal} {Phys. Rev. Lett.}\ }\textbf {\bibinfo {volume} {109}},\
  \bibinfo {pages} {177206} (\bibinfo {year} {2012})}\BibitemShut {NoStop}%
\bibitem [{\citenamefont {Chitra}\ and\ \citenamefont
  {Giamarchi}(1997)}]{PhysRevB.55.5816}%
  \BibitemOpen
  \bibfield  {author} {\bibinfo {author} {\bibfnamefont {R.}~\bibnamefont
  {Chitra}}\ and\ \bibinfo {author} {\bibfnamefont {T.}~\bibnamefont
  {Giamarchi}},\ }\bibfield  {title} {\bibinfo {title} {Critical properties of
  gapped spin-chains and ladders in a magnetic field},\ }\href
  {https://doi.org/10.1103/PhysRevB.55.5816} {\bibfield  {journal} {\bibinfo
  {journal} {Phys. Rev. B}\ }\textbf {\bibinfo {volume} {55}},\ \bibinfo
  {pages} {5816} (\bibinfo {year} {1997})}\BibitemShut {NoStop}%
\bibitem [{\citenamefont {Klanj\ifmmode~\check{s}\else \v{s}\fi{}ek}\ \emph
  {et~al.}(2008)\citenamefont {Klanj\ifmmode~\check{s}\else \v{s}\fi{}ek},
  \citenamefont {Mayaffre}, \citenamefont {Berthier}, \citenamefont
  {Horvati\ifmmode~\acute{c}\else \'{c}\fi{}}, \citenamefont {Chiari},
  \citenamefont {Piovesana}, \citenamefont {Bouillot}, \citenamefont {Kollath},
  \citenamefont {Orignac}, \citenamefont {Citro},\ and\ \citenamefont
  {Giamarchi}}]{PhysRevLett.101.137207}%
  \BibitemOpen
  \bibfield  {author} {\bibinfo {author} {\bibfnamefont {M.}~\bibnamefont
  {Klanj\ifmmode~\check{s}\else \v{s}\fi{}ek}}, \bibinfo {author}
  {\bibfnamefont {H.}~\bibnamefont {Mayaffre}}, \bibinfo {author}
  {\bibfnamefont {C.}~\bibnamefont {Berthier}}, \bibinfo {author}
  {\bibfnamefont {M.}~\bibnamefont {Horvati\ifmmode~\acute{c}\else
  \'{c}\fi{}}}, \bibinfo {author} {\bibfnamefont {B.}~\bibnamefont {Chiari}},
  \bibinfo {author} {\bibfnamefont {O.}~\bibnamefont {Piovesana}}, \bibinfo
  {author} {\bibfnamefont {P.}~\bibnamefont {Bouillot}}, \bibinfo {author}
  {\bibfnamefont {C.}~\bibnamefont {Kollath}}, \bibinfo {author} {\bibfnamefont
  {E.}~\bibnamefont {Orignac}}, \bibinfo {author} {\bibfnamefont
  {R.}~\bibnamefont {Citro}},\ and\ \bibinfo {author} {\bibfnamefont
  {T.}~\bibnamefont {Giamarchi}},\ }\bibfield  {title} {\bibinfo {title}
  {Controlling luttinger liquid physics in spin ladders under a magnetic
  field},\ }\href {https://doi.org/10.1103/PhysRevLett.101.137207} {\bibfield
  {journal} {\bibinfo  {journal} {Phys. Rev. Lett.}\ }\textbf {\bibinfo
  {volume} {101}},\ \bibinfo {pages} {137207} (\bibinfo {year}
  {2008})}\BibitemShut {NoStop}%
\bibitem [{\citenamefont {Thielemann}\ \emph {et~al.}(2009)\citenamefont
  {Thielemann}, \citenamefont {R\"uegg}, \citenamefont {Kiefer}, \citenamefont
  {R\o{}nnow}, \citenamefont {Normand}, \citenamefont {Bouillot}, \citenamefont
  {Kollath}, \citenamefont {Orignac}, \citenamefont {Citro}, \citenamefont
  {Giamarchi}, \citenamefont {L\"auchli}, \citenamefont {Biner}, \citenamefont
  {Kr\"amer}, \citenamefont {Wolff-Fabris}, \citenamefont {Zapf}, \citenamefont
  {Jaime}, \citenamefont {Stahn}, \citenamefont {Christensen}, \citenamefont
  {Grenier}, \citenamefont {McMorrow},\ and\ \citenamefont
  {Mesot}}]{PhysRevB.79.020408}%
  \BibitemOpen
  \bibfield  {author} {\bibinfo {author} {\bibfnamefont {B.}~\bibnamefont
  {Thielemann}}, \bibinfo {author} {\bibfnamefont {C.}~\bibnamefont {R\"uegg}},
  \bibinfo {author} {\bibfnamefont {K.}~\bibnamefont {Kiefer}}, \bibinfo
  {author} {\bibfnamefont {H.~M.}\ \bibnamefont {R\o{}nnow}}, \bibinfo {author}
  {\bibfnamefont {B.}~\bibnamefont {Normand}}, \bibinfo {author} {\bibfnamefont
  {P.}~\bibnamefont {Bouillot}}, \bibinfo {author} {\bibfnamefont
  {C.}~\bibnamefont {Kollath}}, \bibinfo {author} {\bibfnamefont
  {E.}~\bibnamefont {Orignac}}, \bibinfo {author} {\bibfnamefont
  {R.}~\bibnamefont {Citro}}, \bibinfo {author} {\bibfnamefont
  {T.}~\bibnamefont {Giamarchi}}, \bibinfo {author} {\bibfnamefont {A.~M.}\
  \bibnamefont {L\"auchli}}, \bibinfo {author} {\bibfnamefont {D.}~\bibnamefont
  {Biner}}, \bibinfo {author} {\bibfnamefont {K.~W.}\ \bibnamefont {Kr\"amer}},
  \bibinfo {author} {\bibfnamefont {F.}~\bibnamefont {Wolff-Fabris}}, \bibinfo
  {author} {\bibfnamefont {V.~S.}\ \bibnamefont {Zapf}}, \bibinfo {author}
  {\bibfnamefont {M.}~\bibnamefont {Jaime}}, \bibinfo {author} {\bibfnamefont
  {J.}~\bibnamefont {Stahn}}, \bibinfo {author} {\bibfnamefont {N.~B.}\
  \bibnamefont {Christensen}}, \bibinfo {author} {\bibfnamefont
  {B.}~\bibnamefont {Grenier}}, \bibinfo {author} {\bibfnamefont {D.~F.}\
  \bibnamefont {McMorrow}},\ and\ \bibinfo {author} {\bibfnamefont
  {J.}~\bibnamefont {Mesot}},\ }\bibfield  {title} {\bibinfo {title}
  {Field-controlled magnetic order in the quantum spin-ladder system
  (\text{Hpip})$_{2}${CuBr}$_{4}$},\ }\href
  {https://doi.org/10.1103/PhysRevB.79.020408} {\bibfield  {journal} {\bibinfo
  {journal} {Phys. Rev. B}\ }\textbf {\bibinfo {volume} {79}},\ \bibinfo
  {pages} {020408} (\bibinfo {year} {2009})}\BibitemShut {NoStop}%
\bibitem [{\citenamefont {Aczel}\ \emph {et~al.}(2009)\citenamefont {Aczel},
  \citenamefont {Kohama}, \citenamefont {Marcenat}, \citenamefont {Weickert},
  \citenamefont {Jaime}, \citenamefont {Ayala-Valenzuela}, \citenamefont
  {McDonald}, \citenamefont {Selesnic}, \citenamefont {Dabkowska},\ and\
  \citenamefont {Luke}}]{PhysRevLett.103.207203}%
  \BibitemOpen
  \bibfield  {author} {\bibinfo {author} {\bibfnamefont {A.~A.}\ \bibnamefont
  {Aczel}}, \bibinfo {author} {\bibfnamefont {Y.}~\bibnamefont {Kohama}},
  \bibinfo {author} {\bibfnamefont {C.}~\bibnamefont {Marcenat}}, \bibinfo
  {author} {\bibfnamefont {F.}~\bibnamefont {Weickert}}, \bibinfo {author}
  {\bibfnamefont {M.}~\bibnamefont {Jaime}}, \bibinfo {author} {\bibfnamefont
  {O.~E.}\ \bibnamefont {Ayala-Valenzuela}}, \bibinfo {author} {\bibfnamefont
  {R.~D.}\ \bibnamefont {McDonald}}, \bibinfo {author} {\bibfnamefont {S.~D.}\
  \bibnamefont {Selesnic}}, \bibinfo {author} {\bibfnamefont {H.~A.}\
  \bibnamefont {Dabkowska}},\ and\ \bibinfo {author} {\bibfnamefont {G.~M.}\
  \bibnamefont {Luke}},\ }\bibfield  {title} {\bibinfo {title} {Field-induced
  bose-einstein condensation of triplons up to 8 k in
  $\text{Sr}_3\text{Cr}_2\text{O}_8$},\ }\href
  {https://doi.org/10.1103/PhysRevLett.103.207203} {\bibfield  {journal}
  {\bibinfo  {journal} {Phys. Rev. Lett.}\ }\textbf {\bibinfo {volume} {103}},\
  \bibinfo {pages} {207203} (\bibinfo {year} {2009})}\BibitemShut {NoStop}%
\bibitem [{\citenamefont {Hong}\ \emph {et~al.}(2010)\citenamefont {Hong},
  \citenamefont {Kim}, \citenamefont {Hotta}, \citenamefont {Takano},
  \citenamefont {Tremelling}, \citenamefont {Turnbull}, \citenamefont {Landee},
  \citenamefont {Kang}, \citenamefont {Christensen}, \citenamefont {Lefmann},
  \citenamefont {Schmidt}, \citenamefont {Uhrig},\ and\ \citenamefont
  {Broholm}}]{PhysRevLett.105.137207}%
  \BibitemOpen
  \bibfield  {author} {\bibinfo {author} {\bibfnamefont {T.}~\bibnamefont
  {Hong}}, \bibinfo {author} {\bibfnamefont {Y.~H.}\ \bibnamefont {Kim}},
  \bibinfo {author} {\bibfnamefont {C.}~\bibnamefont {Hotta}}, \bibinfo
  {author} {\bibfnamefont {Y.}~\bibnamefont {Takano}}, \bibinfo {author}
  {\bibfnamefont {G.}~\bibnamefont {Tremelling}}, \bibinfo {author}
  {\bibfnamefont {M.~M.}\ \bibnamefont {Turnbull}}, \bibinfo {author}
  {\bibfnamefont {C.~P.}\ \bibnamefont {Landee}}, \bibinfo {author}
  {\bibfnamefont {H.-J.}\ \bibnamefont {Kang}}, \bibinfo {author}
  {\bibfnamefont {N.~B.}\ \bibnamefont {Christensen}}, \bibinfo {author}
  {\bibfnamefont {K.}~\bibnamefont {Lefmann}}, \bibinfo {author} {\bibfnamefont
  {K.~P.}\ \bibnamefont {Schmidt}}, \bibinfo {author} {\bibfnamefont {G.~S.}\
  \bibnamefont {Uhrig}},\ and\ \bibinfo {author} {\bibfnamefont
  {C.}~\bibnamefont {Broholm}},\ }\bibfield  {title} {\bibinfo {title}
  {Field-induced tomonaga-luttinger liquid phase of a two-leg spin-1/2 ladder
  with strong leg interactions},\ }\href
  {https://doi.org/10.1103/PhysRevLett.105.137207} {\bibfield  {journal}
  {\bibinfo  {journal} {Phys. Rev. Lett.}\ }\textbf {\bibinfo {volume} {105}},\
  \bibinfo {pages} {137207} (\bibinfo {year} {2010})}\BibitemShut {NoStop}%
\bibitem [{\citenamefont {Willenberg}\ \emph {et~al.}(2015)\citenamefont
  {Willenberg}, \citenamefont {Ryll}, \citenamefont {Kiefer}, \citenamefont
  {Tennant}, \citenamefont {Groitl}, \citenamefont {Rolfs}, \citenamefont
  {Manuel}, \citenamefont {Khalyavin}, \citenamefont {Rule}, \citenamefont
  {Wolter},\ and\ \citenamefont {S\"ullow}}]{PhysRevB.91.060407}%
  \BibitemOpen
  \bibfield  {author} {\bibinfo {author} {\bibfnamefont {B.}~\bibnamefont
  {Willenberg}}, \bibinfo {author} {\bibfnamefont {H.}~\bibnamefont {Ryll}},
  \bibinfo {author} {\bibfnamefont {K.}~\bibnamefont {Kiefer}}, \bibinfo
  {author} {\bibfnamefont {D.~A.}\ \bibnamefont {Tennant}}, \bibinfo {author}
  {\bibfnamefont {F.}~\bibnamefont {Groitl}}, \bibinfo {author} {\bibfnamefont
  {K.}~\bibnamefont {Rolfs}}, \bibinfo {author} {\bibfnamefont
  {P.}~\bibnamefont {Manuel}}, \bibinfo {author} {\bibfnamefont
  {D.}~\bibnamefont {Khalyavin}}, \bibinfo {author} {\bibfnamefont {K.~C.}\
  \bibnamefont {Rule}}, \bibinfo {author} {\bibfnamefont {A.~U.~B.}\
  \bibnamefont {Wolter}},\ and\ \bibinfo {author} {\bibfnamefont
  {S.}~\bibnamefont {S\"ullow}},\ }\bibfield  {title} {\bibinfo {title}
  {Luttinger liquid behavior in the alternating spin-chain system copper
  nitrate},\ }\href {https://doi.org/10.1103/PhysRevB.91.060407} {\bibfield
  {journal} {\bibinfo  {journal} {Phys. Rev. B}\ }\textbf {\bibinfo {volume}
  {91}},\ \bibinfo {pages} {060407} (\bibinfo {year} {2015})}\BibitemShut
  {NoStop}%
\bibitem [{\citenamefont {Tsirlin}\ \emph {et~al.}(2011)\citenamefont
  {Tsirlin}, \citenamefont {Nath}, \citenamefont {Sichelschmidt}, \citenamefont
  {Skourski}, \citenamefont {Geibel},\ and\ \citenamefont
  {Rosner}}]{PhysRevB.83.144412}%
  \BibitemOpen
  \bibfield  {author} {\bibinfo {author} {\bibfnamefont {A.~A.}\ \bibnamefont
  {Tsirlin}}, \bibinfo {author} {\bibfnamefont {R.}~\bibnamefont {Nath}},
  \bibinfo {author} {\bibfnamefont {J.}~\bibnamefont {Sichelschmidt}}, \bibinfo
  {author} {\bibfnamefont {Y.}~\bibnamefont {Skourski}}, \bibinfo {author}
  {\bibfnamefont {C.}~\bibnamefont {Geibel}},\ and\ \bibinfo {author}
  {\bibfnamefont {H.}~\bibnamefont {Rosner}},\ }\bibfield  {title} {\bibinfo
  {title} {Frustrated couplings between alternating spin-$\frac{1}{2}$ chains
  in $\text{AgVOAsO}_{4}$},\ }\href
  {https://doi.org/10.1103/PhysRevB.83.144412} {\bibfield  {journal} {\bibinfo
  {journal} {Phys. Rev. B}\ }\textbf {\bibinfo {volume} {83}},\ \bibinfo
  {pages} {144412} (\bibinfo {year} {2011})}\BibitemShut {NoStop}%
\bibitem [{\citenamefont {Weickert}\ \emph {et~al.}(2019)\citenamefont
  {Weickert}, \citenamefont {Aczel}, \citenamefont {Stone}, \citenamefont
  {Garlea}, \citenamefont {Dong}, \citenamefont {Kohama}, \citenamefont
  {Movshovich}, \citenamefont {Demuer}, \citenamefont {Harrison}, \citenamefont
  {Gam\ifmmode~\dot{z}\else \.{z}\fi{}a}, \citenamefont {Steppke},
  \citenamefont {Brando}, \citenamefont {Rosner},\ and\ \citenamefont
  {Tsirlin}}]{PhysRevB.100.104422}%
  \BibitemOpen
  \bibfield  {author} {\bibinfo {author} {\bibfnamefont {F.}~\bibnamefont
  {Weickert}}, \bibinfo {author} {\bibfnamefont {A.~A.}\ \bibnamefont {Aczel}},
  \bibinfo {author} {\bibfnamefont {M.~B.}\ \bibnamefont {Stone}}, \bibinfo
  {author} {\bibfnamefont {V.~O.}\ \bibnamefont {Garlea}}, \bibinfo {author}
  {\bibfnamefont {C.}~\bibnamefont {Dong}}, \bibinfo {author} {\bibfnamefont
  {Y.}~\bibnamefont {Kohama}}, \bibinfo {author} {\bibfnamefont
  {R.}~\bibnamefont {Movshovich}}, \bibinfo {author} {\bibfnamefont
  {A.}~\bibnamefont {Demuer}}, \bibinfo {author} {\bibfnamefont
  {N.}~\bibnamefont {Harrison}}, \bibinfo {author} {\bibfnamefont {M.~B.}\
  \bibnamefont {Gam\ifmmode~\dot{z}\else \.{z}\fi{}a}}, \bibinfo {author}
  {\bibfnamefont {A.}~\bibnamefont {Steppke}}, \bibinfo {author} {\bibfnamefont
  {M.}~\bibnamefont {Brando}}, \bibinfo {author} {\bibfnamefont
  {H.}~\bibnamefont {Rosner}},\ and\ \bibinfo {author} {\bibfnamefont {A.~A.}\
  \bibnamefont {Tsirlin}},\ }\bibfield  {title} {\bibinfo {title}
  {Field-induced double dome and bose-einstein condensation in the crossing
  quantum spin chain system $\text{AgVOAsO}_{4}$},\ }\href
  {https://doi.org/10.1103/PhysRevB.100.104422} {\bibfield  {journal} {\bibinfo
   {journal} {Phys. Rev. B}\ }\textbf {\bibinfo {volume} {100}},\ \bibinfo
  {pages} {104422} (\bibinfo {year} {2019})}\BibitemShut {NoStop}%
\bibitem [{\citenamefont {Arjun}\ \emph {et~al.}(2019)\citenamefont {Arjun},
  \citenamefont {Ranjith}, \citenamefont {Koo}, \citenamefont {Sichelschmidt},
  \citenamefont {Skourski}, \citenamefont {Baenitz}, \citenamefont {Tsirlin},\
  and\ \citenamefont {Nath}}]{PhysRevB.99.014421}%
  \BibitemOpen
  \bibfield  {author} {\bibinfo {author} {\bibfnamefont {U.}~\bibnamefont
  {Arjun}}, \bibinfo {author} {\bibfnamefont {K.~M.}\ \bibnamefont {Ranjith}},
  \bibinfo {author} {\bibfnamefont {B.}~\bibnamefont {Koo}}, \bibinfo {author}
  {\bibfnamefont {J.}~\bibnamefont {Sichelschmidt}}, \bibinfo {author}
  {\bibfnamefont {Y.}~\bibnamefont {Skourski}}, \bibinfo {author}
  {\bibfnamefont {M.}~\bibnamefont {Baenitz}}, \bibinfo {author} {\bibfnamefont
  {A.~A.}\ \bibnamefont {Tsirlin}},\ and\ \bibinfo {author} {\bibfnamefont
  {R.}~\bibnamefont {Nath}},\ }\bibfield  {title} {\bibinfo {title} {Singlet
  ground state in the alternating spin-$\frac{1}{2}$ chain compound
  $\text{NaVOAsO}_{4}$},\ }\href {https://doi.org/10.1103/PhysRevB.99.014421}
  {\bibfield  {journal} {\bibinfo  {journal} {Phys. Rev. B}\ }\textbf {\bibinfo
  {volume} {99}},\ \bibinfo {pages} {014421} (\bibinfo {year}
  {2019})}\BibitemShut {NoStop}%
\bibitem [{\citenamefont {Mukharjee}\ \emph {et~al.}(2019)\citenamefont
  {Mukharjee}, \citenamefont {Ranjith}, \citenamefont {Koo}, \citenamefont
  {Sichelschmidt}, \citenamefont {Baenitz}, \citenamefont {Skourski},
  \citenamefont {Inagaki}, \citenamefont {Furukawa}, \citenamefont {Tsirlin},\
  and\ \citenamefont {Nath}}]{PhysRevB.100.144433}%
  \BibitemOpen
  \bibfield  {author} {\bibinfo {author} {\bibfnamefont {P.~K.}\ \bibnamefont
  {Mukharjee}}, \bibinfo {author} {\bibfnamefont {K.~M.}\ \bibnamefont
  {Ranjith}}, \bibinfo {author} {\bibfnamefont {B.}~\bibnamefont {Koo}},
  \bibinfo {author} {\bibfnamefont {J.}~\bibnamefont {Sichelschmidt}}, \bibinfo
  {author} {\bibfnamefont {M.}~\bibnamefont {Baenitz}}, \bibinfo {author}
  {\bibfnamefont {Y.}~\bibnamefont {Skourski}}, \bibinfo {author}
  {\bibfnamefont {Y.}~\bibnamefont {Inagaki}}, \bibinfo {author} {\bibfnamefont
  {Y.}~\bibnamefont {Furukawa}}, \bibinfo {author} {\bibfnamefont {A.~A.}\
  \bibnamefont {Tsirlin}},\ and\ \bibinfo {author} {\bibfnamefont
  {R.}~\bibnamefont {Nath}},\ }\bibfield  {title} {\bibinfo {title}
  {Bose-einstein condensation of triplons close to the quantum critical point
  in the quasi-one-dimensional spin-$\frac{1}{2}$ antiferromagnet
  $\text{NaVOPO}_{4}$},\ }\href {https://doi.org/10.1103/PhysRevB.100.144433}
  {\bibfield  {journal} {\bibinfo  {journal} {Phys. Rev. B}\ }\textbf {\bibinfo
  {volume} {100}},\ \bibinfo {pages} {144433} (\bibinfo {year}
  {2019})}\BibitemShut {NoStop}%
\bibitem [{\citenamefont {Lii}\ \emph {et~al.}(1991)\citenamefont {Lii},
  \citenamefont {Li}, \citenamefont {Chen},\ and\ \citenamefont
  {Wang}}]{+1991+67+73}%
  \BibitemOpen
  \bibfield  {author} {\bibinfo {author} {\bibfnamefont {K.}~\bibnamefont
  {Lii}}, \bibinfo {author} {\bibfnamefont {C.}~\bibnamefont {Li}}, \bibinfo
  {author} {\bibfnamefont {T.-M.}\ \bibnamefont {Chen}},\ and\ \bibinfo
  {author} {\bibfnamefont {S.}~\bibnamefont {Wang}},\ }\bibfield  {title}
  {\bibinfo {title} {Synthesis and structural characterization of sodium
  vanadyl (iv) orthophosphate $\text{NaVOPO}_4$},\ }\href
  {https://doi.org/doi:10.1524/zkri.1991.197.1-2.67} {\bibfield  {journal}
  {\bibinfo  {journal} {Zeitschrift f{\"u}r Kristallographie}\ }\textbf
  {\bibinfo {volume} {197}},\ \bibinfo {pages} {67} (\bibinfo {year}
  {1991})}\BibitemShut {NoStop}%
\bibitem [{\citenamefont {Benhamada}\ \emph {et~al.}(1992)\citenamefont
  {Benhamada}, \citenamefont {Grandin}, \citenamefont {Borel}, \citenamefont
  {Leclaire},\ and\ \citenamefont {Raveau}}]{benhamada1992synthese}%
  \BibitemOpen
  \bibfield  {author} {\bibinfo {author} {\bibfnamefont {L.}~\bibnamefont
  {Benhamada}}, \bibinfo {author} {\bibfnamefont {A.}~\bibnamefont {Grandin}},
  \bibinfo {author} {\bibfnamefont {M.-M.}\ \bibnamefont {Borel}}, \bibinfo
  {author} {\bibfnamefont {A.}~\bibnamefont {Leclaire}},\ and\ \bibinfo
  {author} {\bibfnamefont {B.}~\bibnamefont {Raveau}},\ }\bibfield  {title}
  {\bibinfo {title} {Synth{\`e}se et structure cristalline d'un nouveau
  phosphate de vanadium (iv), $\text{NaVPO}_5$},\ }\href
  {https://inis.iaea.org/search/search.aspx?orig_q=RN:24000100} {\bibfield
  {journal} {\bibinfo  {journal} {Comptes rendus de l'Acad{\'e}mie des
  sciences. S{\'e}rie 2, M{\'e}canique, Physique, Chimie, Sciences de
  l'univers, Sciences de la Terre}\ }\textbf {\bibinfo {volume} {314}},\
  \bibinfo {pages} {585} (\bibinfo {year} {1992})}\BibitemShut {NoStop}%
\bibitem [{\citenamefont {Iffer}\ \emph {et~al.}(2021)\citenamefont {Iffer},
  \citenamefont {Belaiche}, \citenamefont {Ferdi}, \citenamefont {Elansary},
  \citenamefont {Sunar}, \citenamefont {Wang},\ and\ \citenamefont
  {Cao}}]{cathoda_appl}%
  \BibitemOpen
  \bibfield  {author} {\bibinfo {author} {\bibfnamefont {E.~a.}\ \bibnamefont
  {Iffer}}, \bibinfo {author} {\bibfnamefont {M.}~\bibnamefont {Belaiche}},
  \bibinfo {author} {\bibfnamefont {C.~A.}\ \bibnamefont {Ferdi}}, \bibinfo
  {author} {\bibfnamefont {M.}~\bibnamefont {Elansary}}, \bibinfo {author}
  {\bibfnamefont {A.~K.}\ \bibnamefont {Sunar}}, \bibinfo {author}
  {\bibfnamefont {Y.}~\bibnamefont {Wang}},\ and\ \bibinfo {author}
  {\bibfnamefont {Y.}~\bibnamefont {Cao}},\ }\bibfield  {title} {\bibinfo
  {title} {Monoclinic $\alpha$-{NaVOPO$_4$} as cathode materials for
  sodium-ions batteries: Experimental and {DFT} investigation},\ }\href
  {https://doi.org/https://doi.org/10.1002/er.5835} {\bibfield  {journal}
  {\bibinfo  {journal} {International Journal of Energy Research}\ }\textbf
  {\bibinfo {volume} {45}},\ \bibinfo {pages} {1703} (\bibinfo {year}
  {2021})},\ \Eprint
  {https://arxiv.org/abs/https://onlinelibrary.wiley.com/doi/pdf/10.1002/er.5835}
  {https://onlinelibrary.wiley.com/doi/pdf/10.1002/er.5835} \BibitemShut
  {NoStop}%
\bibitem [{\citenamefont {Zapf}\ \emph {et~al.}(2014)\citenamefont {Zapf},
  \citenamefont {Jaime},\ and\ \citenamefont {Batista}}]{RevModPhys.86.563}%
  \BibitemOpen
  \bibfield  {author} {\bibinfo {author} {\bibfnamefont {V.}~\bibnamefont
  {Zapf}}, \bibinfo {author} {\bibfnamefont {M.}~\bibnamefont {Jaime}},\ and\
  \bibinfo {author} {\bibfnamefont {C.~D.}\ \bibnamefont {Batista}},\
  }\bibfield  {title} {\bibinfo {title} {Bose-einstein condensation in quantum
  magnets},\ }\href {https://doi.org/10.1103/RevModPhys.86.563} {\bibfield
  {journal} {\bibinfo  {journal} {Rev. Mod. Phys.}\ }\textbf {\bibinfo {volume}
  {86}},\ \bibinfo {pages} {563} (\bibinfo {year} {2014})}\BibitemShut
  {NoStop}%
\bibitem [{\citenamefont {Nikuni}\ \emph {et~al.}(2000)\citenamefont {Nikuni},
  \citenamefont {Oshikawa}, \citenamefont {Oosawa},\ and\ \citenamefont
  {Tanaka}}]{PhysRevLett.84.5868}%
  \BibitemOpen
  \bibfield  {author} {\bibinfo {author} {\bibfnamefont {T.}~\bibnamefont
  {Nikuni}}, \bibinfo {author} {\bibfnamefont {M.}~\bibnamefont {Oshikawa}},
  \bibinfo {author} {\bibfnamefont {A.}~\bibnamefont {Oosawa}},\ and\ \bibinfo
  {author} {\bibfnamefont {H.}~\bibnamefont {Tanaka}},\ }\bibfield  {title}
  {\bibinfo {title} {Bose-einstein condensation of dilute magnons in
  ${\mathrm{tlcucl}}_{3}$},\ }\href
  {https://doi.org/10.1103/PhysRevLett.84.5868} {\bibfield  {journal} {\bibinfo
   {journal} {Phys. Rev. Lett.}\ }\textbf {\bibinfo {volume} {84}},\ \bibinfo
  {pages} {5868} (\bibinfo {year} {2000})}\BibitemShut {NoStop}%
\bibitem [{\citenamefont {Giamarchi}\ \emph {et~al.}(2008)\citenamefont
  {Giamarchi}, \citenamefont {R{\"u}egg},\ and\ \citenamefont
  {Tchernyshyov}}]{giamarchi2008bose}%
  \BibitemOpen
  \bibfield  {author} {\bibinfo {author} {\bibfnamefont {T.}~\bibnamefont
  {Giamarchi}}, \bibinfo {author} {\bibfnamefont {C.}~\bibnamefont
  {R{\"u}egg}},\ and\ \bibinfo {author} {\bibfnamefont {O.}~\bibnamefont
  {Tchernyshyov}},\ }\bibfield  {title} {\bibinfo {title} {Bose–einstein
  condensation in magnetic insulators},\ }\href
  {https://doi.org/10.1038/nphys893} {\bibfield  {journal} {\bibinfo  {journal}
  {Nature Physics}\ }\textbf {\bibinfo {volume} {4}},\ \bibinfo {pages} {198}
  (\bibinfo {year} {2008})}\BibitemShut {NoStop}%
\bibitem [{\citenamefont {Rice}(2002)}]{rice2002condense}%
  \BibitemOpen
  \bibfield  {author} {\bibinfo {author} {\bibfnamefont {T.}~\bibnamefont
  {Rice}},\ }\bibfield  {title} {\bibinfo {title} {To condense or not to
  condense},\ }\href {DOI: 10.1126/science.1078819} {\bibfield  {journal}
  {\bibinfo  {journal} {Science}\ }\textbf {\bibinfo {volume} {298}},\ \bibinfo
  {pages} {760} (\bibinfo {year} {2002})}\BibitemShut {NoStop}%
\bibitem [{\citenamefont {Garrett}\ \emph {et~al.}(1997)\citenamefont
  {Garrett}, \citenamefont {Nagler}, \citenamefont {Tennant}, \citenamefont
  {Sales},\ and\ \citenamefont {Barnes}}]{PhysRevLett.79.745}%
  \BibitemOpen
  \bibfield  {author} {\bibinfo {author} {\bibfnamefont {A.~W.}\ \bibnamefont
  {Garrett}}, \bibinfo {author} {\bibfnamefont {S.~E.}\ \bibnamefont {Nagler}},
  \bibinfo {author} {\bibfnamefont {D.~A.}\ \bibnamefont {Tennant}}, \bibinfo
  {author} {\bibfnamefont {B.~C.}\ \bibnamefont {Sales}},\ and\ \bibinfo
  {author} {\bibfnamefont {T.}~\bibnamefont {Barnes}},\ }\bibfield  {title}
  {\bibinfo {title} {Magnetic excitations in the
  $\mathit{S}\phantom{\rule{0ex}{0ex}}=\phantom{\rule{0ex}{0ex}}1/2$
  alternating chain compound $\text{(VO)}_{2}\text{P}_{2}\text{O}_{7}$},\
  }\href {https://doi.org/10.1103/PhysRevLett.79.745} {\bibfield  {journal}
  {\bibinfo  {journal} {Phys. Rev. Lett.}\ }\textbf {\bibinfo {volume} {79}},\
  \bibinfo {pages} {745} (\bibinfo {year} {1997})}\BibitemShut {NoStop}%
\bibitem [{\citenamefont {Nishimoto}\ \emph {et~al.}(2012)\citenamefont
  {Nishimoto}, \citenamefont {Drechsler}, \citenamefont {Kuzian}, \citenamefont
  {Richter},\ and\ \citenamefont {van~den Brink}}]{Nishimoto_2012}%
  \BibitemOpen
  \bibfield  {author} {\bibinfo {author} {\bibfnamefont {S.}~\bibnamefont
  {Nishimoto}}, \bibinfo {author} {\bibfnamefont {S.-L.}\ \bibnamefont
  {Drechsler}}, \bibinfo {author} {\bibfnamefont {R.~O.}\ \bibnamefont
  {Kuzian}}, \bibinfo {author} {\bibfnamefont {J.}~\bibnamefont {Richter}},\
  and\ \bibinfo {author} {\bibfnamefont {J.}~\bibnamefont {van~den Brink}},\
  }\bibfield  {title} {\bibinfo {title} {The effect of antiferromagnetic
  interchain coupling on multipolar phases in quasi-1d quantum helimagnets},\
  }\href {https://doi.org/10.1088/1742-6596/400/3/032069} {\bibfield  {journal}
  {\bibinfo  {journal} {Journal of Physics: Conference Series}\ }\textbf
  {\bibinfo {volume} {400}},\ \bibinfo {pages} {032069} (\bibinfo {year}
  {2012})}\BibitemShut {NoStop}%
\bibitem [{\citenamefont {Nishimoto}\ \emph {et~al.}(2015)\citenamefont
  {Nishimoto}, \citenamefont {Drechsler}, \citenamefont {Kuzian}, \citenamefont
  {Richter},\ and\ \citenamefont {van~den Brink}}]{2015Satoshi}%
  \BibitemOpen
  \bibfield  {author} {\bibinfo {author} {\bibfnamefont {S.}~\bibnamefont
  {Nishimoto}}, \bibinfo {author} {\bibfnamefont {S.-L.}\ \bibnamefont
  {Drechsler}}, \bibinfo {author} {\bibfnamefont {R.}~\bibnamefont {Kuzian}},
  \bibinfo {author} {\bibfnamefont {J.}~\bibnamefont {Richter}},\ and\ \bibinfo
  {author} {\bibfnamefont {J.}~\bibnamefont {van~den Brink}},\ }\bibfield
  {title} {\bibinfo {title} {Interplay of interchain interactions and exchange
  anisotropy: Stability and fragility of multipolar states in
  spin-$\frac{1}{2}$ quasi-one-dimensional frustrated helimagnets},\ }\href
  {https://doi.org/10.1103/PhysRevB.92.214415} {\bibfield  {journal} {\bibinfo
  {journal} {Phys. Rev. B}\ }\textbf {\bibinfo {volume} {92}},\ \bibinfo
  {pages} {214415} (\bibinfo {year} {2015})}\BibitemShut {NoStop}%
\bibitem [{\citenamefont {Sudan}\ \emph {et~al.}(2009)\citenamefont {Sudan},
  \citenamefont {L\"uscher},\ and\ \citenamefont {L\"auchli}}]{2009Sudan}%
  \BibitemOpen
  \bibfield  {author} {\bibinfo {author} {\bibfnamefont {J.}~\bibnamefont
  {Sudan}}, \bibinfo {author} {\bibfnamefont {A.}~\bibnamefont {L\"uscher}},\
  and\ \bibinfo {author} {\bibfnamefont {A.~M.}\ \bibnamefont {L\"auchli}},\
  }\bibfield  {title} {\bibinfo {title} {Emergent multipolar spin correlations
  in a fluctuating spiral: The frustrated ferromagnetic spin-$\frac{1}{2}$
  heisenberg chain in a magnetic field},\ }\href
  {https://doi.org/10.1103/PhysRevB.80.140402} {\bibfield  {journal} {\bibinfo
  {journal} {Phys. Rev. B}\ }\textbf {\bibinfo {volume} {80}},\ \bibinfo
  {pages} {140402} (\bibinfo {year} {2009})}\BibitemShut {NoStop}%
\bibitem [{\citenamefont {Hikihara}\ \emph {et~al.}(2008)\citenamefont
  {Hikihara}, \citenamefont {Kecke}, \citenamefont {Momoi},\ and\ \citenamefont
  {Furusaki}}]{2008Hikihara}%
  \BibitemOpen
  \bibfield  {author} {\bibinfo {author} {\bibfnamefont {T.}~\bibnamefont
  {Hikihara}}, \bibinfo {author} {\bibfnamefont {L.}~\bibnamefont {Kecke}},
  \bibinfo {author} {\bibfnamefont {T.}~\bibnamefont {Momoi}},\ and\ \bibinfo
  {author} {\bibfnamefont {A.}~\bibnamefont {Furusaki}},\ }\bibfield  {title}
  {\bibinfo {title} {Vector chiral and multipolar orders in the
  spin-$\frac{1}{2}$ frustrated ferromagnetic chain in magnetic field},\ }\href
  {https://doi.org/10.1103/PhysRevB.78.144404} {\bibfield  {journal} {\bibinfo
  {journal} {Phys. Rev. B}\ }\textbf {\bibinfo {volume} {78}},\ \bibinfo
  {pages} {144404} (\bibinfo {year} {2008})}\BibitemShut {NoStop}%
\bibitem [{\citenamefont {Parvej}\ and\ \citenamefont
  {Kumar}(2017)}]{2018Parvej}%
  \BibitemOpen
  \bibfield  {author} {\bibinfo {author} {\bibfnamefont {A.}~\bibnamefont
  {Parvej}}\ and\ \bibinfo {author} {\bibfnamefont {M.}~\bibnamefont {Kumar}},\
  }\bibfield  {title} {\bibinfo {title} {Multipolar phase in frustrated
  spin-1/2 and spin-1 chains},\ }\href
  {https://doi.org/10.1103/PhysRevB.96.054413} {\bibfield  {journal} {\bibinfo
  {journal} {Phys. Rev. B}\ }\textbf {\bibinfo {volume} {96}},\ \bibinfo
  {pages} {054413} (\bibinfo {year} {2017})}\BibitemShut {NoStop}%
\bibitem [{\citenamefont {Kecke}\ \emph {et~al.}(2007)\citenamefont {Kecke},
  \citenamefont {Momoi},\ and\ \citenamefont {Furusaki}}]{PhysRevB.76.060407}%
  \BibitemOpen
  \bibfield  {author} {\bibinfo {author} {\bibfnamefont {L.}~\bibnamefont
  {Kecke}}, \bibinfo {author} {\bibfnamefont {T.}~\bibnamefont {Momoi}},\ and\
  \bibinfo {author} {\bibfnamefont {A.}~\bibnamefont {Furusaki}},\ }\bibfield
  {title} {\bibinfo {title} {Multimagnon bound states in the frustrated
  ferromagnetic one-dimensional chain},\ }\href
  {https://doi.org/10.1103/PhysRevB.76.060407} {\bibfield  {journal} {\bibinfo
  {journal} {Phys. Rev. B}\ }\textbf {\bibinfo {volume} {76}},\ \bibinfo
  {pages} {060407} (\bibinfo {year} {2007})}\BibitemShut {NoStop}%
\bibitem [{\citenamefont {\textcolor{black}{Aslam Parvej and Manoranjan
  Kumar}}(2016)}]{PARVEJ201696}%
  \BibitemOpen
  \bibfield  {author} {\bibinfo {author} {\bibnamefont {\textcolor{black}{Aslam
  Parvej and Manoranjan Kumar}}},\ }\bibfield  {title} {\bibinfo {title}
  {\textcolor{black}{Degeneracies and exotic phases in an isotropic frustrated
  spin-1/2 chain}},\ }\href
  {https://doi.org/https://doi.org/10.1016/j.jmmm.2015.10.017} {\bibfield
  {journal} {\bibinfo  {journal} {Journal of Magnetism and Magnetic Materials}\
  }\textbf {\bibinfo {volume} {401}},\ \bibinfo {pages} {96} (\bibinfo {year}
  {2016})}\BibitemShut {NoStop}%
\bibitem [{\citenamefont {\textcolor{black}{Shannon, Nic and Momoi, Tsutomu and
  Sindzingre, Philippe}}(2006)}]{Shanon_2006_square_lattice}%
  \BibitemOpen
  \bibfield  {author} {\bibinfo {author} {\bibnamefont
  {\textcolor{black}{Shannon, Nic and Momoi, Tsutomu and Sindzingre,
  Philippe}}},\ }\bibfield  {title} {\bibinfo {title}
  {\textcolor{black}{Nematic Order in Square Lattice Frustrated
  Ferromagnets}},\ }\href {https://doi.org/10.1103/PhysRevLett.96.027213}
  {\bibfield  {journal} {\bibinfo  {journal} {Phys. Rev. Lett.}\ }\textbf
  {\bibinfo {volume} {96}},\ \bibinfo {pages} {027213} (\bibinfo {year}
  {2006})}\BibitemShut {NoStop}%
\bibitem [{\citenamefont {\textcolor{black}{Momoi, Tsutomu and Sindzingre,
  Philippe and Shannon, Nic}}(2006)}]{Shanon_2006_triangular_lattice}%
  \BibitemOpen
  \bibfield  {author} {\bibinfo {author} {\bibnamefont
  {\textcolor{black}{Momoi, Tsutomu and Sindzingre, Philippe and Shannon,
  Nic}}},\ }\bibfield  {title} {\bibinfo {title} {\textcolor{black}{Octupolar
  Order in the Multiple Spin Exchange Model on a Triangular Lattice}},\ }\href
  {https://doi.org/10.1103/PhysRevLett.97.257204} {\bibfield  {journal}
  {\bibinfo  {journal} {Phys. Rev. Lett.}\ }\textbf {\bibinfo {volume} {97}},\
  \bibinfo {pages} {257204} (\bibinfo {year} {2006})}\BibitemShut {NoStop}%
\bibitem [{\citenamefont {Perdew}\ \emph
  {et~al.}(1996{\natexlab{a}})\citenamefont {Perdew}, \citenamefont {Burke},\
  and\ \citenamefont {Ernzerhof}}]{perdew1996generalized}%
  \BibitemOpen
  \bibfield  {author} {\bibinfo {author} {\bibfnamefont {J.~P.}\ \bibnamefont
  {Perdew}}, \bibinfo {author} {\bibfnamefont {K.}~\bibnamefont {Burke}},\ and\
  \bibinfo {author} {\bibfnamefont {M.}~\bibnamefont {Ernzerhof}},\ }\bibfield
  {title} {\bibinfo {title} {Generalized gradient approximation made simple},\
  }\href {https://doi.org/10.1103/PhysRevLett.77.3865} {\bibfield  {journal}
  {\bibinfo  {journal} {Physical review letters}\ }\textbf {\bibinfo {volume}
  {77}},\ \bibinfo {pages} {3865} (\bibinfo {year}
  {1996}{\natexlab{a}})}\BibitemShut {NoStop}%
\bibitem [{\citenamefont {Bl\"ochl}(1994)}]{PhysRevB.50.17953}%
  \BibitemOpen
  \bibfield  {author} {\bibinfo {author} {\bibfnamefont {P.~E.}\ \bibnamefont
  {Bl\"ochl}},\ }\bibfield  {title} {\bibinfo {title} {Projector augmented-wave
  method},\ }\href {https://doi.org/10.1103/PhysRevB.50.17953} {\bibfield
  {journal} {\bibinfo  {journal} {Phys. Rev. B}\ }\textbf {\bibinfo {volume}
  {50}},\ \bibinfo {pages} {17953} (\bibinfo {year} {1994})}\BibitemShut
  {NoStop}%
\bibitem [{\citenamefont {Tackett}\ \emph {et~al.}(2001)\citenamefont
  {Tackett}, \citenamefont {Holzwarth},\ and\ \citenamefont
  {Matthews}}]{TACKETT2001348}%
  \BibitemOpen
  \bibfield  {author} {\bibinfo {author} {\bibfnamefont {A.}~\bibnamefont
  {Tackett}}, \bibinfo {author} {\bibfnamefont {N.}~\bibnamefont {Holzwarth}},\
  and\ \bibinfo {author} {\bibfnamefont {G.}~\bibnamefont {Matthews}},\
  }\bibfield  {title} {\bibinfo {title} {A projector augmented wave ({PAW})
  code for electronic structure calculations, part {II}: pwpaw for periodic
  solids in a plane wave basis},\ }\href
  {https://doi.org/https://doi.org/10.1016/S0010-4655(00)00241-1} {\bibfield
  {journal} {\bibinfo  {journal} {Computer Physics Communications}\ }\textbf
  {\bibinfo {volume} {135}},\ \bibinfo {pages} {348} (\bibinfo {year}
  {2001})}\BibitemShut {NoStop}%
\bibitem [{\citenamefont {Paier}\ \emph {et~al.}(2005)\citenamefont {Paier},
  \citenamefont {Hirschl}, \citenamefont {Marsman},\ and\ \citenamefont
  {Kresse}}]{10.1063/1.1926272}%
  \BibitemOpen
  \bibfield  {author} {\bibinfo {author} {\bibfnamefont {J.}~\bibnamefont
  {Paier}}, \bibinfo {author} {\bibfnamefont {R.}~\bibnamefont {Hirschl}},
  \bibinfo {author} {\bibfnamefont {M.}~\bibnamefont {Marsman}},\ and\ \bibinfo
  {author} {\bibfnamefont {G.}~\bibnamefont {Kresse}},\ }\bibfield  {title}
  {\bibinfo {title} {{The {P}erdew–{B}urke–{E}rnzerhof exchange-correlation
  functional applied to the {G}2-1 test set using a plane-wave basis set}},\
  }\href {https://doi.org/10.1063/1.1926272} {\bibfield  {journal} {\bibinfo
  {journal} {The Journal of Chemical Physics}\ }\textbf {\bibinfo {volume}
  {122}},\ \bibinfo {pages} {234102} (\bibinfo {year} {2005})}\BibitemShut
  {NoStop}%
\bibitem [{\citenamefont {Perdew}\ \emph
  {et~al.}(1996{\natexlab{b}})\citenamefont {Perdew}, \citenamefont {Burke},\
  and\ \citenamefont {Ernzerhof}}]{PhysRevLett.77.3865}%
  \BibitemOpen
  \bibfield  {author} {\bibinfo {author} {\bibfnamefont {J.~P.}\ \bibnamefont
  {Perdew}}, \bibinfo {author} {\bibfnamefont {K.}~\bibnamefont {Burke}},\ and\
  \bibinfo {author} {\bibfnamefont {M.}~\bibnamefont {Ernzerhof}},\ }\bibfield
  {title} {\bibinfo {title} {Generalized gradient approximation made simple},\
  }\href {https://doi.org/10.1103/PhysRevLett.77.3865} {\bibfield  {journal}
  {\bibinfo  {journal} {Phys. Rev. Lett.}\ }\textbf {\bibinfo {volume} {77}},\
  \bibinfo {pages} {3865} (\bibinfo {year} {1996}{\natexlab{b}})}\BibitemShut
  {NoStop}%
\bibitem [{\citenamefont {Dudarev}\ \emph {et~al.}(1998)\citenamefont
  {Dudarev}, \citenamefont {Botton}, \citenamefont {Savrasov}, \citenamefont
  {Humphreys},\ and\ \citenamefont {Sutton}}]{dudarev1998electron}%
  \BibitemOpen
  \bibfield  {author} {\bibinfo {author} {\bibfnamefont {S.~L.}\ \bibnamefont
  {Dudarev}}, \bibinfo {author} {\bibfnamefont {G.~A.}\ \bibnamefont {Botton}},
  \bibinfo {author} {\bibfnamefont {S.~Y.}\ \bibnamefont {Savrasov}}, \bibinfo
  {author} {\bibfnamefont {C.}~\bibnamefont {Humphreys}},\ and\ \bibinfo
  {author} {\bibfnamefont {A.~P.}\ \bibnamefont {Sutton}},\ }\bibfield  {title}
  {\bibinfo {title} {{Electron-energy-loss spectra and the structural stability
  of nickel oxide: An LSDA+ U study}},\ }\href
  {https://doi.org/10.1103/PhysRevB.57.1505} {\bibfield  {journal} {\bibinfo
  {journal} {Physical Review B}\ }\textbf {\bibinfo {volume} {57}},\ \bibinfo
  {pages} {1505} (\bibinfo {year} {1998})}\BibitemShut {NoStop}%
\bibitem [{\citenamefont {Andersen}\ and\ \citenamefont
  {Saha-Dasgupta}(2000)}]{andersen2000muffin}%
  \BibitemOpen
  \bibfield  {author} {\bibinfo {author} {\bibfnamefont {O.}~\bibnamefont
  {Andersen}}\ and\ \bibinfo {author} {\bibfnamefont {T.}~\bibnamefont
  {Saha-Dasgupta}},\ }\bibfield  {title} {\bibinfo {title} {Muffin-tin orbitals
  of arbitrary order},\ }\href {https://doi.org/10.1103/PhysRevB.62.R16219}
  {\bibfield  {journal} {\bibinfo  {journal} {Physical Review B}\ }\textbf
  {\bibinfo {volume} {62}},\ \bibinfo {pages} {R16219} (\bibinfo {year}
  {2000})}\BibitemShut {NoStop}%
\bibitem [{\citenamefont {Andersen}\ and\ \citenamefont
  {Jepsen}(1984)}]{andersen1984explicit}%
  \BibitemOpen
  \bibfield  {author} {\bibinfo {author} {\bibfnamefont {O.~K.}\ \bibnamefont
  {Andersen}}\ and\ \bibinfo {author} {\bibfnamefont {O.}~\bibnamefont
  {Jepsen}},\ }\bibfield  {title} {\bibinfo {title} {Explicit, first-principles
  tight-binding theory},\ }\href {https://doi.org/10.1103/PhysRevLett.53.2571}
  {\bibfield  {journal} {\bibinfo  {journal} {Physical Review Letters}\
  }\textbf {\bibinfo {volume} {53}},\ \bibinfo {pages} {2571} (\bibinfo {year}
  {1984})}\BibitemShut {NoStop}%
\bibitem [{\citenamefont {Davidson}(1975)}]{DAVIDSON197587}%
  \BibitemOpen
  \bibfield  {author} {\bibinfo {author} {\bibfnamefont {E.~R.}\ \bibnamefont
  {Davidson}},\ }\bibfield  {title} {\bibinfo {title} {The iterative
  calculation of a few of the lowest eigenvalues and corresponding eigenvectors
  of large real-symmetric matrices},\ }\href
  {https://doi.org/https://doi.org/10.1016/0021-9991(75)90065-0} {\bibfield
  {journal} {\bibinfo  {journal} {Journal of Computational Physics}\ }\textbf
  {\bibinfo {volume} {17}},\ \bibinfo {pages} {87} (\bibinfo {year}
  {1975})}\BibitemShut {NoStop}%
\bibitem [{\citenamefont {Davidson}\ and\ \citenamefont
  {Thompson}(1993)}]{davidson1993monster}%
  \BibitemOpen
  \bibfield  {author} {\bibinfo {author} {\bibfnamefont {E.~R.}\ \bibnamefont
  {Davidson}}\ and\ \bibinfo {author} {\bibfnamefont {W.~J.}\ \bibnamefont
  {Thompson}},\ }\bibfield  {title} {\bibinfo {title} {Monster matrices: their
  eigenvalues and eigenvectors},\ }\href
  {https://doi.org/https://doi.org/10.1063/1.4823212} {\bibfield  {journal}
  {\bibinfo  {journal} {Computers in Physics}\ }\textbf {\bibinfo {volume}
  {7}},\ \bibinfo {pages} {519} (\bibinfo {year} {1993})}\BibitemShut {NoStop}%
\bibitem [{\citenamefont {Murray}\ \emph {et~al.}(1992)\citenamefont {Murray},
  \citenamefont {Racine},\ and\ \citenamefont {Davidson}}]{MURRAY1992382}%
  \BibitemOpen
  \bibfield  {author} {\bibinfo {author} {\bibfnamefont {C.~W.}\ \bibnamefont
  {Murray}}, \bibinfo {author} {\bibfnamefont {S.~C.}\ \bibnamefont {Racine}},\
  and\ \bibinfo {author} {\bibfnamefont {E.~R.}\ \bibnamefont {Davidson}},\
  }\bibfield  {title} {\bibinfo {title} {Improved algorithms for the lowest few
  eigenvalues and associated eigenvectors of large matrices},\ }\href
  {https://doi.org/https://doi.org/10.1016/0021-9991(92)90409-R} {\bibfield
  {journal} {\bibinfo  {journal} {Journal of Computational Physics}\ }\textbf
  {\bibinfo {volume} {103}},\ \bibinfo {pages} {382} (\bibinfo {year}
  {1992})}\BibitemShut {NoStop}%
\bibitem [{\citenamefont {White}(1992)}]{white1992density}%
  \BibitemOpen
  \bibfield  {author} {\bibinfo {author} {\bibfnamefont {S.~R.}\ \bibnamefont
  {White}},\ }\bibfield  {title} {\bibinfo {title} {Density matrix formulation
  for quantum renormalization groups},\ }\href
  {https://doi.org/10.1103/PhysRevLett.69.2863} {\bibfield  {journal} {\bibinfo
   {journal} {Phys. Rev. Lett.}\ }\textbf {\bibinfo {volume} {69}},\ \bibinfo
  {pages} {2863} (\bibinfo {year} {1992})}\BibitemShut {NoStop}%
\bibitem [{\citenamefont {White}(1993)}]{white1993density}%
  \BibitemOpen
  \bibfield  {author} {\bibinfo {author} {\bibfnamefont {S.~R.}\ \bibnamefont
  {White}},\ }\bibfield  {title} {\bibinfo {title} {Density-matrix algorithms
  for quantum renormalization groups},\ }\href
  {https://doi.org/10.1103/PhysRevB.48.10345} {\bibfield  {journal} {\bibinfo
  {journal} {Phys. Rev. B}\ }\textbf {\bibinfo {volume} {48}},\ \bibinfo
  {pages} {10345} (\bibinfo {year} {1993})}\BibitemShut {NoStop}%
\bibitem [{\citenamefont {Schollw\"ock}(2005)}]{schollwock2005density}%
  \BibitemOpen
  \bibfield  {author} {\bibinfo {author} {\bibfnamefont {U.}~\bibnamefont
  {Schollw\"ock}},\ }\bibfield  {title} {\bibinfo {title} {The density-matrix
  renormalization group},\ }\href {https://doi.org/10.1103/RevModPhys.77.259}
  {\bibfield  {journal} {\bibinfo  {journal} {Rev. Mod. Phys.}\ }\textbf
  {\bibinfo {volume} {77}},\ \bibinfo {pages} {259} (\bibinfo {year}
  {2005})}\BibitemShut {NoStop}%
\bibitem [{\citenamefont {Kumar}\ \emph {et~al.}(2010)\citenamefont {Kumar},
  \citenamefont {Soos}, \citenamefont {Sen},\ and\ \citenamefont
  {Ramasesha}}]{Mkumar2010}%
  \BibitemOpen
  \bibfield  {author} {\bibinfo {author} {\bibfnamefont {M.}~\bibnamefont
  {Kumar}}, \bibinfo {author} {\bibfnamefont {Z.~G.}\ \bibnamefont {Soos}},
  \bibinfo {author} {\bibfnamefont {D.}~\bibnamefont {Sen}},\ and\ \bibinfo
  {author} {\bibfnamefont {S.}~\bibnamefont {Ramasesha}},\ }\bibfield  {title}
  {\bibinfo {title} {Modified density matrix renormalization group algorithm
  for the zigzag spin-$\frac{1}{2}$ chain with frustrated antiferromagnetic
  exchange: Comparison with field theory at large {${J}_{2}/{J}_{1}$}},\ }\href
  {https://doi.org/10.1103/PhysRevB.81.104406} {\bibfield  {journal} {\bibinfo
  {journal} {Phys. Rev. B}\ }\textbf {\bibinfo {volume} {81}},\ \bibinfo
  {pages} {104406} (\bibinfo {year} {2010})}\BibitemShut {NoStop}%
\bibitem [{SM()}]{SM}%
  \BibitemOpen
  \href@noop {} {}\bibinfo {note} {The supplementary material contains
  additional information about DFT and DMRG calculations.}\BibitemShut {Stop}%
\bibitem [{\citenamefont {Kugel'}\ and\ \citenamefont
  {Khomskiĭ}(1982)}]{Kliment}%
  \BibitemOpen
  \bibfield  {author} {\bibinfo {author} {\bibfnamefont {K.~I.}\ \bibnamefont
  {Kugel'}}\ and\ \bibinfo {author} {\bibfnamefont {D.~I.}\ \bibnamefont
  {Khomskiĭ}},\ }\bibfield  {title} {\bibinfo {title} {The jahn-teller effect
  and magnetism: transition metal compounds},\ }\href
  {https://doi.org/10.1070/PU1982v025n04ABEH004537} {\bibfield  {journal}
  {\bibinfo  {journal} {Soviet Physics Uspekhi}\ }\textbf {\bibinfo {volume}
  {25}},\ \bibinfo {pages} {231} (\bibinfo {year} {1982})}\BibitemShut
  {NoStop}%
\bibitem [{\citenamefont {Mazurenko}\ \emph {et~al.}(2006)\citenamefont
  {Mazurenko}, \citenamefont {Mila},\ and\ \citenamefont
  {Anisimov}}]{PhysRevB.73.014418}%
  \BibitemOpen
  \bibfield  {author} {\bibinfo {author} {\bibfnamefont {V.~V.}\ \bibnamefont
  {Mazurenko}}, \bibinfo {author} {\bibfnamefont {F.}~\bibnamefont {Mila}},\
  and\ \bibinfo {author} {\bibfnamefont {V.~I.}\ \bibnamefont {Anisimov}},\
  }\bibfield  {title} {\bibinfo {title} {Electronic structure and exchange
  interactions of ${\mathrm{na}}_{2}{\mathrm{v}}_{3}{\mathrm{o}}_{7}$},\ }\href
  {https://doi.org/10.1103/PhysRevB.73.014418} {\bibfield  {journal} {\bibinfo
  {journal} {Phys. Rev. B}\ }\textbf {\bibinfo {volume} {73}},\ \bibinfo
  {pages} {014418} (\bibinfo {year} {2006})}\BibitemShut {NoStop}%
\bibitem [{\citenamefont {Das}\ \emph {et~al.}(2008)\citenamefont {Das},
  \citenamefont {Waghmare}, \citenamefont {Saha-Dasgupta},\ and\ \citenamefont
  {Sarma}}]{PhysRevLett.100.186402}%
  \BibitemOpen
  \bibfield  {author} {\bibinfo {author} {\bibfnamefont {H.}~\bibnamefont
  {Das}}, \bibinfo {author} {\bibfnamefont {U.~V.}\ \bibnamefont {Waghmare}},
  \bibinfo {author} {\bibfnamefont {T.}~\bibnamefont {Saha-Dasgupta}},\ and\
  \bibinfo {author} {\bibfnamefont {D.~D.}\ \bibnamefont {Sarma}},\ }\bibfield
  {title} {\bibinfo {title} {Electronic structure, phonons, and dielectric
  anomaly in ferromagnetic insulating double pervoskite
  ${\mathrm{la}}_{2}{\mathrm{nimno}}_{6}$},\ }\href
  {https://doi.org/10.1103/PhysRevLett.100.186402} {\bibfield  {journal}
  {\bibinfo  {journal} {Phys. Rev. Lett.}\ }\textbf {\bibinfo {volume} {100}},\
  \bibinfo {pages} {186402} (\bibinfo {year} {2008})}\BibitemShut {NoStop}%
\bibitem [{\citenamefont {Kumar}\ \emph {et~al.}(2007)\citenamefont {Kumar},
  \citenamefont {Ramasesha}, \citenamefont {Sen},\ and\ \citenamefont
  {Soos}}]{PhysRevB.75.052404}%
  \BibitemOpen
  \bibfield  {author} {\bibinfo {author} {\bibfnamefont {M.}~\bibnamefont
  {Kumar}}, \bibinfo {author} {\bibfnamefont {S.}~\bibnamefont {Ramasesha}},
  \bibinfo {author} {\bibfnamefont {D.}~\bibnamefont {Sen}},\ and\ \bibinfo
  {author} {\bibfnamefont {Z.~G.}\ \bibnamefont {Soos}},\ }\bibfield  {title}
  {\bibinfo {title} {Scaling exponents in spin-$\frac{1}{2}$ heisenberg chains
  with dimerization and frustration studied with the density-matrix
  renormalization group},\ }\href {https://doi.org/10.1103/PhysRevB.75.052404}
  {\bibfield  {journal} {\bibinfo  {journal} {Phys. Rev. B}\ }\textbf {\bibinfo
  {volume} {75}},\ \bibinfo {pages} {052404} (\bibinfo {year}
  {2007})}\BibitemShut {NoStop}%
\bibitem [{foo()}]{footnote}%
  \BibitemOpen
  \bibfield  {title} {\bibinfo {title} {\textcolor{black}{We prefer to describe
  the quasi-particles in our system as magnons rather than triplons. In the
  present case, as alternation is very weak, each of the weakly dimerized chain
  has very small singlet-triplet gap and higher excitations forming continuum.
  The resultant quasi-particles thus have very well defined dispersion curve,
  even in absence of inter-chain interactions. Therefore, the quasi-particle in
  the discussed case, following the spirit of the earlier papers (Phys. Rev.
  Lett. 77, 3649 (1996); Phys. Rev. B 55, 15048, (1997), Physical Review B 67,
  054414 (2003)), should be called magnon.}},\ }\href@noop {} {\ }\BibitemShut
  {NoStop}%
\bibitem [{\citenamefont {Islam}\ \emph {et~al.}(2024)\citenamefont {Islam},
  \citenamefont {Mukharjee}, \citenamefont {Biswas}, \citenamefont {Telling},
  \citenamefont {Skourski}, \citenamefont {Ranjith}, \citenamefont {Baenitz},
  \citenamefont {Inagaki}, \citenamefont {Furukawa}, \citenamefont {Tsirlin},\
  and\ \citenamefont {Nath}}]{PhysRevB.109.L060406}%
  \BibitemOpen
  \bibfield  {author} {\bibinfo {author} {\bibfnamefont {S.~S.}\ \bibnamefont
  {Islam}}, \bibinfo {author} {\bibfnamefont {P.~K.}\ \bibnamefont
  {Mukharjee}}, \bibinfo {author} {\bibfnamefont {P.~K.}\ \bibnamefont
  {Biswas}}, \bibinfo {author} {\bibfnamefont {M.}~\bibnamefont {Telling}},
  \bibinfo {author} {\bibfnamefont {Y.}~\bibnamefont {Skourski}}, \bibinfo
  {author} {\bibfnamefont {K.~M.}\ \bibnamefont {Ranjith}}, \bibinfo {author}
  {\bibfnamefont {M.}~\bibnamefont {Baenitz}}, \bibinfo {author} {\bibfnamefont
  {Y.}~\bibnamefont {Inagaki}}, \bibinfo {author} {\bibfnamefont
  {Y.}~\bibnamefont {Furukawa}}, \bibinfo {author} {\bibfnamefont {A.~A.}\
  \bibnamefont {Tsirlin}},\ and\ \bibinfo {author} {\bibfnamefont
  {R.}~\bibnamefont {Nath}},\ }\bibfield  {title} {\bibinfo {title} {Repulsive
  tomonaga-luttinger liquid in the quasi-one-dimensional alternating
  spin-$\frac{1}{2}$ antiferromagnet ${\mathrm{navopo}}_{4}$},\ }\href
  {https://doi.org/10.1103/PhysRevB.109.L060406} {\bibfield  {journal}
  {\bibinfo  {journal} {Phys. Rev. B}\ }\textbf {\bibinfo {volume} {109}},\
  \bibinfo {pages} {L060406} (\bibinfo {year} {2024})}\BibitemShut {NoStop}%
\bibitem [{\citenamefont {Joseph W.~Bennett}\ and\ \citenamefont
  {Mason}(2019)}]{Uvalue}%
  \BibitemOpen
  \bibfield  {author} {\bibinfo {author} {\bibfnamefont {I.~K. M. D. L. S. S.
  Q.~C.}\ \bibnamefont {Joseph W.~Bennett}, \bibfnamefont {Blake G.~Hudson}}\
  and\ \bibinfo {author} {\bibfnamefont {S.~E.}\ \bibnamefont {Mason}},\
  }\bibfield  {title} {\bibinfo {title} {A systematic determination of hubbard
  u using the gbrv ultrasoft pseudopotential set},\ }\href
  {https://doi.org/10.1016/j.commatsci.2019.109137} {\bibfield  {journal}
  {\bibinfo  {journal} {Computational Materials Science}\ }\textbf {\bibinfo
  {volume} {170}},\ \bibinfo {pages} {109137} (\bibinfo {year}
  {2019})}\BibitemShut {NoStop}%
\end{thebibliography}%
\end{document}


\title{Supplemental Material: Microscopic Modeling, Multi-magnon condensation and Strain effect in chain compound NaVOPO$_4$}

\author{Manoj Gupta$^{a}$}
\thanks{$a$ The authors contributed equally to this work.}
 \affiliation{Department of Condensed Matter and Materials Physics,
S. N. Bose National Centre for Basic Sciences, Kolkata 700106, India}
 \author{Manodip Routh$^{a}$}
 \affiliation{Department of Condensed Matter and Materials Physics,
S. N. Bose National Centre for Basic Sciences, Kolkata 700106, India}
 \author{Manoranjan Kumar}
 \email{manoranjan.kumar@bose.res.in}
 \affiliation{Department of Condensed Matter and Materials Physics,
S. N. Bose National Centre for Basic Sciences, Kolkata 700106, India}
  \author{Tanusri Saha Dasgupta}
 \email{t.sahadasgupta@gmail.com}
 \affiliation{Department of Condensed Matter and Materials Physics,
S. N. Bose National Centre for Basic Sciences, Kolkata 700106, India}
\date{\today}

\maketitle
\section{GGA+$U$ Electronic structure}
 \begin{figure*}
		\centering
  		\includegraphics[width=1.0\columnwidth]{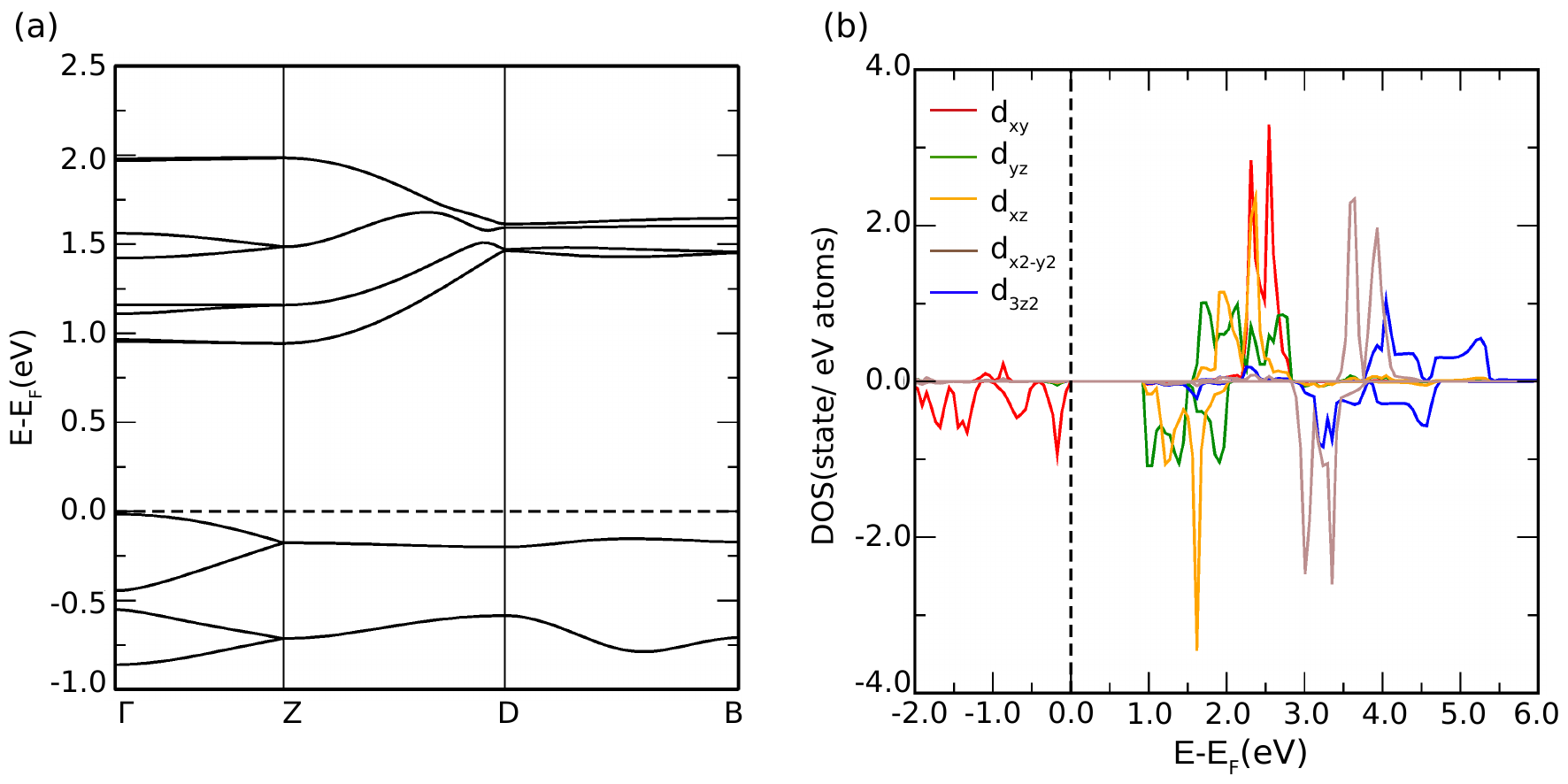}
		\caption{GGA+U band-structure (a) and density of states (b) for $U$ = 4eV and 
  J$_H$ = 1eV}
		\label{fig:GGA+U}
\end{figure*} 

Figs ~\ref{fig:GGA+U}(a) and (b) show the band-structure and density of states calculated with choice of exchange-correlation functional GGA supplemented by Hubbard $U$. Upon application of $U$, in comparison of GGA bands, the $d_{xy}$ bands get pushed below the Fermi-energy and becomes fully occupied, while the $d_{xz}/d_{yz}$ bands get the moved above the Fermi-level, becoming fully unoccupied. This leads to a insulating solution with a gap of $\sim$ 1 eV for
$U$= 4 eV.

\section{$t_{2g}$ Hopping processes}
	
 \begin{figure*}
		\centering
  		\includegraphics[width=0.8\columnwidth]{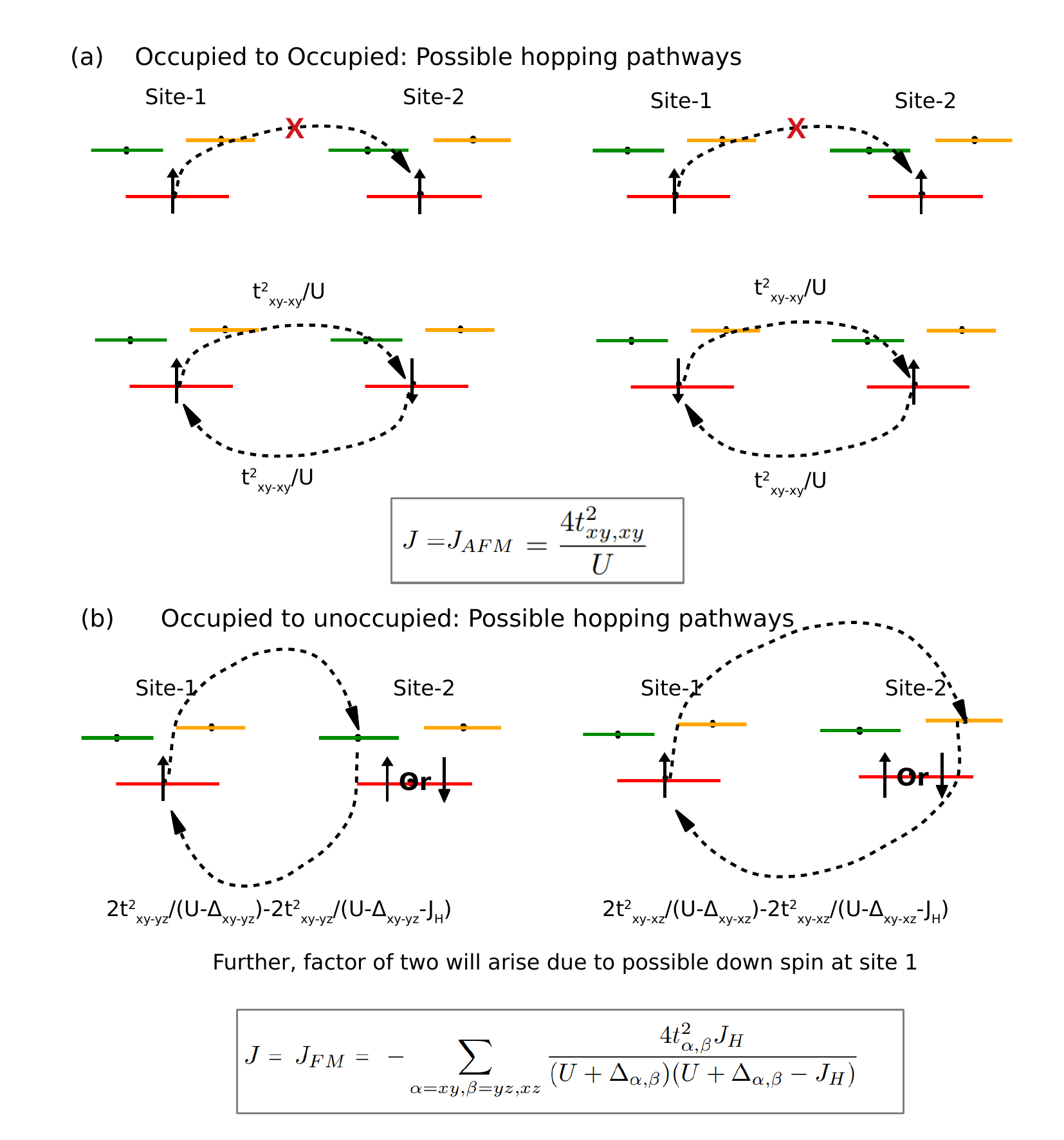}
		\caption{Hopping processes. Red, green and orange represent $d_{xy}$, $d_{yz}$, and $d_{xz}$ energy levels, respectively.}
		\label{fig:hooping_process}
\end{figure*} 

Our minimal model suggest that $t_{2g}$ bands are sufficient to capture the low energy physics. The fitting of these bands give us the tight-binding Hamiltonian and hopping matrices, thus a quantitative estimates of electronic hopping from one $V^{+4}$ site to neighbouring $V^{+4}$ site. Employing the second order perturbation theory, we calculated the exchange-integrals which are just equal to the energy difference arising due to possible electronic hopping processes. Electrons in $d_{xy}$ level can hop in two possible ways (1) Occupied level to occupied level (2) Occupied to unoccupied level as presented schematically in Fig~\ref{fig:hooping_process}.
The first case will only be allowed only if two sites are aligned in antiferromagnetically (AFM) due to Pauli-exclusion principle leading to AFM type of interaction. This will cost Coulomb energy $U$ and gain in energy by $t^2_{xy-xy}$. Whereas for the latter case, electron can hop with parallel and anti-parallel spin alignment from $d_{xy}$ level to $d_{yz}$ and $d_{yz}$ orbital. The anti-parallel and parallel alignment will cost $U$ coulomb energy and charge-transfer energy $\Delta$. The parallel spin-alignment will be favoured by Hund's coupling with energy J$_{H}$ leading to a overall FM type of exchange-interaction.    

\newpage
\section{Spin gap from DMRG}

The spin gap is defined as $\Delta = E(S^Z = 1) - E(S^Z = 0)$, where $E(S^Z=0)$ is the ground state energy and $E(S^Z=1)$ is the first excited state energy. 

\begin{figure*}[h]
		\centering
  		\includegraphics[width=0.9\columnwidth]{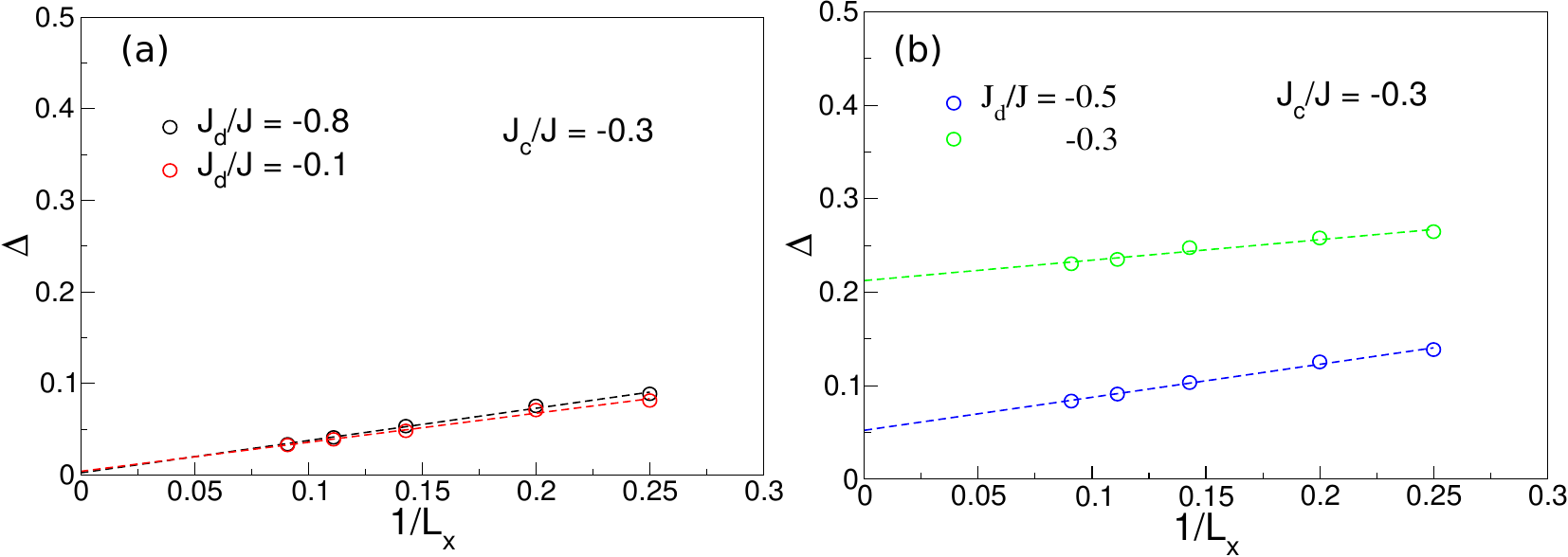}
		\caption{Extrapolation of spin gap ($\Delta$) for fixed $J^{'}/J=0.92$. (a) In staggered phase at $J_c/J=-0.3$, $J_d/J = -0.8$ and in stripy phase at $J_c/J=-0.3$, $J_d/J = -0.1$. (b) In the disorder phase. The extrapolated spin gap is $\approx 0.2 J$ for $J_c/J=-0.3$, $J_d/J = -0.3$, and $\approx 0.05 J$ for $J_c/J=-0.3$, $J_d/J = -0.5$. }
		\label{fig:spingap}
\end{figure*}

Fig.~\ref{fig:spingap}(a) shows the system size dependence of spin gap in the staggered phase at $J_c/J=-0.3$, $J_d/J = -0.8$ and in the stripy phase at $J_c/J=-0.3$, $J_d/J = -0.1$ for fixed $J^{'}/J=0.92$. The extrapolated gap values in these phases are found to be $\approx 0.001 J$. These extrapolations imply that these two phases are gapless in nature at the thermodynamic limit. The spin gap extrapolation is performed on the YC geometry with a fixed $L_{y}=4$ while varying $L_x$, as the long-range FM order forms along the diagonal or Y direction in staggered or stripy phases, respectively. Fig.~\ref{fig:spingap}(b) shows the system size dependence of spin gap in the disorder phase. As opposed to staggered and stripy phases, the extrapolated spin gap is found to be finite, $\approx 0.2 J$ for $J_c/J=-0.3$, $J_d/J = -0.3$, and $\approx 0.05 J$ for $J_c/J=-0.3$, $J_d/J = -0.5$ with a fixed $J^{'}/J=0.92$. Thus the disorder phase exhibits gapped nature at the thermodynamic limit. The spin gap extrapolation is performed on the XC geometry with a fixed $L_{y}=4$ while varying $L_x$ as this phase is highly disordered along Y and diagonal direction and has a short range AFM order along $J-J^{'}$ chain. It is also noteworthy that in this phase, the spin gap reaches its maximum at the $J_c = J_d$ line, and decreases as we move away from $J_c=J_d$.

